\documentclass[extra]{geopros}

\usepackage{amsmath}
\usepackage{amssymb}
\usepackage{amsfonts}
\usepackage{epsfig, color, ulem}
\usepackage{graphicx}

\usepackage{setspace}

\def\rev#1{\textcolor{black}{#1}}

\newlength{\dinwidth}
\newlength{\dinmargin}
\def\refname{REFERENCES}
\usepackage{lineno}

\begin{document}
\bibliographystyle{geopros}

\begin{spacing}{1}

\title{Multiple reflections on Huygens' principle}
\author{\small Kees Wapenaar \,\,\\
Department of Geoscience and Engineering, Delft University of Technology,  The Netherlands}

\date{\today}

\maketitle

\begin{abstract}{\small
According to Huygens' principle, all points on a wave front act as secondary sources emitting spherical waves, and the \rev{envelope} 
of these spherical waves forms a new wave front.
In the mathematical formulation of Huygens' principle, the waves emitted by the secondary sources are represented by Green's functions.
In many present-day applications of Huygens' principle, these Green's functions are replaced by their time-reversed versions, thus
 forming a basis for backpropagation, imaging, inversion, seismic interferometry, etc. 
However, when the input wave field is available only on a single open boundary, this approach has its limitations. In particular, it does not properly account for \rev{multiply reflected waves}.
This is remedied by a modified form of Huygens' principle, in which the Green's functions are replaced by focusing functions.
The modified Huygens' principle forms a basis for imaging, inverse scattering, monitoring of induced sources, etc., thereby properly taking \rev{multiply reflected waves} into account.}
\end{abstract}


\section{Introduction}
Dutch mathematician, physicist and astronomer Christiaan Huygens (1629 -- 1695) described light as a longitudinal mechanical wave, propagating through an ether medium.
Even though, centuries later, Maxwell proposed light as a transverse electromagnetic wave and Einstein showed it doesn't need an ether to support its propagation,
the early wave theoretical approach of Huygens appeared very effective in the analysis of the propagation and reflection of light.
In his book ``Trait\'e de la Lumi\`ere'' (Treatise on Light, 
published in 1690), 
he explains that around each undulating particle of the matter through which a wave propagates, a spherical wave is formed of which this particle is the center.
\rev{The common tangent (or envelope) of these spherical waves forms a new wave front.}
This is in a nutshell Huygens' principle, and it applies to light as well as to other wave phenomena.
For an extensive discussion of the work of Huygens and his important role in bridging ancient and modern science, see \citet{Moser2024Book}.

\rev{In the early nineteenth century, French physicist Augustin-Jean Fresnel (1788 -- 1827) added the theory of interference to Huygens' principle. 
With this extension, the new wave front along the envelope of aforementioned spherical waves can be explained as the result of constructive interference of these spherical waves.
In the following, when we speak ``Huygens' principle'', we mean the original theory of Huygens, extended with the theory of interference.}

\begin{figure}
\centerline{\hspace{2cm}\epsfysize=4.9 cm\epsfbox{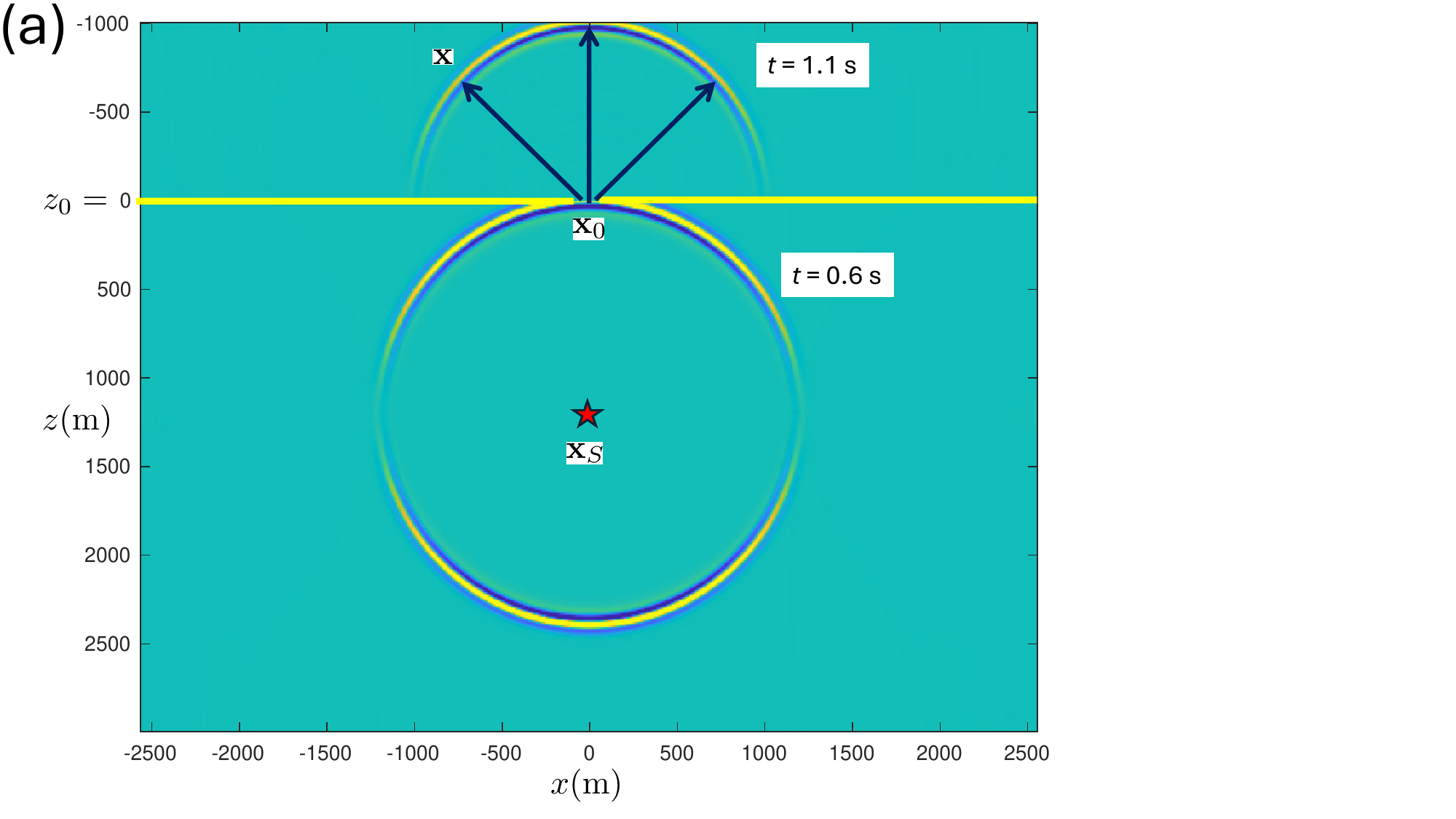}}
\centerline{\hspace{2cm}\epsfysize=4.9 cm\epsfbox{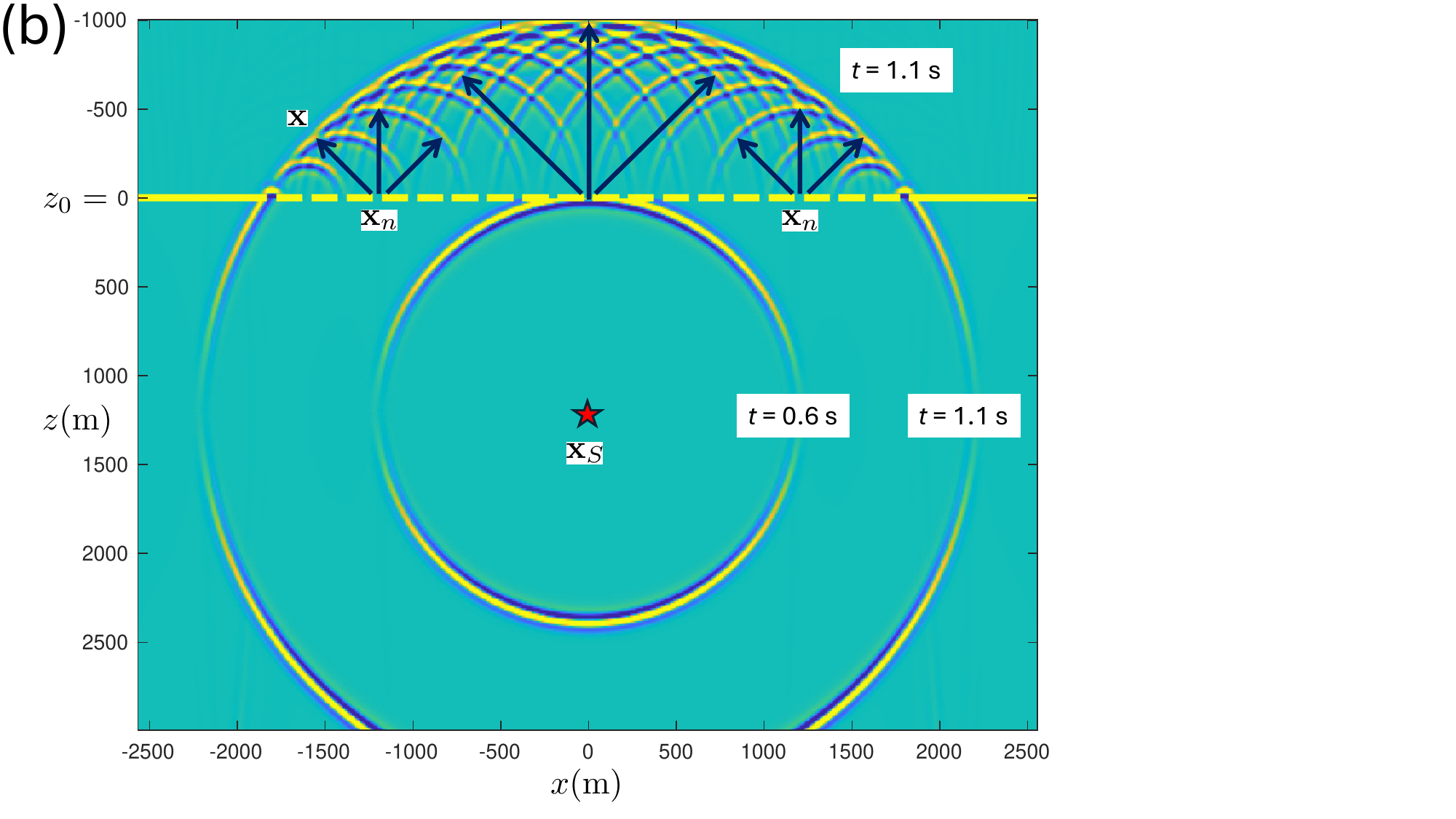}}
\centerline{\hspace{2cm}\epsfysize=4.9 cm\epsfbox{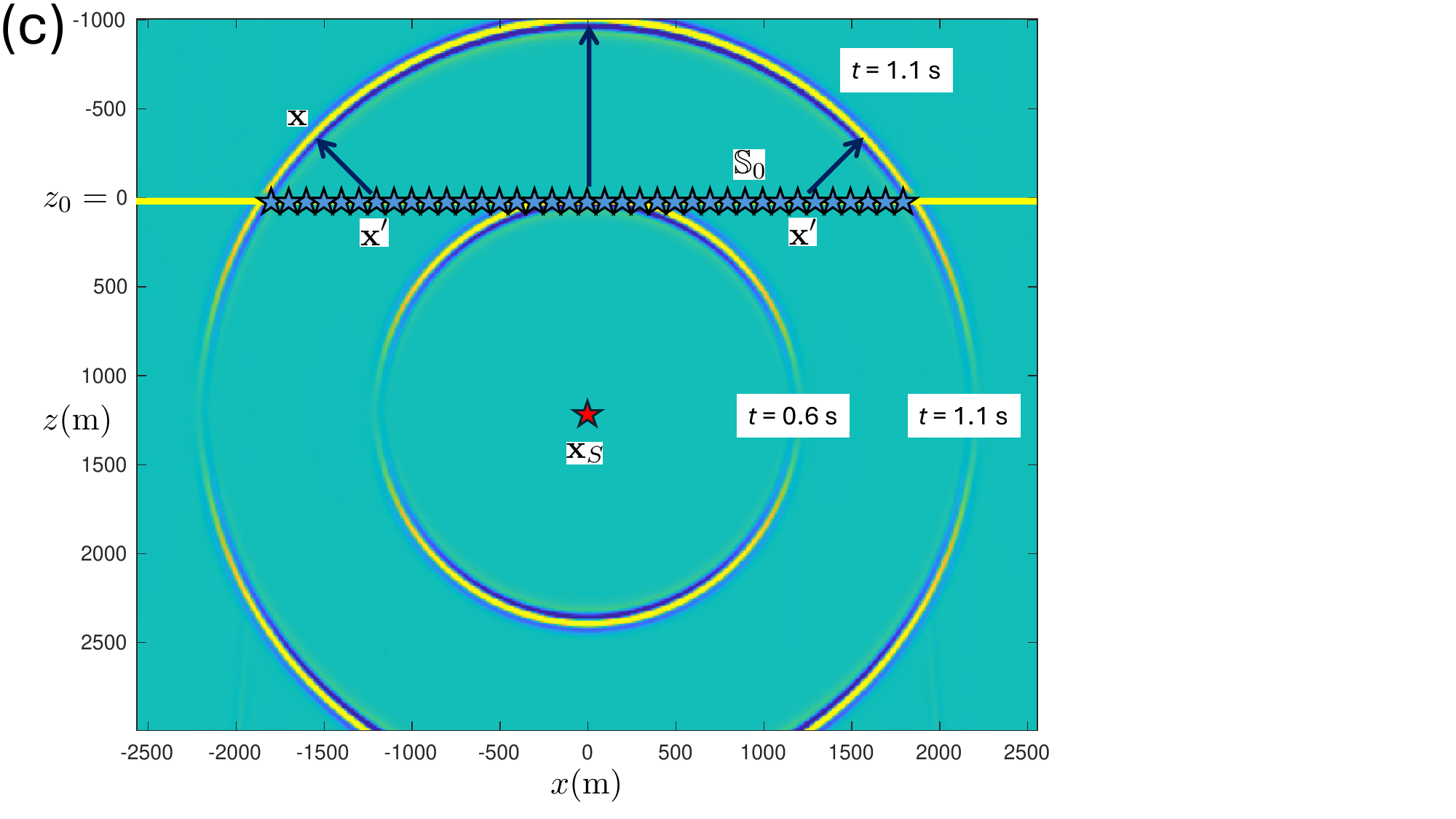}}
\caption{\it  Illustration of Huygens' principle for acoustic waves in a homogeneous medium.
\rev{(a) A source at ${\bf x}_S$ emits a circular wave. The wave field in the opening of the screen at ${\bf x}_0$ acts as a secondary source, emitting a secondary circular wave into the upper half-space. 
(b) Many openings in the screen, acting as secondary sources. The envelope of the superposition of the secondary 
circular waves in the upper half-space approaches the circular wave, originating from the source at ${\bf x}_S$.
(c) All points in one large opening in the screen act as secondary sources (indicated by the blue stars). 
The superposition of the secondary circular waves in the upper half-space has converged to the circular wave, originating from the source at ${\bf x}_S$.
Animations of this and other figures can be found in the Supporting Information.}}\label{Fig1}
\end{figure}

Figure \ref{Fig1} is an illustration of Huygens' principle, applied to acoustic waves. A point source, indicated by the red star, 
emits a circular wave which propagates through a medium with a constant propagation velocity (the example is 2D, hence, instead of spherical waves we have circular waves). 
At a certain time this wave reaches a screen with a small opening (Figure \ref{Fig1}a).
The wave field in this opening acts as a secondary source, which emits a circular wave  into the half-space above the screen. 
Figure \ref{Fig1}b shows a similar setup, but this time the screen has many small openings, which all act as secondary sources, 
emitting circular waves at the time the original wave reaches these openings. Hence, the field above the screen consists of a superposition of circular waves. 
The envelope of these superposed waves approximately forms a circular wave, resembling the wave that would be radiated by the original point source 
\rev{into the upper half-space} in absence of the screen. 
In Figure \ref{Fig1}c the screen contains one large opening. All points in this opening act as secondary sources 
(indicated by the dense distribution of blue stars)
and the superposition of the circular waves above the screen has indeed converged to the circular wave radiated by the original source.

Huygens' principle has found many applications in optics, acoustics and other fields in which wave propagation and scattering plays a role. 
In this paper we restrict ourselves to applications in acoustics and geophysics, in particular for wave field extrapolation.
Traditionally, the waves emitted by the secondary sources in Huygens' principle are represented by Green's functions.
\rev{In the section ``Traditional Huygens' principle, using Green's functions'', we review} applications of 
Huygens' principle in forward and inverse wave field extrapolation
\rev{through homogeneous and inhomogeneous media}.
It appears that with the traditional Huygens' principle, \rev{internal multiply reflected waves} are not correctly handled in inverse extrapolation through an inhomogeneous medium.
\rev{In the section ``Modified Huygens' principle, using focusing functions'', we first introduce focusing functions for homogeneous and inhomogeneous media. Next, we} 
discuss a modified version of Huygens' principle, in which the Green's functions are replaced by \rev{these} focusing functions.
We \rev{discuss} applications of this modified Huygens' principle \rev{in forward and inverse wave field extrapolation through an inhomogeneous medium and in the retrieval of the homogeneous Green's function of
an inhomogeneous medium. We show that internal multiply reflected waves are correctly handled in these applications.} 

The style of the main text is informal, with an emphasis on explanations of the different forms of Huygens' principle, using simple mathematics. 
More detailed  derivations can be found in the appendices.

\section{Traditional Huygens' principle, using Green's functions}

\subsection{Forward wave field extrapolation through a homogeneous medium}

We discuss some mathematics behind Huygens' principle, as illustrated in Figure \ref{Fig1}, and use this as a starting point for the discussion of forward wave field extrapolation. 
We define a Cartesian coordinate system, with the $z$-axis pointing downward and coordinate vector ${\bf x}$ denoting position in this system.
For the 3D situation this vector is defined as ${\bf x}=(x,y,z)$. Whereas most of the theory in this paper holds for 3D, the examples are 2D, in which case the coordinate vector
is defined as ${\bf x}=(x,z)$. Time is denoted by $t$.


Let ${\bf x}_S$ denote the position of a monopole source (in Figure \ref{Fig1} it is defined as ${\bf x}_S=(0,1200)$ m).
We define the acoustic Green's function $G({\bf x},{\bf x}_S,t)$ (named after George Green,  1793 -- 1841) 
as the response to an impulsive monopole source at ${\bf x}_S$ and $t=0$, observed at ${\bf x}$ as a function of $t$.
The Green's function is a causal function of time, meaning $G({\bf x},{\bf x}_S,t)=0$ for $t<0$.
The source of the Green's function is a volume-injection rate source, see Appendix A--1 for further details. 
When the source in an actual situation is not an impulse but a transient wavelet $s(t)$, 
then the observed acoustic pressure  is given by the convolution of the Green's function with the wavelet,
according to
\begin{equation}
p({\bf x},t)=\int_0^\infty G({\bf x},{\bf x}_S,t')s(t-t'){\rm d}t'.\label{eqcon1}
\end{equation}
To simplify the notation, here and in subsequent sections, we introduce the convolutional symbol $*$ and replace the integral notation of equation (\ref{eqcon1}) by
\begin{equation}
p({\bf x},t)= G({\bf x},{\bf x}_S,t)*s(t).\label{eqcon2}
\end{equation}
The wave fronts in Figure \ref{Fig1} below the screen are described by this equation. The source function $s(t)$ is a Ricker wavelet with a central frequency of 20 Hz
and $G({\bf x},{\bf x}_S,t)$ is the 2D Green's function in a homogeneous lossless medium with 
propagation velocity $c=2000$ m/s and mass density $\rho=1000$ kg/m$^3$ (hence, the wavelength at the central frequency is 100 m). 
The amplitudes along the wave fronts are tapered at large propagation angles (relative to the vertical axis) and waves reflected by the screen are not shown. 
Let ${\bf x}_0=(x_0,z_0)$ denote the position of the opening in the screen in Figure \ref{Fig1}a, with $z_0$ being the depth level of the screen 
(with $z_0=0$ m here and in subsequent figures).
According to Huygens' principle, the acoustic pressure at this position, $p({\bf x}_0,t)$, acts as a secondary source for the wave 
field above the screen, hence, analogous to equation (\ref{eqcon2}), this is given by 
\begin{equation}
p({\bf x},t)\propto G({\bf x},{\bf x}_0,t)*p({\bf x}_0,t),\label{eqHuyg1}
\end{equation}
where the symbol $\propto$ means ``proportional to''. Next, let ${\bf x}_n=(n\Delta x,z_0)$, 
$n=-N, \dots ,-1, 0,1, \dots , N$, denote the positions of the openings in the screen in Figure \ref{Fig1}b
(with $N=9$ and $\Delta x=200$ m).
Then, according to Huygens' superposition principle, the wave field in the half-space above the screen can be expressed as
\begin{equation}
p({\bf x},t)\propto \sum_{n=-N}^N G({\bf x},{\bf x}_n,t)*p({\bf x}_n,t).\label{eqHuyg2}
\end{equation}
Next, for the situation of one large opening in the screen, as in Figure \ref{Fig1}c, 
we reduce the distance between the secondary sources to $\Delta x=10$ m. Since this is significantly smaller than the central wavelength of 100 m, 
we now have effectively a continuum of 
secondary sources and we replace the summation by an integration, according to
\begin{equation}
p({\bf x},t)\propto \int_{\mathbb{S}_0}G({\bf x},{\bf x}',t)*p({\bf x}',t){\rm d}{\bf x}',\label{eqHuyg3}
\end{equation}
where $\mathbb{S}_0$ denotes the integration boundary (the opening in the screen as in Figure \ref{Fig1}c, or an infinite horizontal boundary in absence of the screen).
For the 2D situation considered here, this is a 1D integral over $x'$; for the 3D situation it is a 2D integral over $x'$ and $y'$. In both cases $z'$ is fixed and equal to $z_0$, 
being the depth of integration boundary $\mathbb{S}_0$.

\begin{figure}
\centerline{\hspace{2cm}\epsfysize=4.9 cm\epsfbox{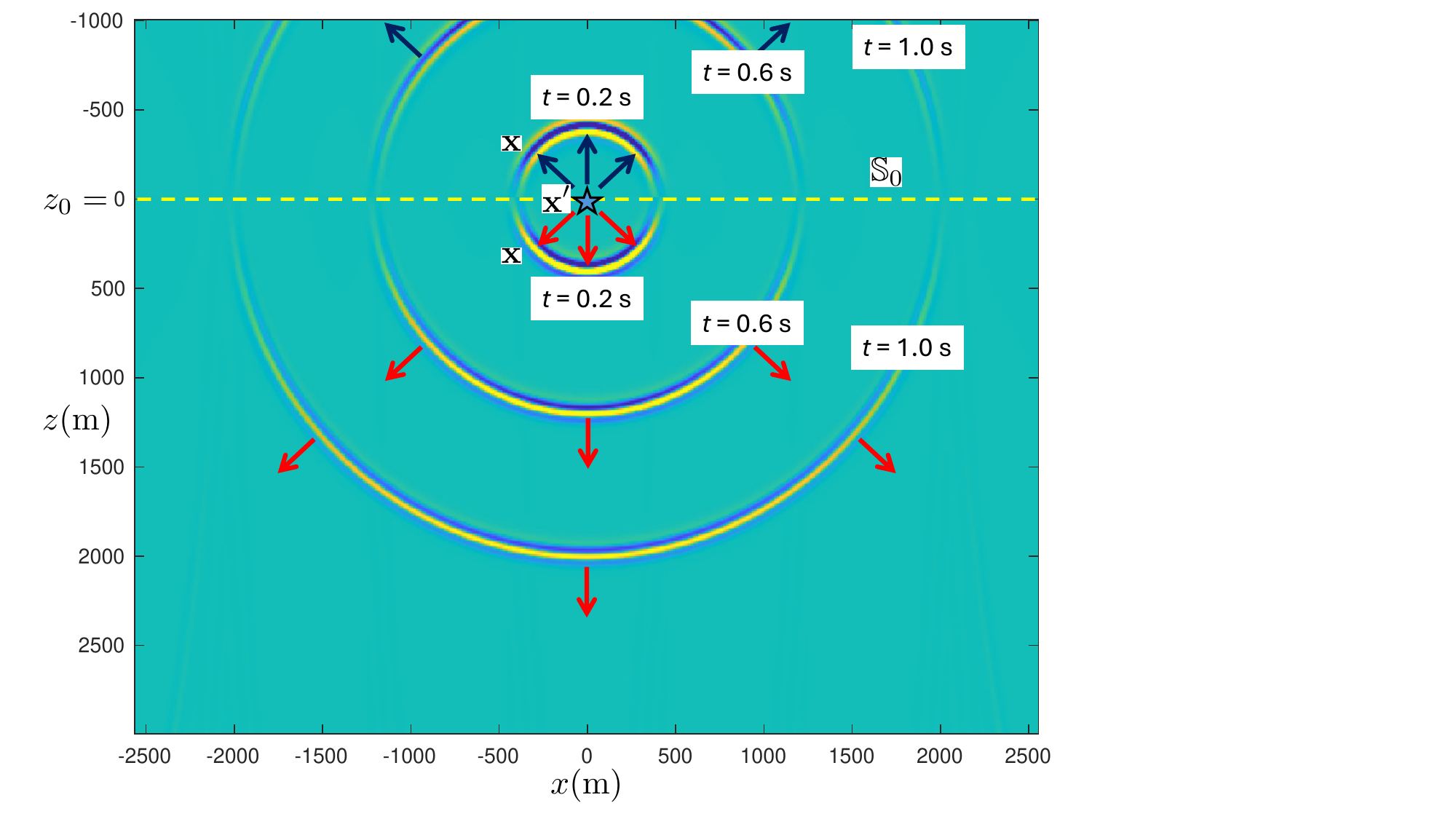}}
\caption{\it  Dipole Green's function $G_{\rm d}({\bf x},{\bf x}',t)$ (convolved with a Ricker wavelet to get a nicer display) 
in a homogeneous medium, for a dipole at ${\bf x}'$ on $\mathbb{S}_0$ at depth $z_0$.
\rev{In the mathematical formulation of 
Huygens' principle, this dipole Green's function describes the propagation from a secondary source at ${\bf x}'$ to an observation point ${\bf x}$, see equation (\ref{eqHuyg4}).}}\label{Fig2}  
\end{figure}

Up to this point, we captured the physical arguments of Huygens \rev{and Fresnel} in mathematical form.
Using only physical arguments, the proportionality factor remains unknown. In the 19th century, Kirchhoff, Helmholtz, Rayleigh and others 
derived expressions which formalize Huygens' principle. In Appendix B we summarize their derivation and obtain the following more precise form of equation (\ref{eqHuyg3})
\begin{equation}
p({\bf x},t)=-2\int_{\mathbb{S}_0}G_{\rm d}({\bf x},{\bf x}',t)*p({\bf x}',t){\rm d}{\bf x}',\label{eqHuyg4}
\end{equation}
for ${\bf x}$ above $\mathbb{S}_0$.
Here $G_{\rm d}({\bf x},{\bf x}',t)$ is the response, \rev{observed at ${\bf x}$ and $t$,} to an impulsive dipole source at ${\bf x}'$ and $t=0$, with ${\bf x}'$ on $\mathbb{S}_0$, see Figure \ref{Fig2}
(actually we already used this dipole Green's function in generating the example in Figure \ref{Fig1}).
This dipole Green's function is further specified in Appendix A--3. 
The minus-sign in equation (\ref{eqHuyg4}) stems from the definition of the dipole (it is oriented with respect to the positive $z$-axis, whereas in equation (\ref{eqHuyg4}) it radiates
in the negative $z$-direction). The factor 2 in equation (\ref{eqHuyg4}) is explained later.

Huygens' wave-theoretical description of light was not immediately accepted. 
One of the reasons was that it does not explain why the secondary sources radiate only forward: 
if each point in a wave field acts as a secondary source, one would expect it to radiate in all directions (like the dipole source of the Green's function in Figure \ref{Fig2}). 
Consequently, the envelope of the superposed waves of all secondary sources on a plane would consist of two contributions: 
one propagating forward, in the direction of the original wave, and one propagating backward, against the direction of the original wave. 
In the time of Huygens it was not clear why the secondary sources do not give rise to this backward propagating wave. 
This was seen as a serious drawback of Huygens' wave-theoretical approach. 
Newton's competing theory (light consisting of particles moving along straight lines) did not have this drawback, but it had other shortcomings, 
such as not explaining diffraction and interference. All in all, Huygens' wave theory has withstood the test of time.

To understand why the secondary sources generate only the forward propagating wave, consider the Kirchhoff-Helmholtz integral in equation ($B$-8). This integral contains
monopole and dipole Green's functions, driven by the particle velocity and acoustic pressure, respectively, at the horizontal boundary $\mathbb{S}_0$. 
Equation ($B$-8) states that the combination of secondary monopole and dipole responses yields the forward propagating wave in the half-space above $\mathbb{S}_0$,
whereas their contributions cancel in the half-space below $\mathbb{S}_0$.
Moreover, for the homogeneous medium configuration of Figure \ref{Fig1}, the secondary monopoles and dipoles give equal contributions to the wave field above $\mathbb{S}_0$, 
so one of the terms can be omitted and the other term doubled, yielding equation (\ref{eqHuyg4}) and explaining the factor 2 in this equation. Whereas equation ($B$-8)
holds for ${\bf x}$ at either side of $\mathbb{S}_0$, equation (\ref{eqHuyg4}) is only valid for ${\bf x}$ above $\mathbb{S}_0$.

Huygens' principle was developed in the pre-industrial age. At the time it was merely meant to explain the physics of wave propagation.
Technological developments in the 20th century enabled many other interesting 
applications of Huygens' principle. For example, in equation (\ref{eqHuyg4}), the wave field $p({\bf x}',t)$ at $\mathbb{S}_0$ 
can be replaced by electric signals that are fed to a dense array of
piezoelectric transducers which emit ultrasound. Equation (\ref{eqHuyg4}) then describes the synthesized wave field emitted by the array
into the half-space above $\mathbb{S}_0$. On the other hand, digitized measurements
of an acoustic or seismic wave field $p({\bf x}',t)$ at $\mathbb{S}_0$ can be fed to a computer and equation (\ref{eqHuyg4}) 
can be evaluated numerically to compute the wave field $p({\bf x},t)$ 
at any position ${\bf x}$ in the half-space above $\mathbb{S}_0$. The latter application is  wave field extrapolation \citep{Berkhout85Book}.
In practice, the convolution along the time coordinate is often replaced by 
a multiplication in the frequency domain, see equation ($B$-9), but for clarity we keep our expressions in the time domain 
because this appeals better to the physics of Huygens' principle.

In equation (\ref{eqHuyg4}), ${\bf x}$ is assumed to be situated in the half-space above $\mathbb{S}_0$, 
whereas the source (or source distribution) of the wave field $p({\bf x}',t)$ resides
 in the half-space below $\mathbb{S}_0$. For this situation we speak of {\it forward} wave field extrapolation, since the direction of extrapolation (upward from $\mathbb{S}_0$ to
 ${\bf x}$ above $\mathbb{S}_0$) corresponds to the direction of the upgoing wave field $p({\bf x}',t)$ at $\mathbb{S}_0$. Since this is the direction 
 of the negative $z$-axis, we can indicate this with a superscript $-$ as follows
\begin{equation}
p^-({\bf x},t)=-2\int_{\mathbb{S}_0}G_{\rm d}({\bf x},{\bf x}',t)*p^-({\bf x}',t){\rm d}{\bf x}',\label{eqHuyg5}
\end{equation}
for ${\bf x}$ above $\mathbb{S}_0$. Similarly, for forward extrapolation of a downgoing field, indicated with a superscript $+$, we can derive in a similar way
\begin{equation}
p^+({\bf x},t)=2\int_{\mathbb{S}_0}G_{\rm d}({\bf x},{\bf x}',t)*p^+({\bf x}',t){\rm d}{\bf x}',\label{eqHuyg6}
\end{equation}
for ${\bf x}$ below $\mathbb{S}_0$ (here the source is assumed to be situated in the half-space above $\mathbb{S}_0$). 
Note that the dipole Green's function
$G_{\rm d}({\bf x},{\bf x}',t)$ for ${\bf x}$ below $\mathbb{S}_0$ has a sign opposite to that for ${\bf x}$ above $\mathbb{S}_0$, see Figure \ref{Fig2}.
This compensates for the different signs in front of the integrals in equations (\ref{eqHuyg5}) and (\ref{eqHuyg6}).

\subsection{Inverse wave field extrapolation through a homogeneous medium}

\begin{figure}
\centerline{\hspace{2cm}\epsfysize=4.9 cm\epsfbox{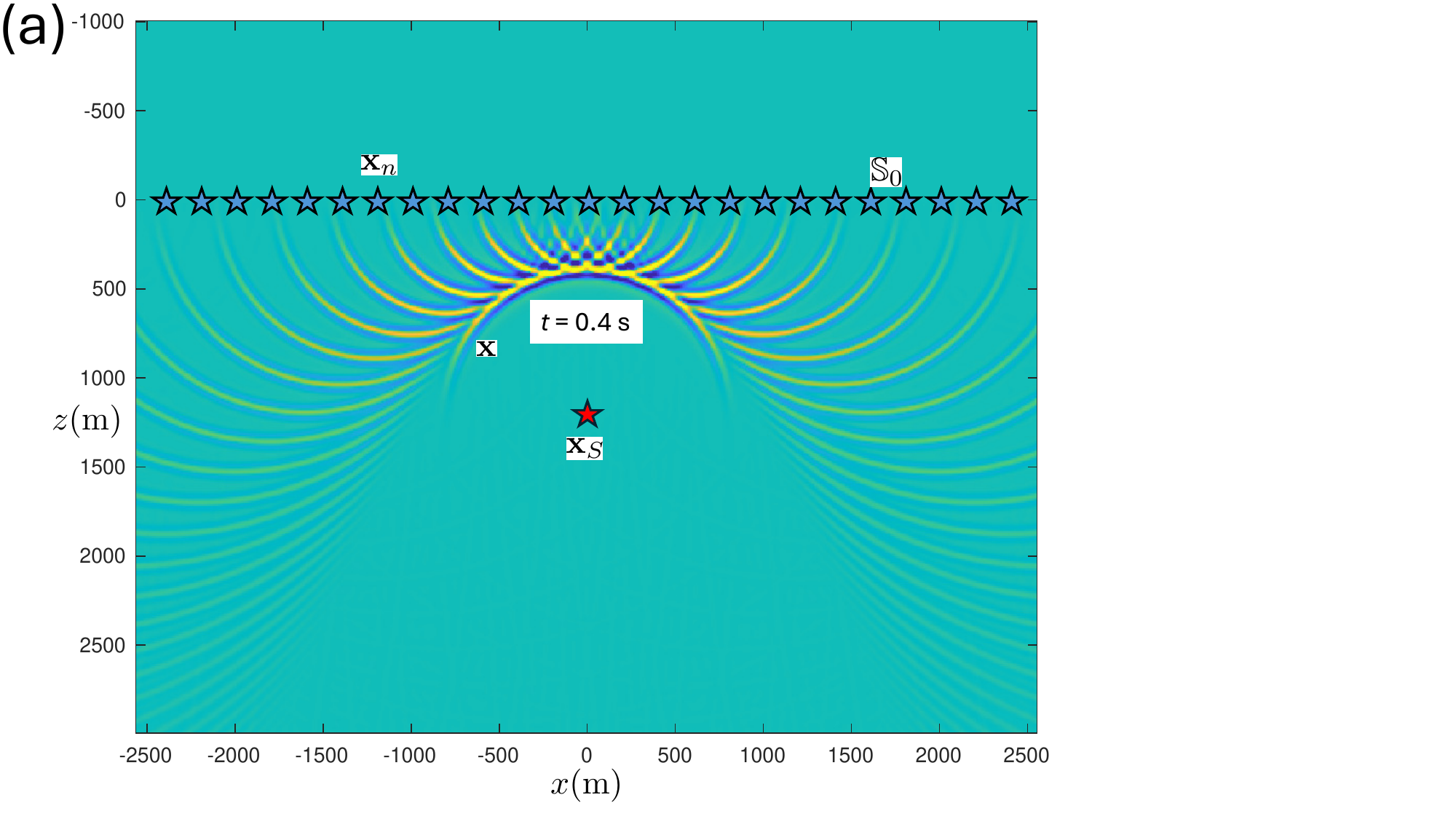}}
\centerline{\hspace{2cm}\epsfysize=4.9 cm\epsfbox{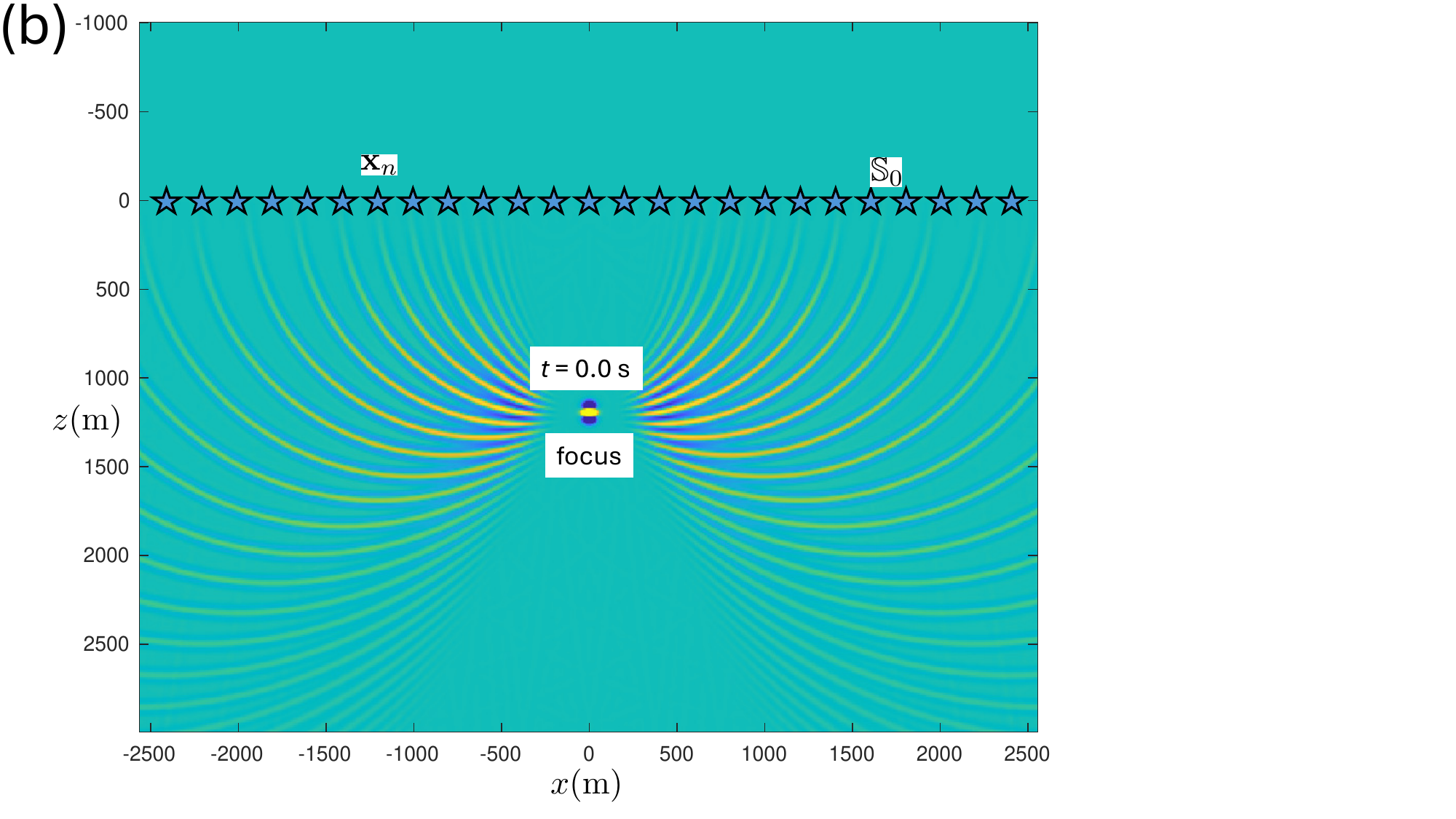}}
\centerline{\hspace{2cm}\epsfysize=4.9 cm\epsfbox{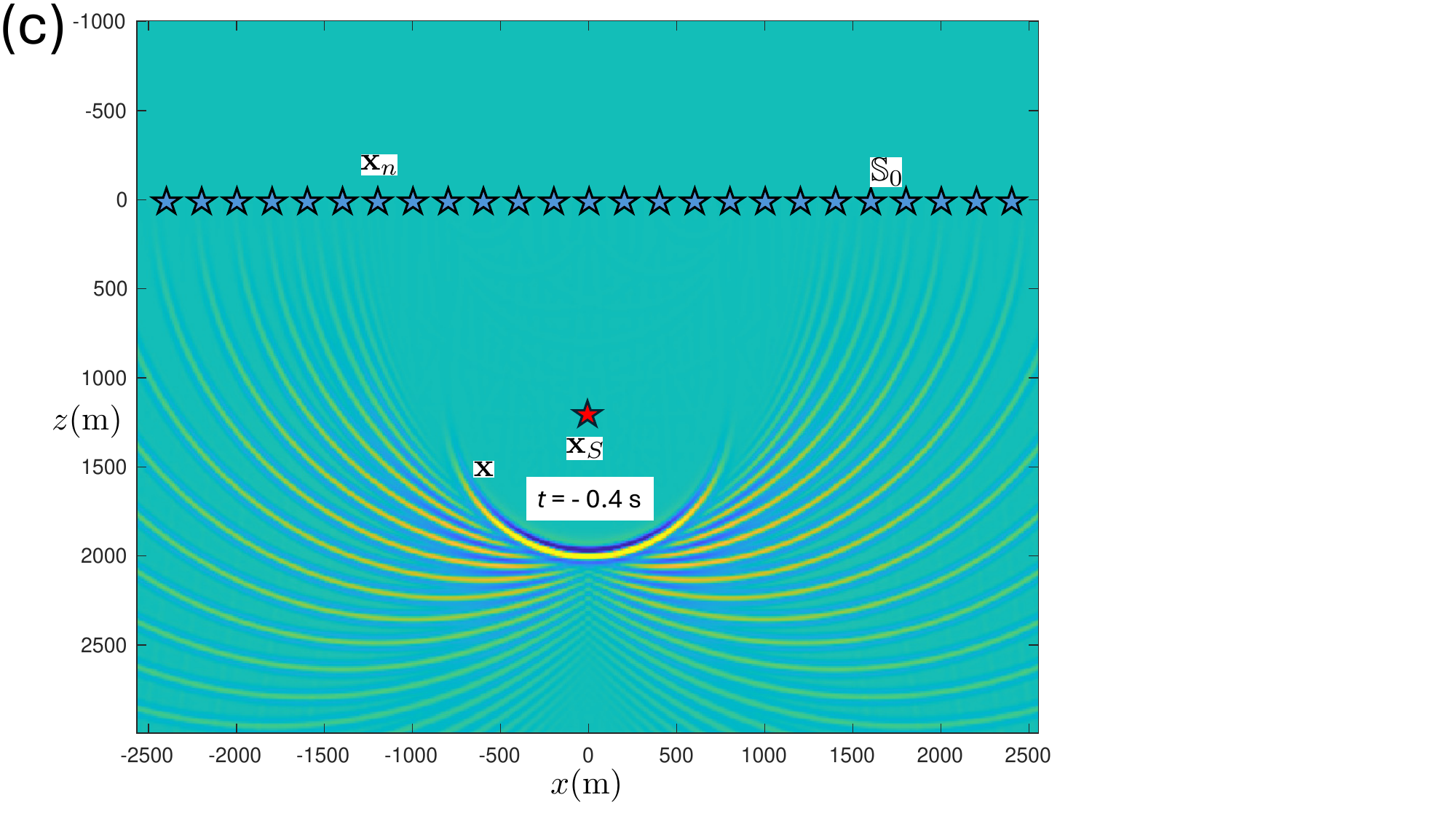}}
\centerline{\hspace{2cm}\epsfysize=4.9 cm\epsfbox{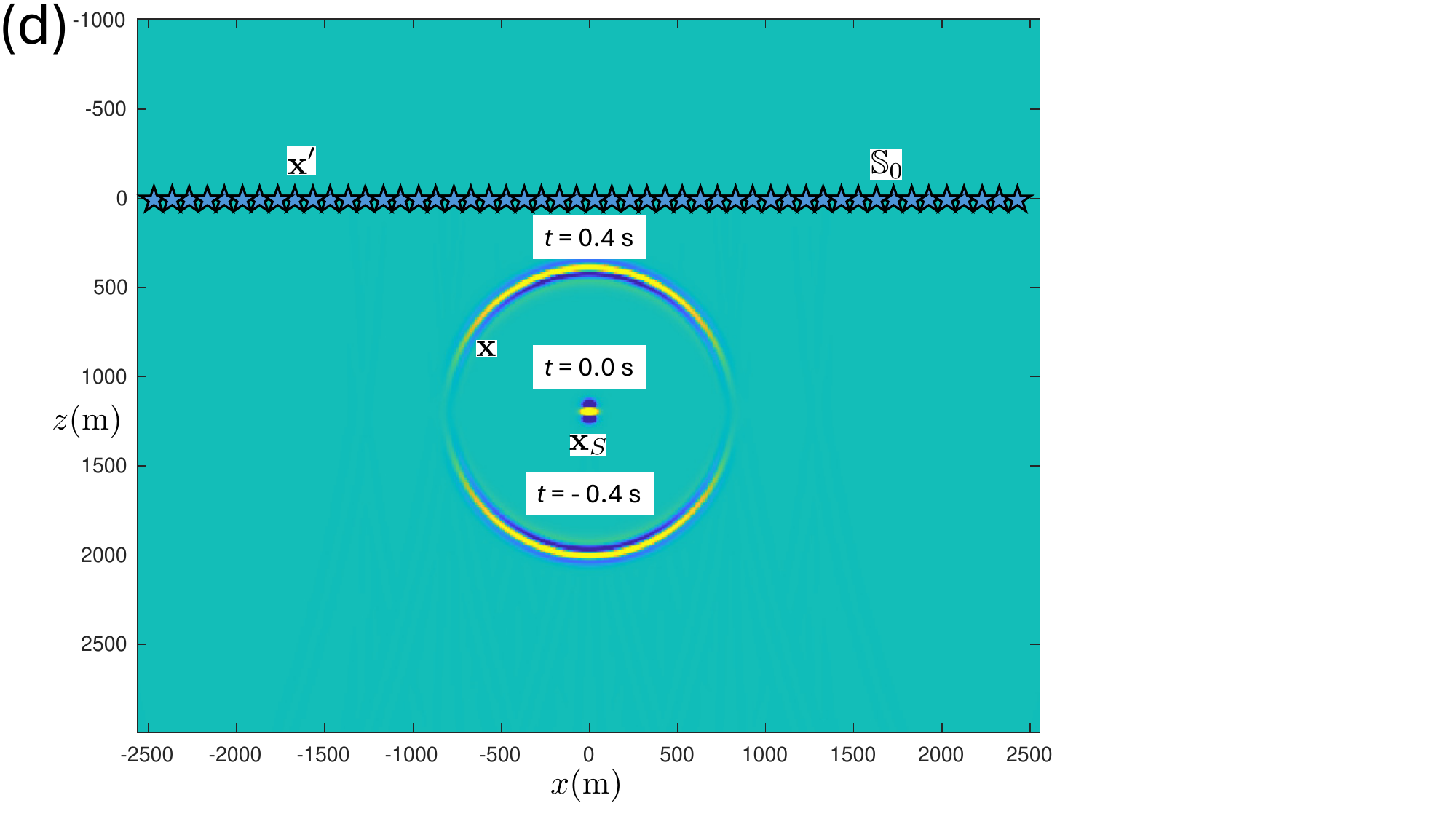}}
\caption{\it (a)--(c) Illustration of Huygens' principle for backpropagation of acoustic waves in a homogeneous medium, according to equation (\ref{eqHuyginv1}), 
\rev{with the time-reversal of the dipole Green's function of Figure \ref{Fig2}}.
 (d) Idem, according to equation (\ref{eqHuyginv2}).}\label{Fig3}
\end{figure}

We start this section with discussing an intuitive modification of Huygens' principle for backpropagation.
We consider again the acoustic pressure $p({\bf x},t)$ in a homogeneous lossless medium, in response to a source in the lower half-space (at ${\bf x}_S=(0,1200)$ m), observed at
${\bf x}_n=(n\Delta x,z_0)$ at $\mathbb{S}_0$, with $n=-N, \dots ,-1, 0,1, \dots , N$ (with $N=50$ and $\Delta x=200$ m, \rev{hence $N\Delta x=10000$ m}).
In  equation (\ref{eqHuyg2}) we replace the Green's function by the time-reversed Green's function $G_{\rm d}({\bf x},{\bf x}_n,-t)$
(i.e., the time-reversal of the dipole Green's function shown in Figure \ref{Fig2}). 
Hence, we evaluate the expression
\begin{equation}
\langle p({\bf x},t)\rangle \propto 2\sum_{n=-N}^N G_{\rm d}({\bf x},{\bf x}_n,-t)*p({\bf x}_n,t),\label{eqHuyginv1}
\end{equation}
for ${\bf x}$ in the half-space below $\mathbb{S}_0$. The notation $\langle p({\bf x},t)\rangle$ means ``estimate of $p({\bf x},t)$.''
Whereas in equation (\ref{eqHuyg2}) the Green's function $G({\bf x},{\bf x}_n,t)$ forward propagates the field of the secondary sources $p({\bf x}_n,t)$
into the half-space above $\mathbb{S}_0$, in equation (\ref{eqHuyginv1}) the time-reversed Green's function $G_{\rm d}({\bf x},{\bf x}_n,-t)$
backpropagates the field of the secondary sources $p({\bf x}_n,t)$
into the half-space below $\mathbb{S}_0$ \citep{Schneider78GEO}. The result $\langle p({\bf x},t)\rangle$ for $t=0.4$ s is shown in Figure \ref{Fig3}a.
This figure illustrates Huygens' superposition principle for backpropagation.
The envelope of the superposed circular waves approximately forms a circular wave, resembling the wave emitted by the original point source at ${\bf x}_S$, 
observed  above this point source at $t=0.4$ s. 
Figure \ref{Fig3}b shows $\langle p({\bf x},t)\rangle$ for $t=0$ s. Here a focus is formed at the position of the original point source. 
Since there is no sink at ${\bf x}_S$ to absorb the focused field, $\langle p({\bf x},t)\rangle$ does not vanish when we continue
the backpropagation to negative times, as is shown in Figure \ref{Fig3}c for $t=-0.4$ s.
Next, we replace the summation in equation (\ref{eqHuyginv1}) by an integration, according to
\begin{equation}
\langle p({\bf x},t)\rangle = 2 \int_{\mathbb{S}_0}G_{\rm d}({\bf x},{\bf x}',-t)*p({\bf x}',t){\rm d}{\bf x}'\label{eqHuyginv2}
\end{equation}
(but in the numerical implementation we actually evaluate equation (\ref{eqHuyginv1}), with $\Delta x$ reduced to $\Delta x=10$ m, 
\rev{which is much smaller than the central wavelength of 100 m,} and  $N=1000$, \rev{so that again $N\Delta x=10000$ m}).
Figure \ref{Fig3}d shows a superposition of snapshots of $\langle p({\bf x},t)\rangle$ for $t=0.4$ s, $t=0$ s and $t=-0.4$ s. 
In the left-hand side of equation (\ref{eqHuyginv2}) we still use the notation $\langle p({\bf x},t)\rangle$, indicating an approximation of $p({\bf x},t)$.
The actual response to a point source at ${\bf x}_S$ is causal and consists of circular wave fronts around ${\bf x}_S$ at positive times only, whereas
Figure \ref{Fig3}d shows an incomplete response at positive time (a half-circle above ${\bf x}_S$), and 
 a non-existing  response at negative time (a half-circle below ${\bf x}_S$).
In Appendix C we review a step-by-step derivation of equation (\ref{eqHuyginv2}). 
Ignoring evanescent waves, we arrive at equation (C-4), which has three terms on the right-hand side.
The first of these terms is the integral in the right-hand side of equation (\ref{eqHuyginv2}),
\rev{explaining Figure \ref{Fig3}d. The second term restores the missing half-circle below ${\bf x}_S$ at positive time and creates a half-circle above ${\bf x}_S$ at negative time.
The third term suppresses the entire acausal response}.
Since the second and third term require measurements  at a boundary below the source and knowledge of the source at ${\bf x}_S$, 
they cannot be evaluated in most practical situations.
By ignoring the second and third term in equation (C-4), we are left with equation (\ref{eqHuyginv2}), with the limitations discussed above.
For an inhomogeneous medium the limitations are  more severe, as we will see in the next section.

Unlike equation (\ref{eqHuyg4}), which formalizes Huygens' explanation of the physics of wave propagation, equation (\ref{eqHuyginv2}) with the time-reversed
Green's function does not describe a physical situation. However, 
it can be used for numerical  wave field extrapolation of an upgoing wave field $p({\bf x}',t)$, measured at $\mathbb{S}_0$, 
to any position below $\mathbb{S}_0$ and above the source.
Here we speak of {\it inverse} wave field extrapolation, since the direction of 
extrapolation (downward from $\mathbb{S}_0$ to ${\bf x}$ below $\mathbb{S}_0$) is opposite to the direction of the upgoing wave field $p({\bf x}',t)$ at $\mathbb{S}_0$.
We indicate upgoing wave fields  again with a superscript $-$ and replace equation (\ref{eqHuyginv2}) by
\begin{equation}
 p^-({\bf x},t) = 2 \int_{\mathbb{S}_0}G_{\rm d}({\bf x},{\bf x}',-t)*p^-({\bf x}',t){\rm d}{\bf x}',\label{eqHuyginv3}
\end{equation}
for ${\bf x}$ below $\mathbb{S}_0$. As long as ${\bf x}$ is above the source in the lower half-space, 
equation  (\ref{eqHuyginv3}) describes the complete upgoing wave field at ${\bf x}$, see Figure \ref{Fig3}
(in this case the only approximation  is the neglection of evanescent waves). When the lower half-space is source-free, 
equation  (\ref{eqHuyginv3}) even holds for the entire lower half-space.
Similarly, for inverse extrapolation of a downgoing wave field we can derive in a similar way
\begin{equation}
 p^+({\bf x},t) = -2 \int_{\mathbb{S}_0}G_{\rm d}({\bf x},{\bf x}',-t)*p^+({\bf x}',t){\rm d}{\bf x}',\label{eqHuyginv4}
\end{equation}
for ${\bf x}$ above $\mathbb{S}_0$. As long as ${\bf x}$ is below the source in 
the upper half-space, equation  (\ref{eqHuyginv4}) describes the complete downgoing wave field at ${\bf x}$.
When the upper half-space is source-free, equation  (\ref{eqHuyginv4})  holds for the entire upper half-space.

Equations  (\ref{eqHuyginv3}) and (\ref{eqHuyginv4}) are the basic expressions for inverse wave field extrapolation through a homogeneous lossless medium, 
as applied in acoustic and seismic imaging methods. They can be implemented in the space-time domain,
as in Kirchhoff migration \citep{Schneider78GEO, Tygel2000IP} and reverse-time migration \citep{Whitmore83SEG, McMechan83GP},
in the space-frequency domain, as in seismic inversion \citep{Cohen86GEO} and seismic migration \citep{Berkhout85Book},
or in the wavenumber-frequency domain, as in migration with the phase-shift method \citep{Gazdag78GEO}.

Another application of equation (\ref{eqHuyginv2}) is obtained when we revert the time coordinate \rev{on the right-hand side}, according to
\begin{equation}
2 \int_{\mathbb{S}_0}G_{\rm d}({\bf x},{\bf x}',t)*p({\bf x}',-t){\rm d}{\bf x}'.\label{eqHuyginv5}
\end{equation}
Note that $G_{\rm d}({\bf x},{\bf x}',t)$ is again the causal response to a dipole at ${\bf x}'$ on $\mathbb{S}_0$,
\rev{similar as in equation (\ref{eqHuyg4}). The expression of equation (\ref{eqHuyginv5})}  underlies the principle of time-reversed acoustics, as advocated by \citet{Fink92IEEE} and coworkers.
In this situation, $p({\bf x}',-t)$ 
represents the time-reversal of measurements at $\mathbb{S}_0$, 
which are fed to a dense array of piezoelectric transducers which emit ultrasound. 
\rev{This expression} thus describes the synthesized wave field emitted by the array into the half-space below $\mathbb{S}_0$.
If we interchange the labels for positive and negative times in Figure \ref{Fig3}, 
this figure shows the synthesized wave field at $t=-0.4$ s converging to the position of the original source,
the focused field at $t=0$ s, and the field at $t=0.4$ s diverging from the focus. Note that in this case the focused field acts as a downward radiating virtual source.  
A further discussion of time-reversed acoustics is beyond the scope of this paper.

\begin{figure}
\centerline{\hspace{2cm}\epsfysize=4.9 cm\epsfbox{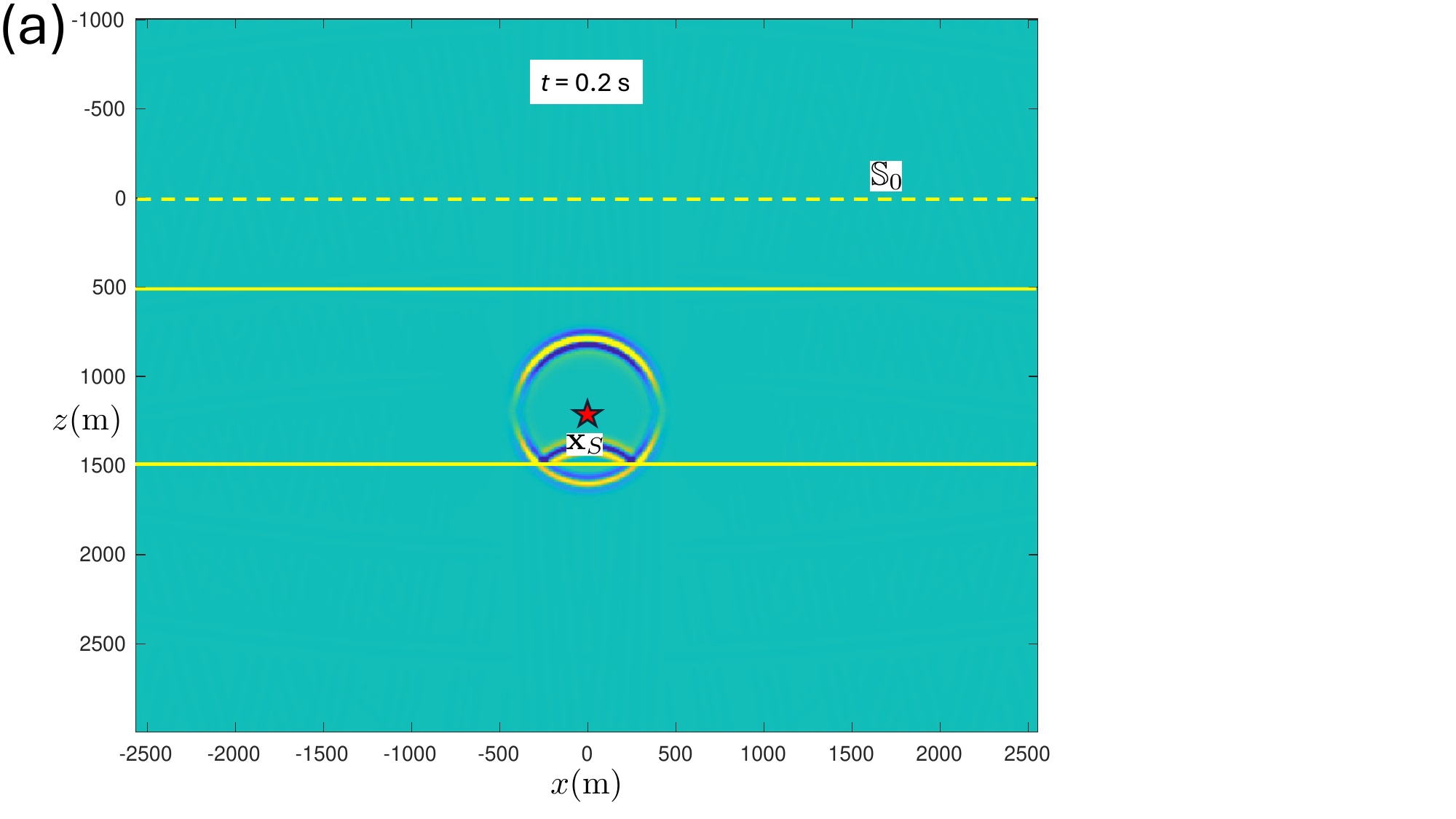}}
\centerline{\hspace{2cm}\epsfysize=4.9 cm\epsfbox{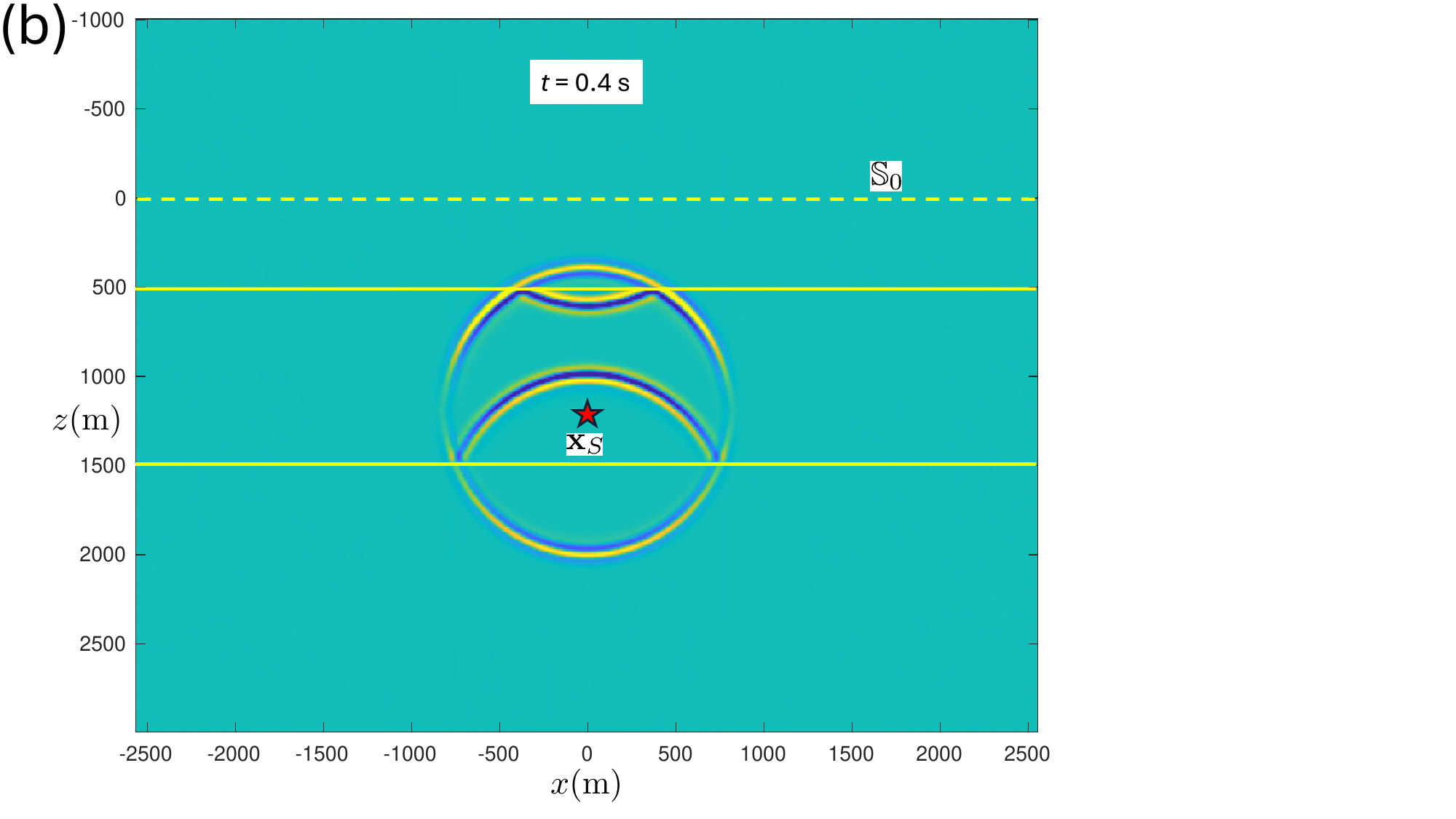}}
\centerline{\hspace{2cm}\epsfysize=4.9 cm\epsfbox{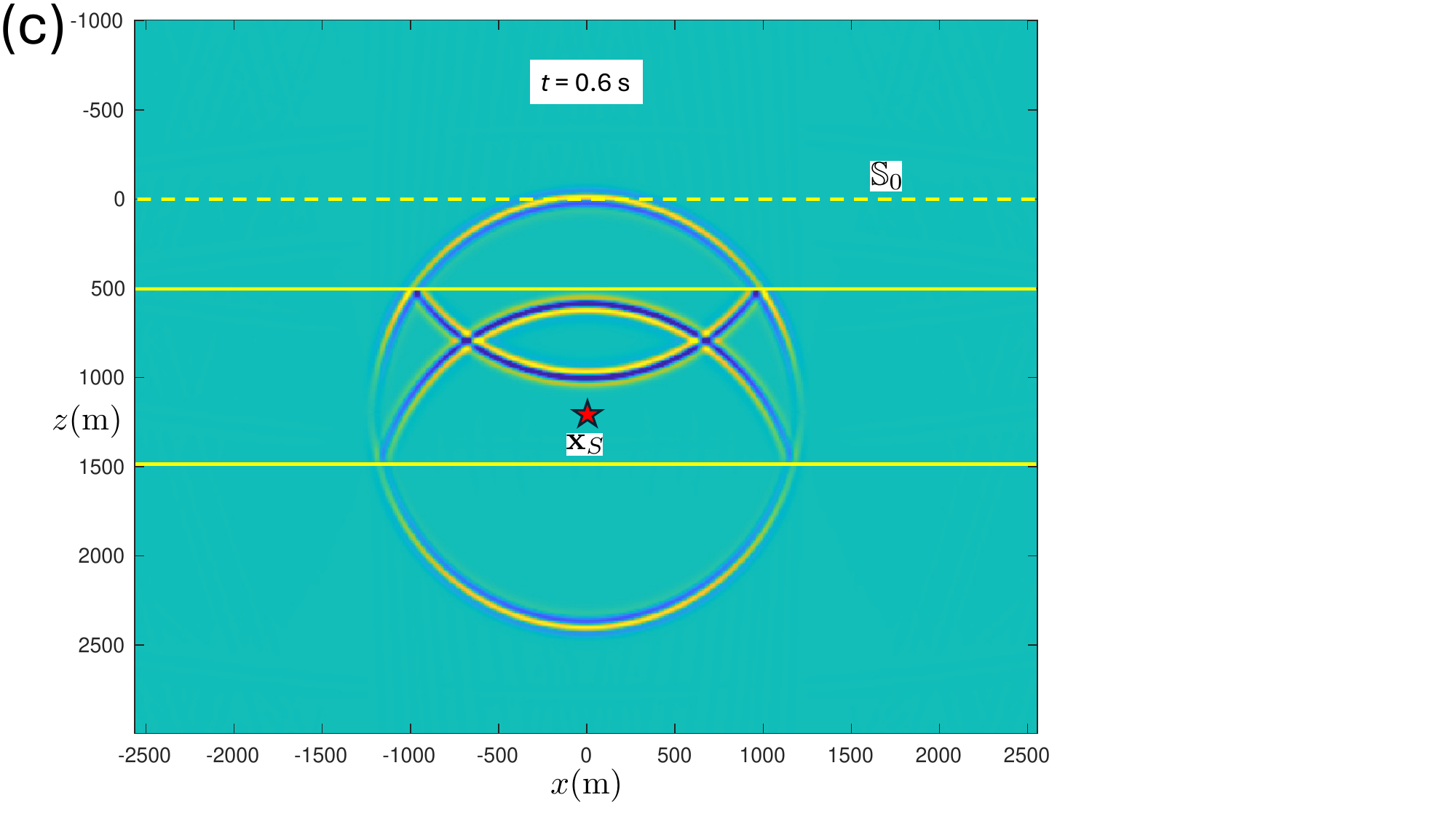}}
\centerline{\hspace{2cm}\epsfysize=4.9 cm\epsfbox{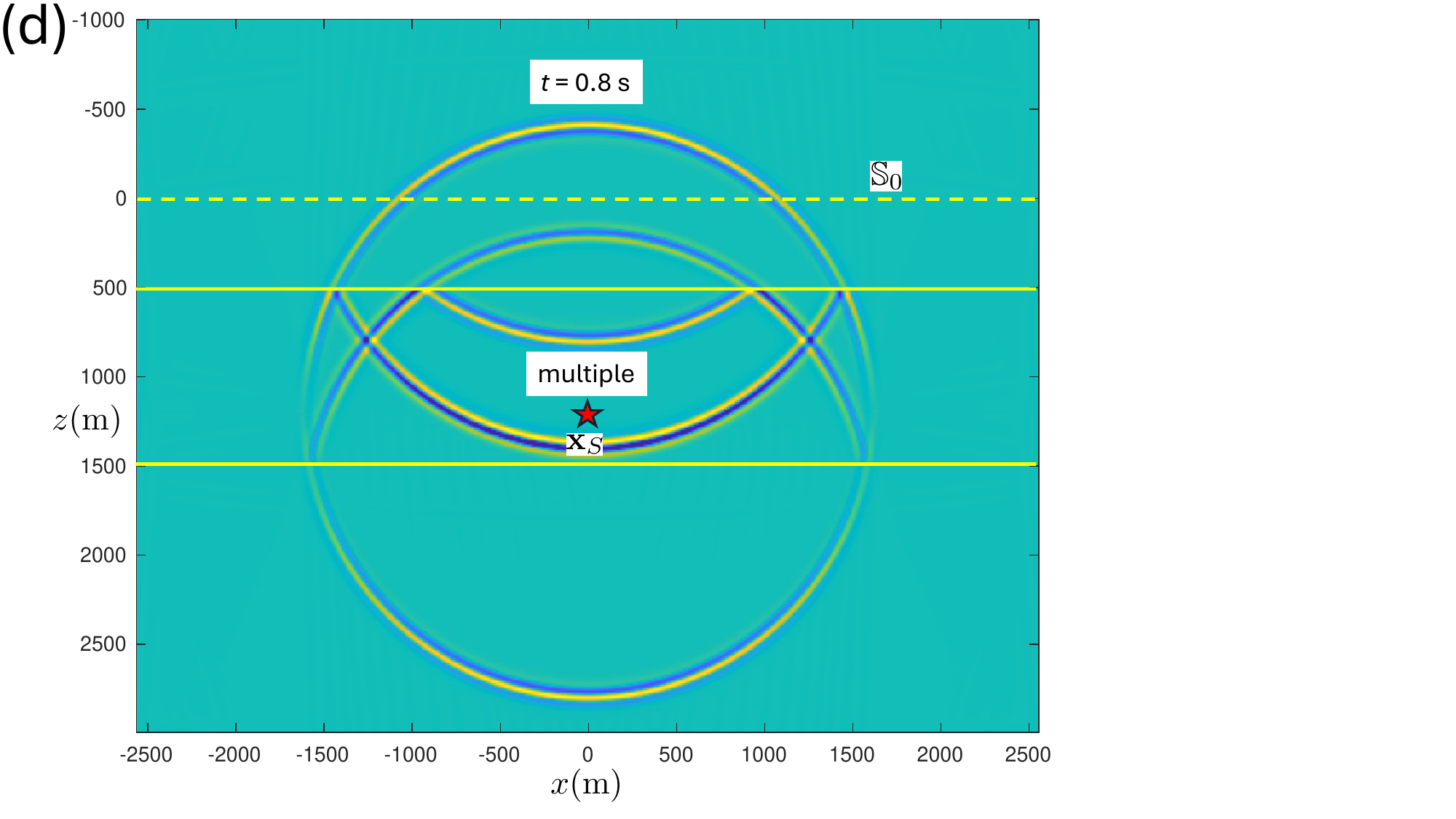}}
\caption{\it  Wave field $p({\bf x},t)=G({\bf x},{\bf x}_S,t)*s(t)$  in a layered medium, for a monopole at ${\bf x}_S$.}\label{Fig4a}  
\end{figure}

\begin{figure}
\centerline{\hspace{2cm}\epsfysize=4.9 cm\epsfbox{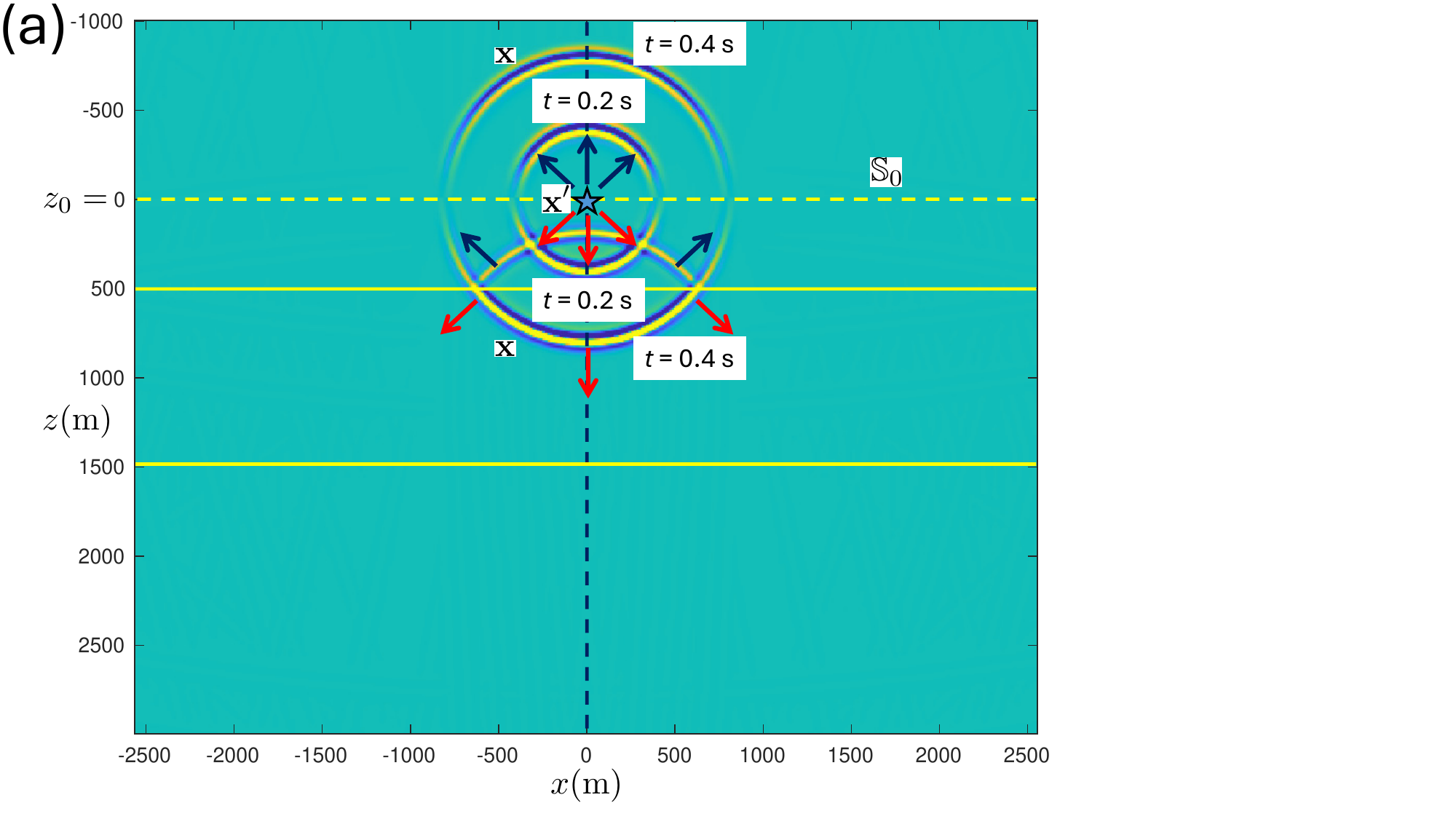}}
\centerline{\hspace{2cm}\epsfysize=4.9 cm\epsfbox{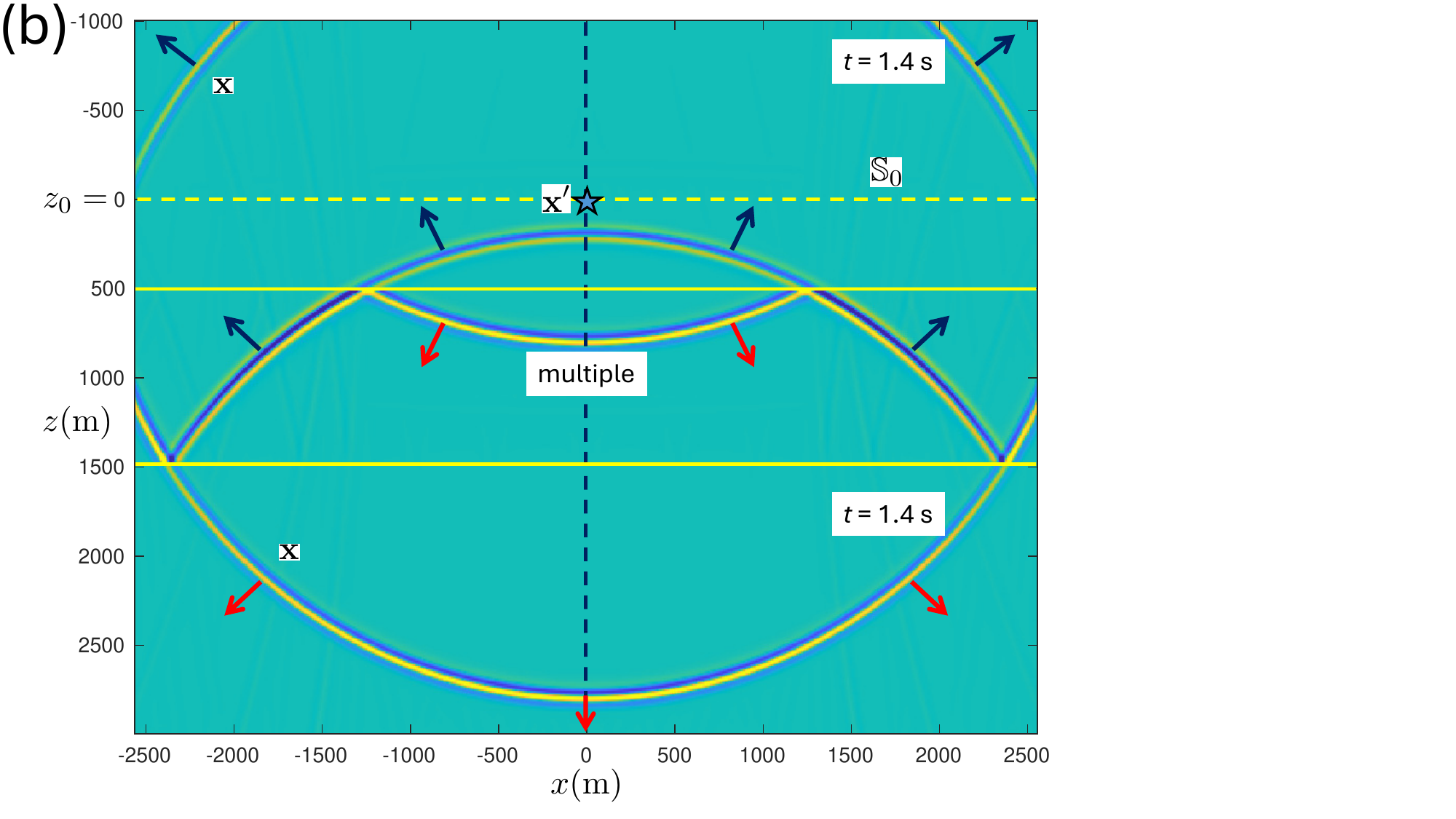}}
\centerline{\hspace{2cm}\epsfysize=4.9 cm\epsfbox{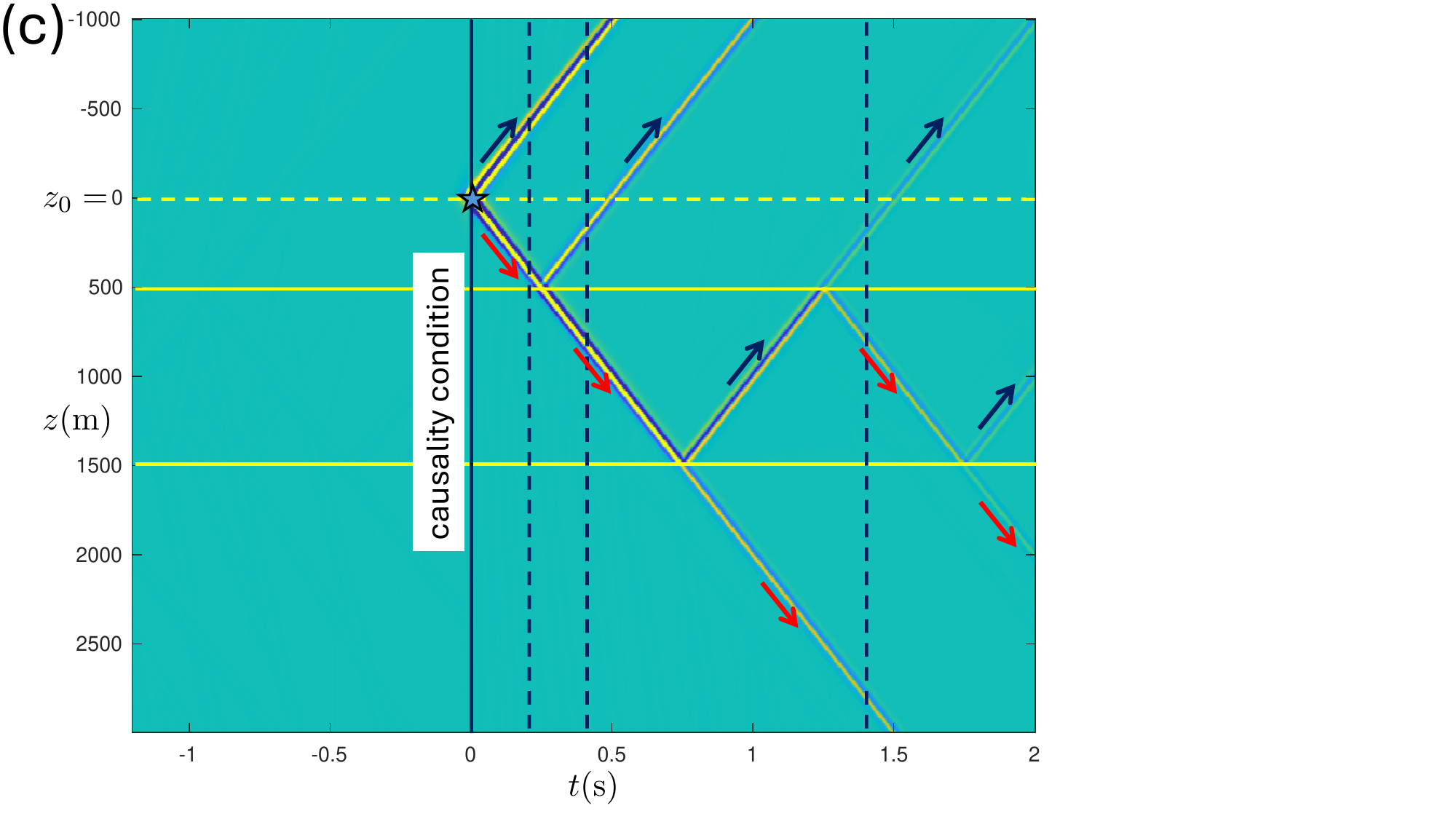}}
\centerline{\hspace{2cm}\epsfysize=4.9 cm\epsfbox{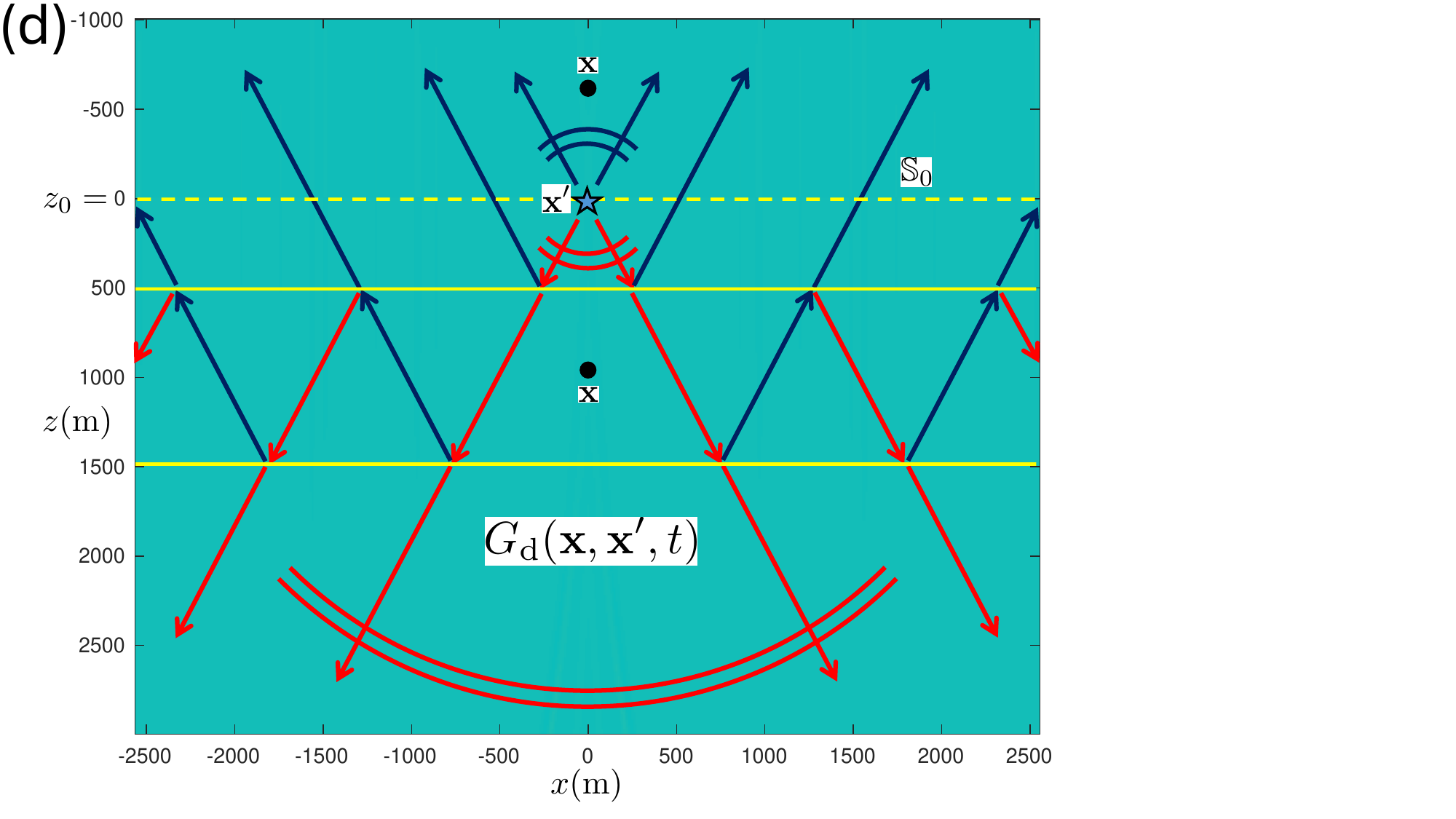}}
\caption{\it  (a),(b) 
Dipole Green's function $G_{\rm d}({\bf x},{\bf x}',t)$ (convolved with a Ricker wavelet) in a layered medium, for a dipole at ${\bf x}'$ on $\mathbb{S}_0$ at depth $z_0$.  
(c) Cross-section at $x=x'=0$. (d) Ray diagram. \rev{Red and blue arrows represent downgoing and upgoing waves, respectively. 
The arcs represent some of the wave fronts at $t=0.2$ {\rm s} and $t=1.4$ {\rm s}.} }\label{Fig4}
\end{figure}

\subsection{Inverse wave field extrapolation through an inhomogeneous medium}

We discuss Huygens' principle for backpropagation through an inhomogeneous lossless medium. 
To show the essence, we consider for simplicity the horizontally layered medium of Figure \ref{Fig4a}, with interfaces 
(indicated by the yellow solid lines) at $z=500$ m and $z=1500$ m.
 The propagation velocity $c$ is taken constant throughout at 2000 m/s.
The mass density $\rho$ in the half-spaces $z<500$ m and $z>1500$ m equals 1000 kg/m$^3$ and in the layer $500<z<1500$ m it equals 4000 kg/m$^3$.
We place again a monopole source at ${\bf x}_S=(0,1200)$ m (indicated by the red star), between the two interfaces.
The source function  is again a Ricker wavelet $s(t)$ with a central frequency of 20 Hz.  
\rev{We use a recursive ``layer-code'' method \citep{Kennett83Book} to model the response to this source.} 
This response, $p({\bf x},t)=G({\bf x},{\bf x}_S,t)*s(t)$, 
is shown in Figures \ref{Fig4a}a -- \ref{Fig4a}d for $t=0.2$ s, $t=0.4$ s, $t=0.6$ s and $t=0.8$ s (note that the amplitudes along the wave fronts are again tapered at large propagation angles).
The interfaces at $z=500$ m and $z=1500$ m partially reflect and partially transmit the waves. Figure \ref{Fig4a}d shows the first multiply reflected wave.

For the same layered medium, the dipole Green's function $G_{\rm d}({\bf x},{\bf x}',t)$, for a dipole at ${\bf x}'$ on $\mathbb{S}_0$, is shown in Figure \ref{Fig4}.
Snapshots of this Green's function for $t=0.2$ s and $t=0.4$ s are shown in Figure \ref{Fig4}a, whereas Figure \ref{Fig4}b shows a snapshot for $t=1.4$ s,
\rev{including the first multiply reflected event}.
Figure \ref{Fig4}c is a cross-section of $G_{\rm d}({\bf x},{\bf x}',t)$ along a vertical line through the dipole source, as a function of depth $z$ and time $t$. 
The vertical dashed lines in this figure at $t=0.2$ s, $t=0.4$ s and $t=1.4$ s correspond to the vertical dashed lines in the snapshots in Figures  \ref{Fig4}a and  \ref{Fig4}b.
The vertical solid line in Figure \ref{Fig4}c at $t=0$ s indicates the causality condition, which states that $G_{\rm d}({\bf x},{\bf x}',t)$ 
is non-zero only after the source at $t=0$ (hence, right of this line). Figure \ref{Fig4}d is a ray diagram of this dipole Green's function.

\begin{figure}
\centerline{\hspace{2cm}\epsfysize=4.9 cm\epsfbox{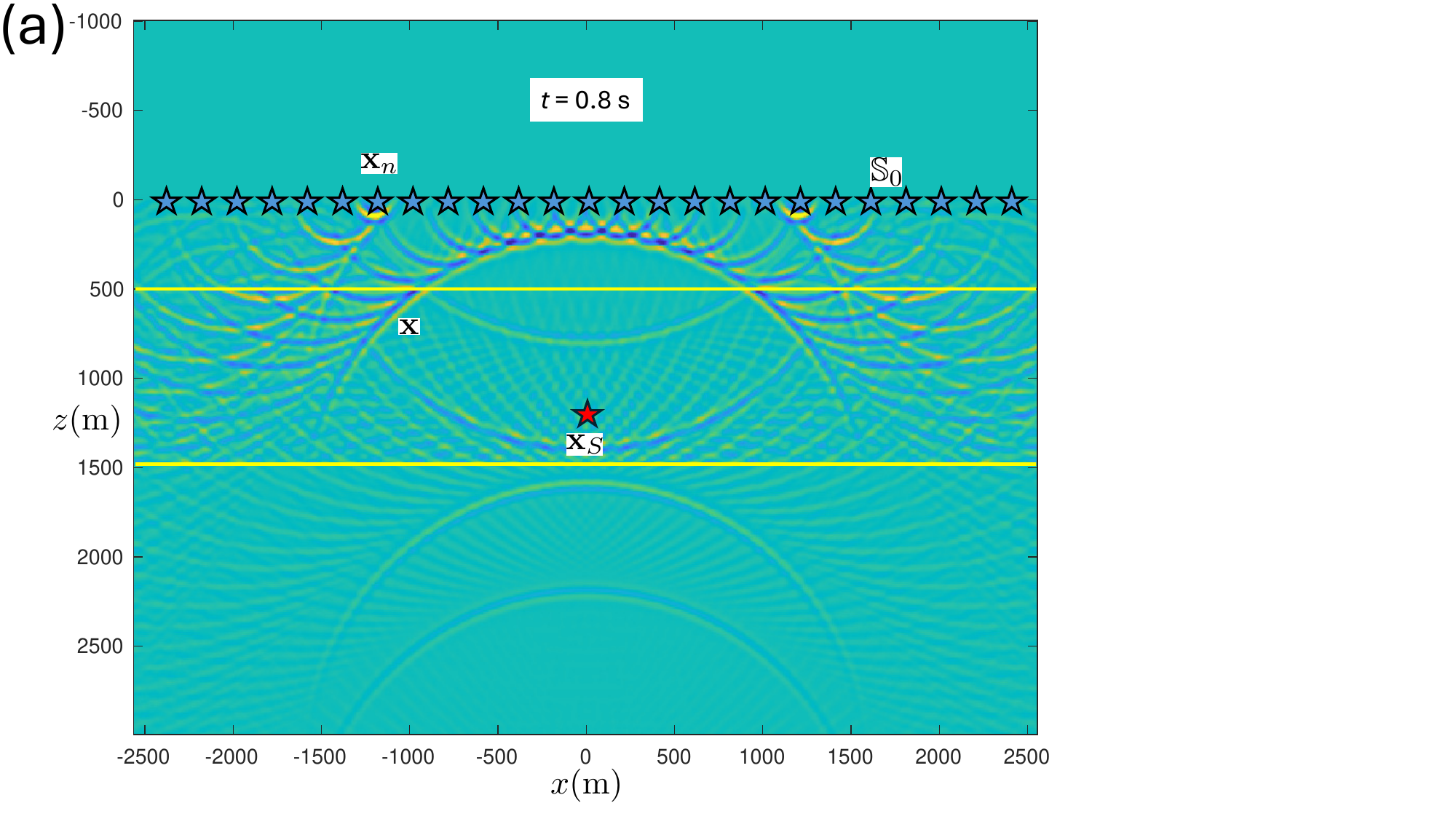}}
\centerline{\hspace{2cm}\epsfysize=4.9 cm\epsfbox{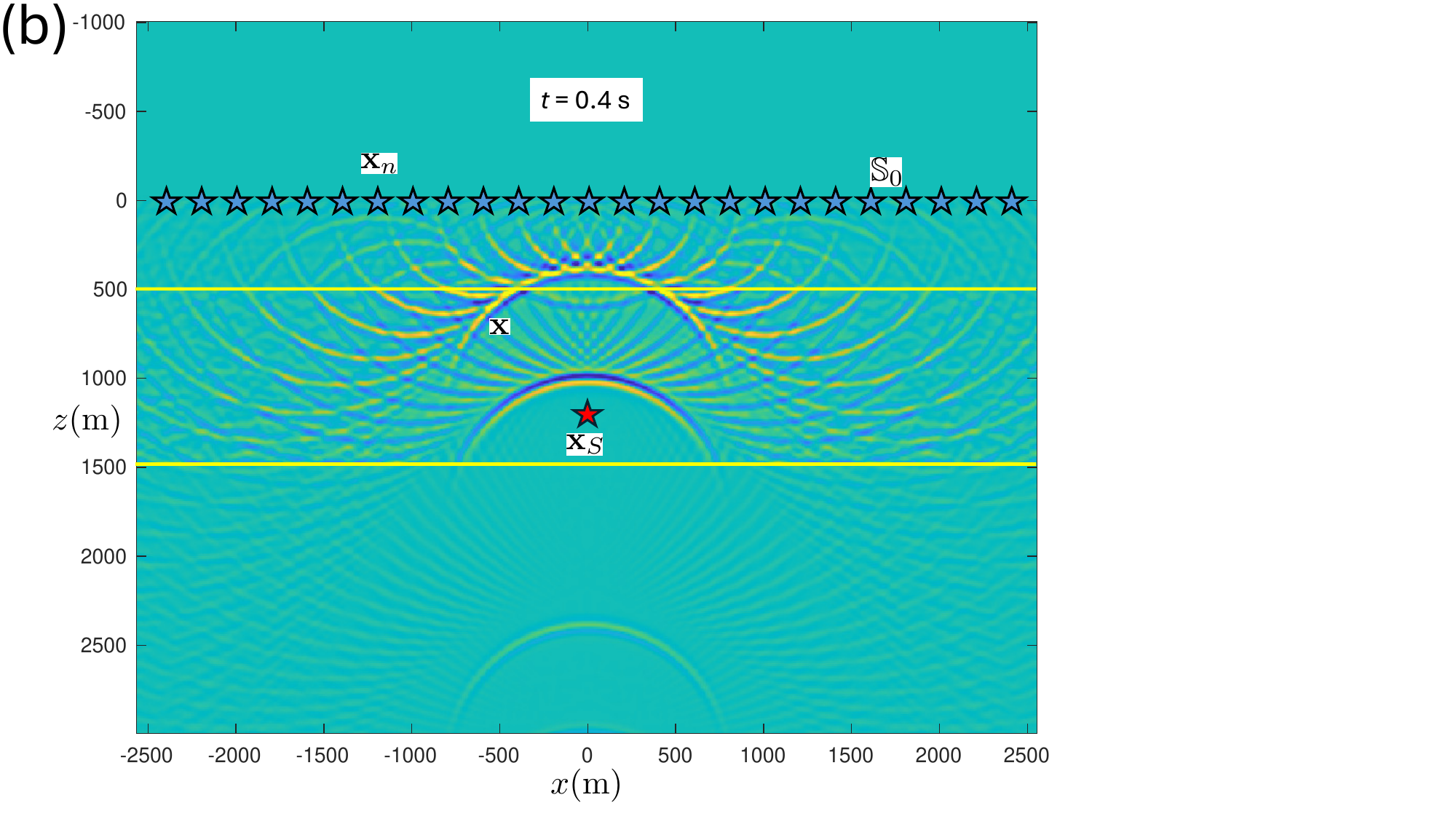}}
\centerline{\hspace{2cm}\epsfysize=4.9 cm\epsfbox{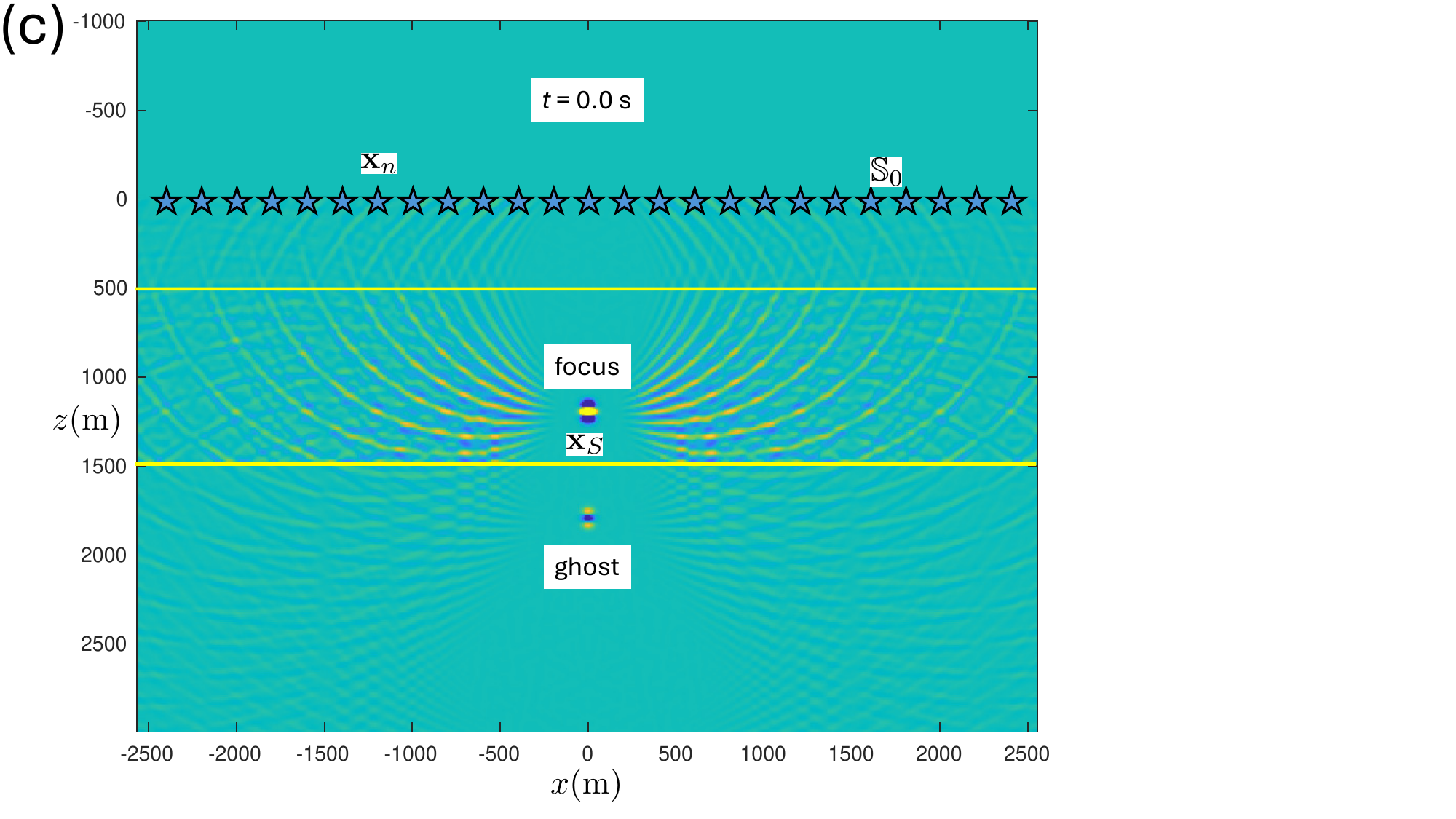}}
\centerline{\hspace{2cm}\epsfysize=4.9 cm\epsfbox{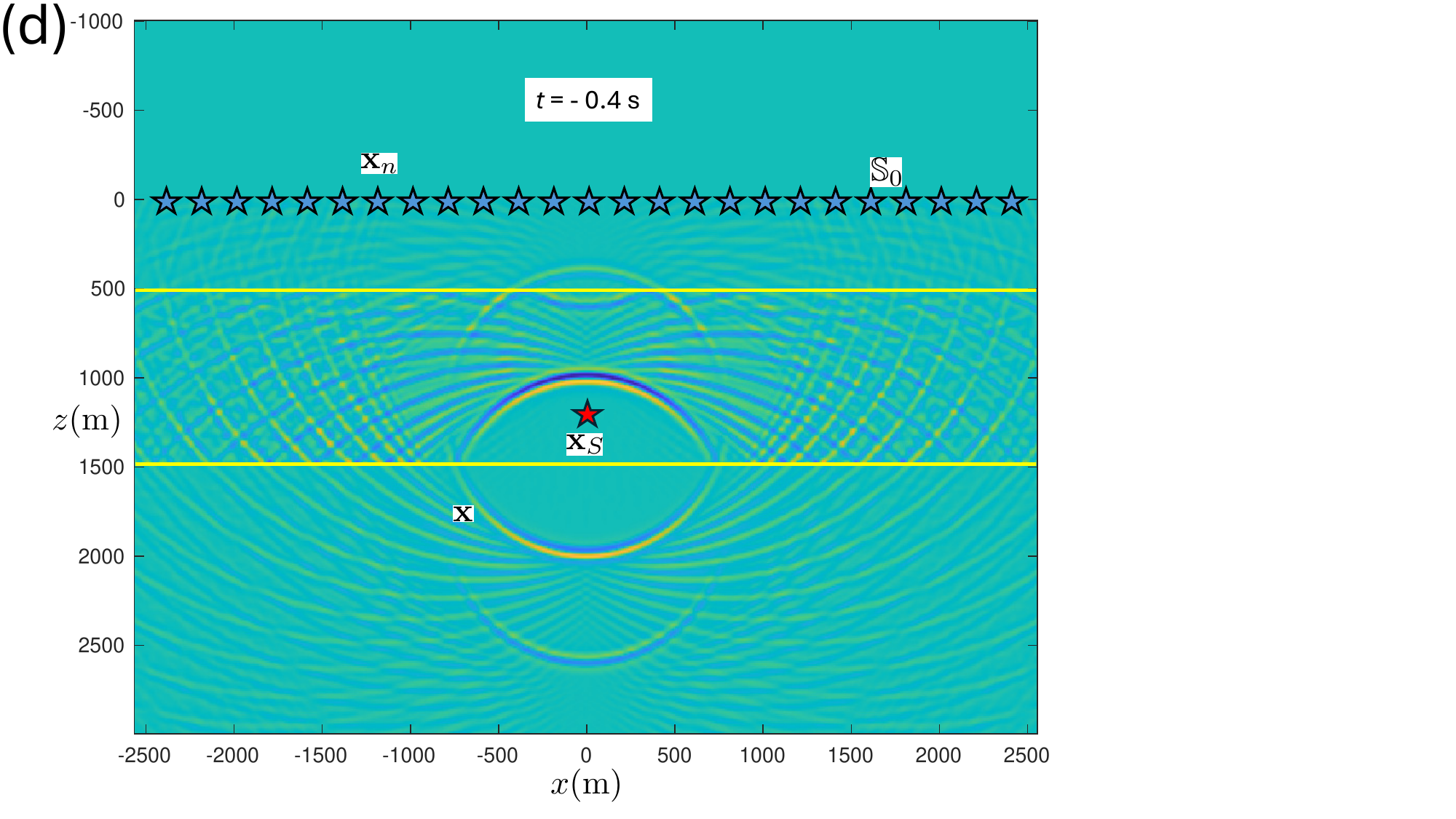}}
\caption{\it Illustration of Huygens' principle for backpropagation of acoustic waves in a layered medium (equation (\ref{eqHuyginv1}), with the \rev{time-reversal of the} 
dipole Green's function of Figure \ref{Fig4}).}\label{Fig5}  
\end{figure}

We use the time-reversal of the dipole Green's function of Figure \ref{Fig4}
 to backpropagate the acoustic pressure wave field of Figure \ref{Fig4a} from $\mathbb{S}_0$ to any point ${\bf x}$ below $\mathbb{S}_0$.
First we use the discretized form of Huygens' principle, as formulated by equation (\ref{eqHuyginv1}), with
${\bf x}_n=(n\Delta x,z_0)$,  $\Delta x=200$ m and $N=50$. 
The results $\langle p({\bf x},t)\rangle$ for $t=0.8$ s and $t=0.4$ s are shown in Figures \ref{Fig5}a and \ref{Fig5}b. 
Compare these figures with Figures \ref{Fig4a}d and \ref{Fig4a}b, which show the desired field $p({\bf x},t)$ at the same time instants. 
It appears that the envelopes of the superposed waves in Figures \ref{Fig5}a and \ref{Fig5}b resemble parts of the desired field,
 but significant parts are missing
(in particular the downgoing field in the lower half-space), the amplitudes of the reflected waves are too low, and  circular ghost events appear in the lower half-space.
Figure \ref{Fig5}c shows $\langle p({\bf x},t)\rangle$ for $t=0$ s. Apart from the focus at the position of the original point source, a ghost focus is formed
below the second interface. Figure \ref{Fig5}d shows  $\langle p({\bf x},t)\rangle$ for $t=-0.4$ s. Clearly $\langle p({\bf x},t)\rangle$ does not vanish for negative times.
The situation is more complex than in Figure \ref{Fig3}, which is the result of applying equation (\ref{eqHuyginv1}) in a homogeneous medium.
In particular,  reflected waves are not correctly backpropagated by the time-reversed dipole Green's function of the layered medium,
\rev{despite the fact that primary and multiply reflected waves are included in this Green's function (see Figure \ref{Fig4})}.

\begin{figure}
\centerline{\hspace{2cm}\epsfysize=4.9 cm\epsfbox{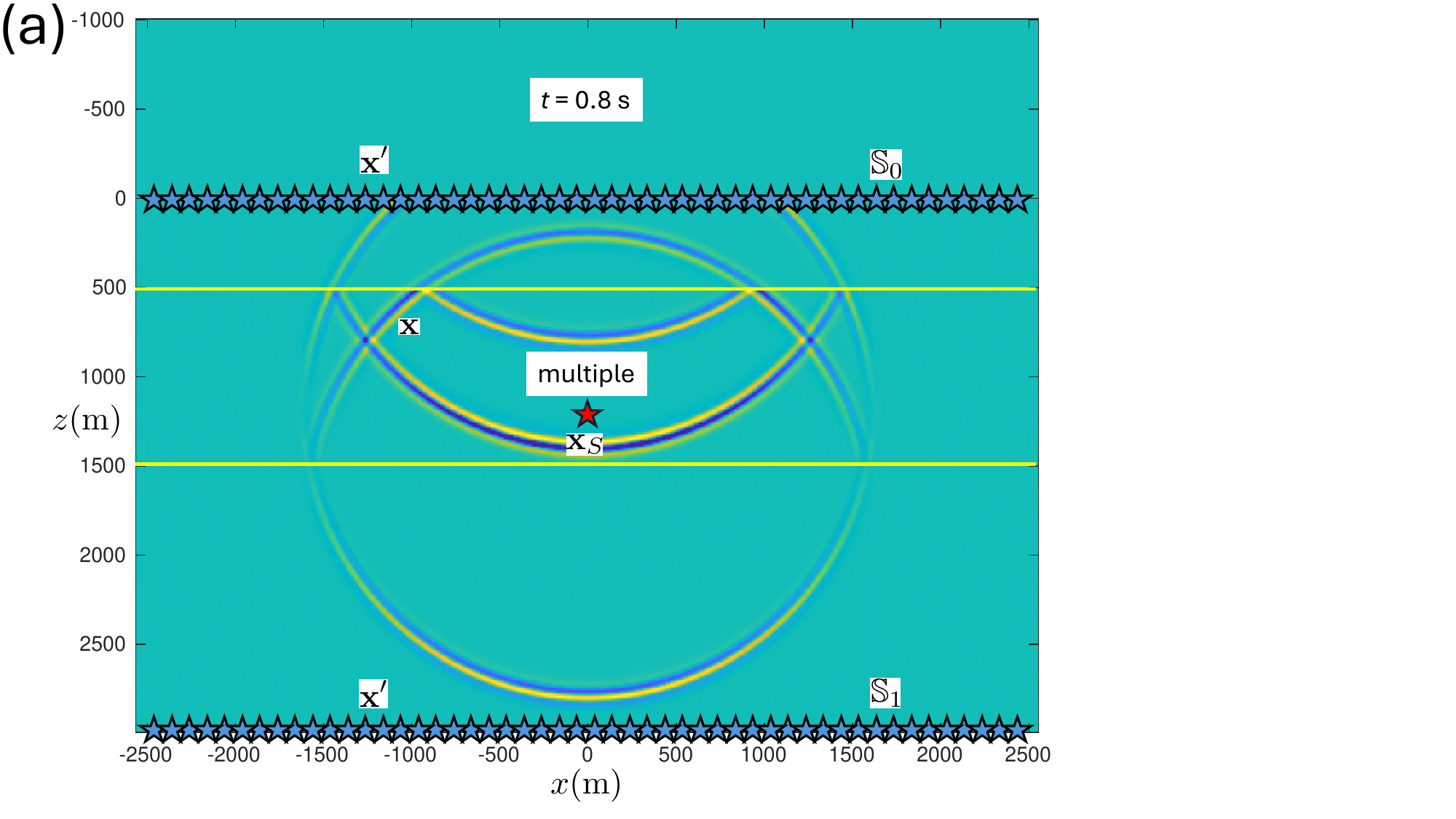}}
\centerline{\hspace{2cm}\epsfysize=4.9 cm\epsfbox{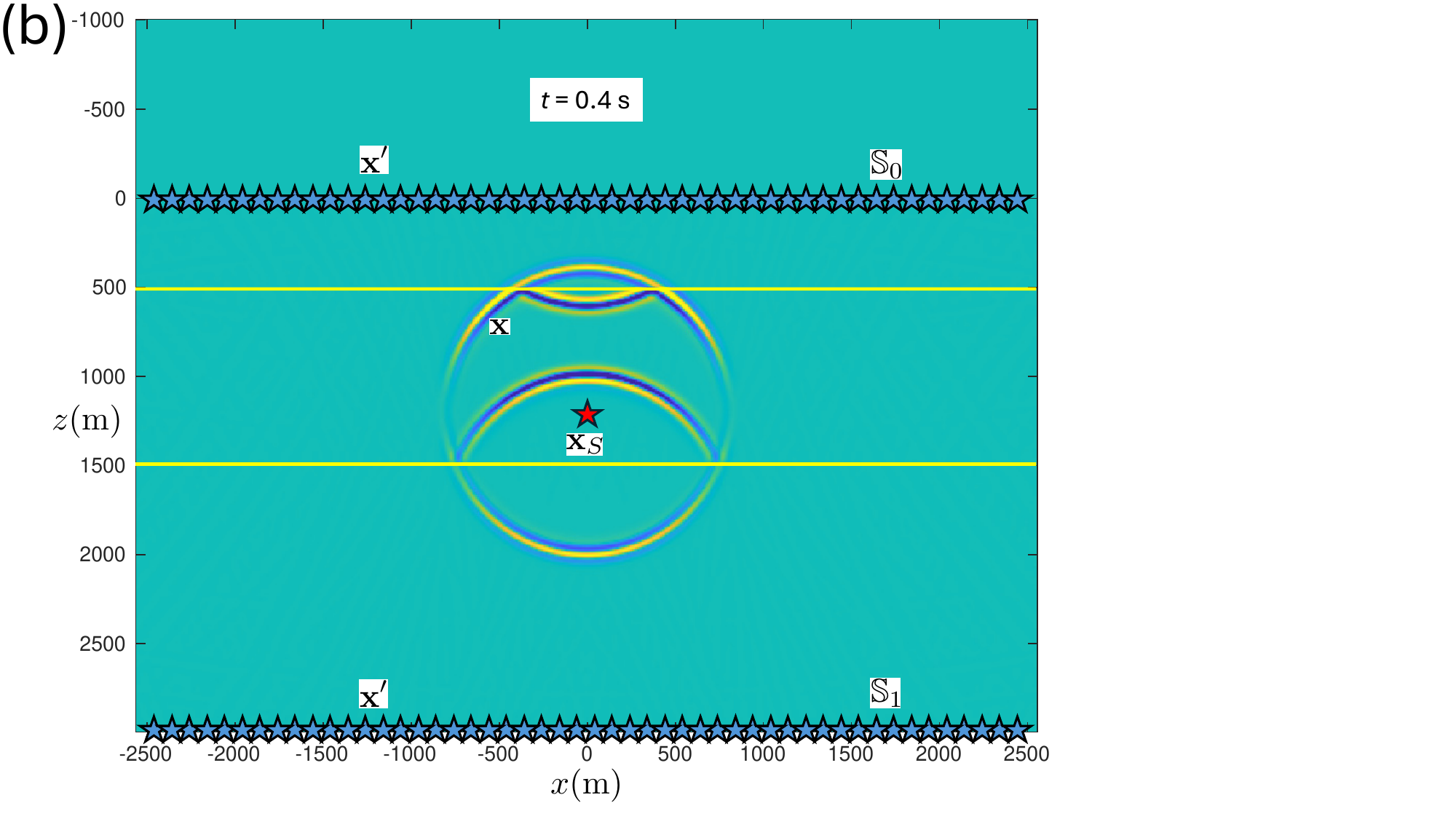}}
\centerline{\hspace{2cm}\epsfysize=4.9 cm\epsfbox{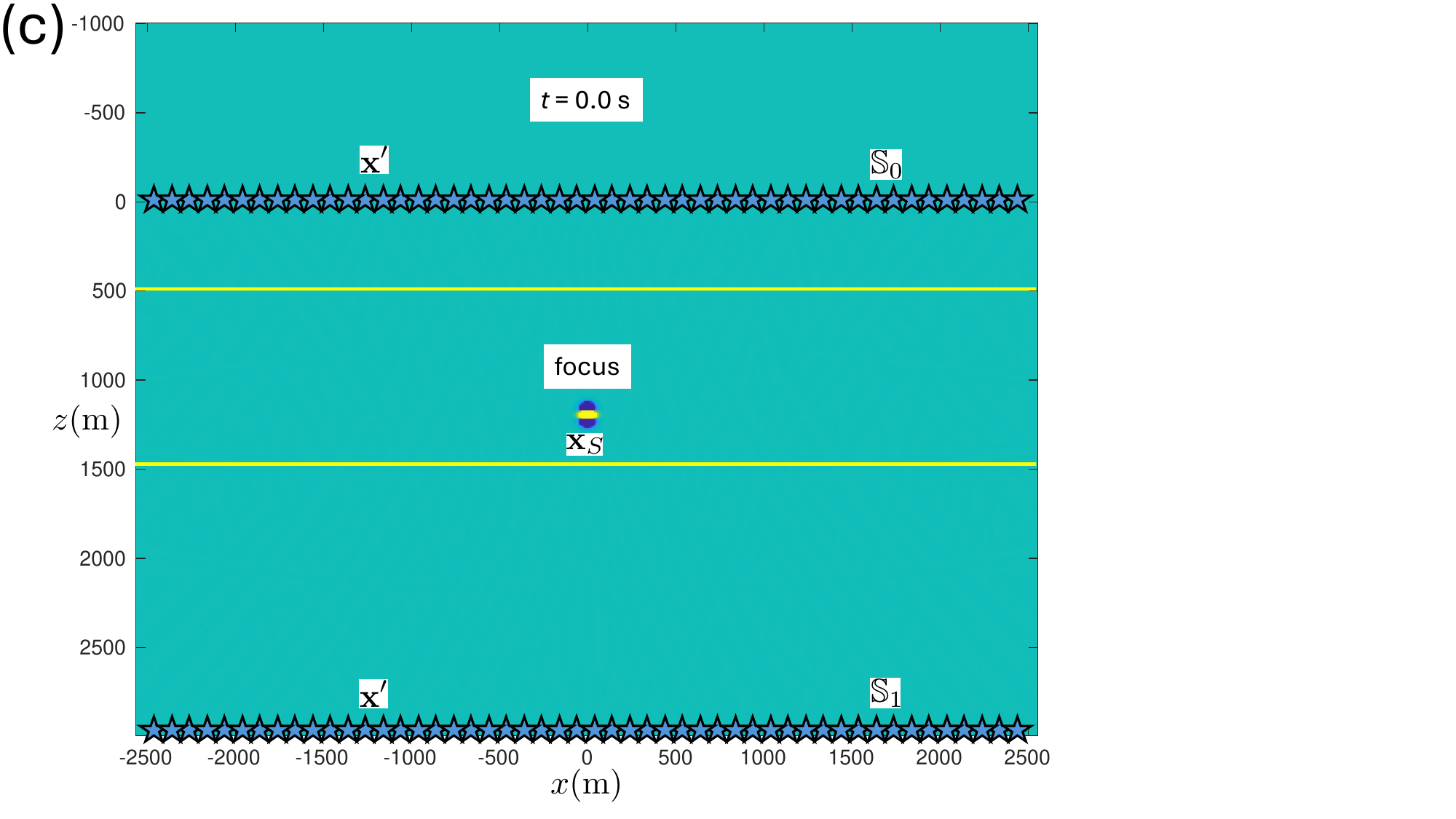}}
\centerline{\hspace{2cm}\epsfysize=4.9 cm\epsfbox{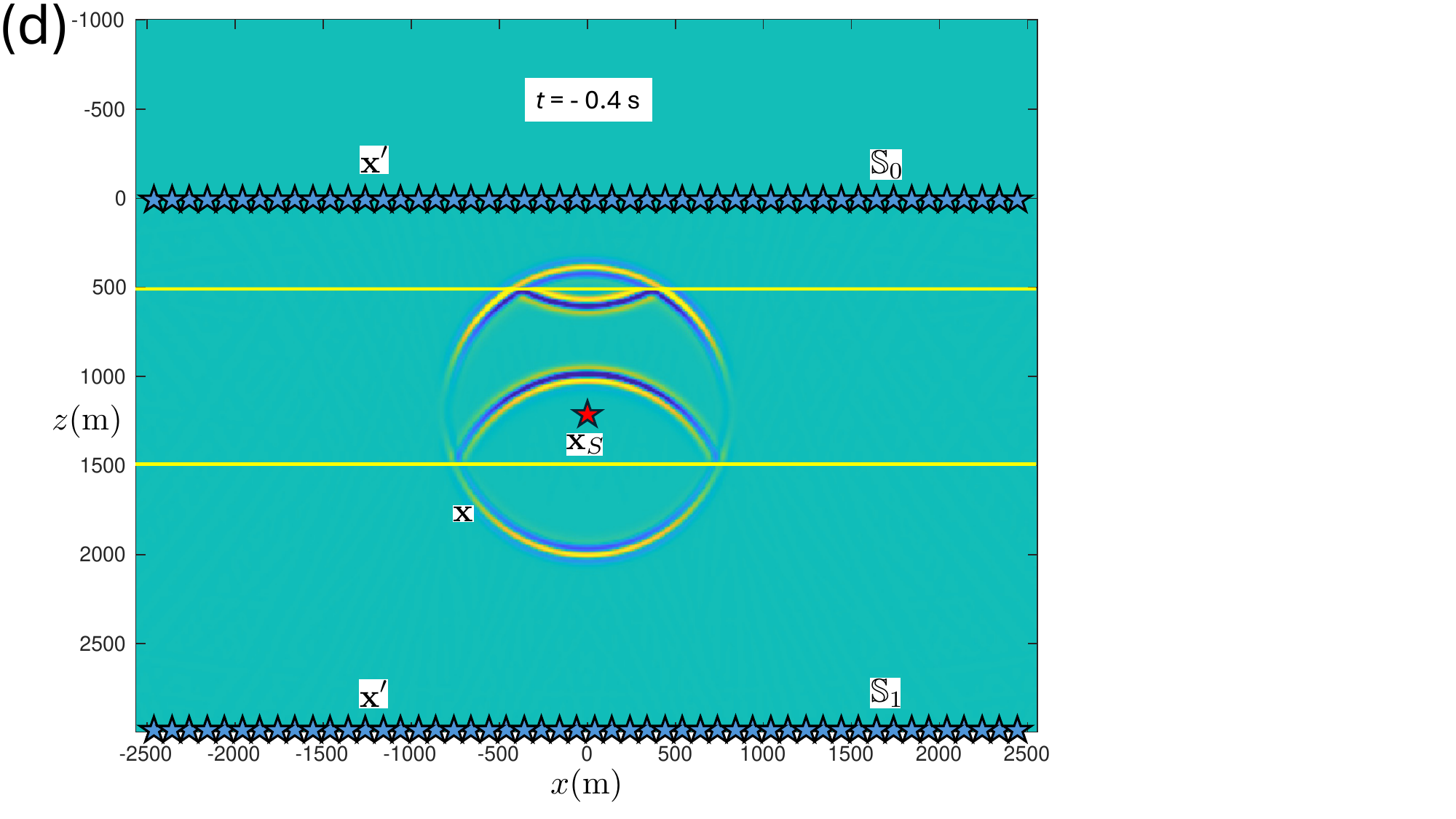}}
\caption{\it  Illustration of Huygens' principle for backpropagation of acoustic waves in a layered medium from two enclosing boundaries (equation (\ref{eqHuyginv6})).}\label{Fig6} 
\end{figure}

Next, we use the integral form of Huygens' principle, as formulated by equation (\ref{eqHuyginv2}), 
which we extend with a second integral over a boundary $\mathbb{S}_1$ (at $z_1=3000$ m) below the source, hence
\begin{eqnarray}
\langle p({\bf x},t)\rangle &=& 2 \int_{\mathbb{S}_0}G_{\rm d}({\bf x},{\bf x}',-t)*p({\bf x}',t){\rm d}{\bf x}'\nonumber\\
&-&2 \int_{\mathbb{S}_1}G_{\rm d}({\bf x},{\bf x}',-t)*p({\bf x}',t){\rm d}{\bf x}',\label{eqHuyginv6}
\end{eqnarray}
for ${\bf x}$ between $\mathbb{S}_0$ and $\mathbb{S}_1$. 
The results $\langle p({\bf x},t)\rangle$ for $t=0.8$ s and $t=0.4$ s are shown in Figures \ref{Fig6}a and  \ref{Fig6}b.
These results accurately resemble the desired field $p({\bf x},t)$, shown in Figures \ref{Fig4a}d and \ref{Fig4a}b, respectively, including the internal \rev{multiply reflected waves}. 
Figure \ref{Fig6}c, which  shows  $\langle p({\bf x},t)\rangle$ for $t=0$ s, contains a single focus at the position of the original point source.
Finally, Figure \ref{Fig6}d shows  $\langle p({\bf x},t)\rangle$ for $t=-0.4$ s, which again appears to be non-zero.
Comparing equation (\ref{eqHuyginv6}) with equation (C-4), we find that $\langle p({\bf x},t)\rangle$ consists of two contributions, according to
\begin{equation}
\langle p({\bf x},t)\rangle= p({\bf x},t)+G({\bf x},{\bf x}_S,-t) * s(t).\label{eqHuyginv7}
\end{equation}
The second term in this expression explains the contribution at negative time in Figure \ref{Fig6}d.

In most practical situations, measurements are available only at a single boundary, say $\mathbb{S}_0$, 
meaning that the second integral in equation  (\ref{eqHuyginv6}) cannot be evaluated. Hence, we are left with the integral along $\mathbb{S}_0$,
as formulated by equation (\ref{eqHuyginv2}), with $G_{\rm d}({\bf x},{\bf x}',-t)$ defined in the inhomogeneous medium.  
Similar as for a homogeneous medium, we can reformulate this again
into equation (\ref{eqHuyginv3}), which then describes approximate 
inverse extrapolation of upgoing waves $p^-({\bf x}',t)$ from $\mathbb{S}_0$ through an inhomogeneous medium, to ${\bf x}$ below $\mathbb{S}_0$ 
and above the source in the lower half-space.
As we have seen above, this approximation does not properly handle \rev{multiply reflected waves}, so it only accounts for primary waves. 
Moreover, even for these primary waves, amplitude errors occur,  
which are proportional to \rev{amplitudes of} \rev{multiply reflected waves} (see \citet{Wapenaar89GEO} for a detailed analysis). Despite these approximations, equation  (\ref{eqHuyginv3}),
with the dipole Green's function defined in the inhomogeneous medium,
forms the basis for many acoustic and seismic imaging schemes, including reverse time migration, time-reversed acoustics, etc.
The approximations are acceptable as long as the contrasts in the medium are sufficiently small so that \rev{internal multiply reflected waves} can be ignored. 
For situations in which \rev{internal multiply reflected waves} cannot be ignored, 
other approaches are needed. One of these approaches is the replacement of the dipole Green's functions
by focusing functions.  This modification of Huygens' principle is the subject of the next section.

We conclude this section by deriving a representation for the homogeneous Green's function from equations (\ref{eqHuyginv6}) and (\ref{eqHuyginv7}).
Using equation (\ref{eqcon2}) for the first term on the right-hand side of equation (\ref{eqHuyginv7}), we observe that this equation 
can be written as $\langle p({\bf x},t)\rangle= G_{\rm h}({\bf x},{\bf x}_S,t) * s(t)$, where $G_{\rm h}({\bf x},{\bf x}_S,t)$ is the homogeneous Green's function, defined as
\begin{equation}
G_{\rm h}({\bf x},{\bf x}_S,t)=G({\bf x},{\bf x}_S,t)+G({\bf x},{\bf x}_S,-t),\label{eqHuyginv9}
\end{equation}
see Appendix A-2 (here ``homogeneous'' refers to the fact that the wave equation for $G_{\rm h}$ has no source term, hence, it is a homogeneous differential equation). 
Next, using this in the left-hand side of equation (\ref{eqHuyginv6}) and using equation (\ref{eqcon2}) for $p({\bf x}',t)$ in the right-hand side of equation (\ref{eqHuyginv6}),
we obtain (after removing the convolution with $s(t)$ on both sides) the following representation for the homogeneous Green's function
in an inhomogeneous lossless medium
\begin{eqnarray}
G_{\rm h}({\bf x},{\bf x}_S,t)&=&2\int_{\mathbb{S}_0} G_{\rm d}({\bf x},{\bf x}',-t) * G({\bf x}',{\bf x}_S,t){\rm d}{\bf x}'\nonumber\\
&-&2\int_{\mathbb{S}_1} G_{\rm d}({\bf x},{\bf x}',-t) * G({\bf x}',{\bf x}_S,t){\rm d}{\bf x}'\label{eqHuyginv8}
\end{eqnarray}
\citep{Porter70JOSA, Oristaglio89IP}. This representation forms the basis for  holographic imaging and inverse scattering methods.
\rev{Moreover, if} we use the source-receiver reciprocity relation $G({\bf x}',{\bf x}_S,t)=G({\bf x}_S,{\bf x}',t)$, then
the integrands on the right-hand side 
describe the cross-correlation of responses at receivers at ${\bf x}_S$ and ${\bf x}$
(both located between $\mathbb{S}_0$ and $\mathbb{S}_1$), due to sources at ${\bf x}'$ at the boundaries $\mathbb{S}_0$ and $\mathbb{S}_1$ \citep{Wapenaar2006GEO}. 
In this form equation (\ref{eqHuyginv8}) is the theoretical basis for Green's function retrieval, also known as seismic interferometry
\citep{Weaver2001PRL, Campillo2003Science, Schuster2004GJI, Snieder2004PRE, Roux2004JASA2, Sabra2007APL, Draganov2009GEO}.
Again, \rev{in many practical situations (for holographic imaging,  interferometry, etc.) receivers or sources} are available only at a single boundary. 
An alternative representation of the homogeneous Green's function, in terms of integrals over a single boundary, 
follows from the modified version of Huygens' principle in the next section.

\section{Modified Huygens' principle, using focusing functions}

\subsection{\rev{Introducing the focusing function}}
\rev{Huygens' principle, formalized by Kirchhoff, Rayleigh and others,  accurately explains the physics of wave propagation.
As we have seen, it can also be used for forward extrapolation of a wave field measured on a plane $\mathbb{S}_0$ into a source-free half-space. 
This is formulated by equation (\ref{eqHuyg4}) and illustrated for a homogeneous upper half-space in Figure \ref{Fig1}.
When the upper half-space is inhomogeneous, equation (\ref{eqHuyg4}) still holds when the dipole Green's function is replaced by that of the inhomogeneous
upper half-space (and the wave field at $\mathbb{S}_0$ by its upgoing part). 
Hence, for these situations there is no need to modify the mathematical formulation of Huygens' principle.}

\rev{For inverse wave field extrapolation (i.e., extrapolation of a wave field measured on a plane $\mathbb{S}_0$ into the half-space containing the source(s) of this wave field),
the dipole Green's function is commonly replaced by its time-reversed version, see equation (\ref{eqHuyginv2}). This is illustrated for a homogeneous lower half-space in Figure \ref{Fig3} and
for an inhomogeneous lower half-space (using the time-reversed dipole Green's function of this inhomogeneous half-space) in Figure \ref{Fig5}.
In the latter case, inverse wave field extrapolation yields reasonable results for the primary waves but it breaks down
for multiply reflected waves, even though these are included in the Green's functions.}

\rev{Huygens' principle 
was meant to explain the physics of wave propagation, 
so the fact that it has limitations for inverse wave field extrapolation is not a shortcoming of this principle in itself. 
Nevertheless, it is worthwhile to modify Huygens' principle for inverse wave field extrapolation, in such a way that it accounts for multiply reflected waves. 
To address this, in} this section we replace the dipole Green's functions 
by focusing functions and, hence, the dipole sources by focal points. In previous work on the Marchenko method we introduced two types of focusing functions:
 $f_1$, which has a focal point inside the inhomogeneous medium, 
and  $f_2$, with its focal point at the boundary between the inhomogeneous lower half-space and the homogeneous upper half-space \citep{Wapenaar2014GEO}.
Since the dipole Green's functions in Huygens' principle  have their sources at the boundary $\mathbb{S}_0$, 
we choose for  focusing functions with their focal points at $\mathbb{S}_0$. Hence, the focusing function $F$ we discuss below is akin
to the focusing function $f_2$. However, it is normalized in a different way. Moreover, whereas $f_2$ is defined in a truncated version of the actual medium, 
the focusing function $F$ is defined in the actual medium
and it is not decomposed into downgoing and upgoing components inside the medium.
Before we discuss this focusing function in an inhomogeneous medium, we start with discussing the focusing function $F$ in a homogeneous medium.

\subsection{Focusing function in a homogeneous medium}

\begin{figure}
\centerline{\hspace{2cm}\epsfysize=4.9 cm\epsfbox{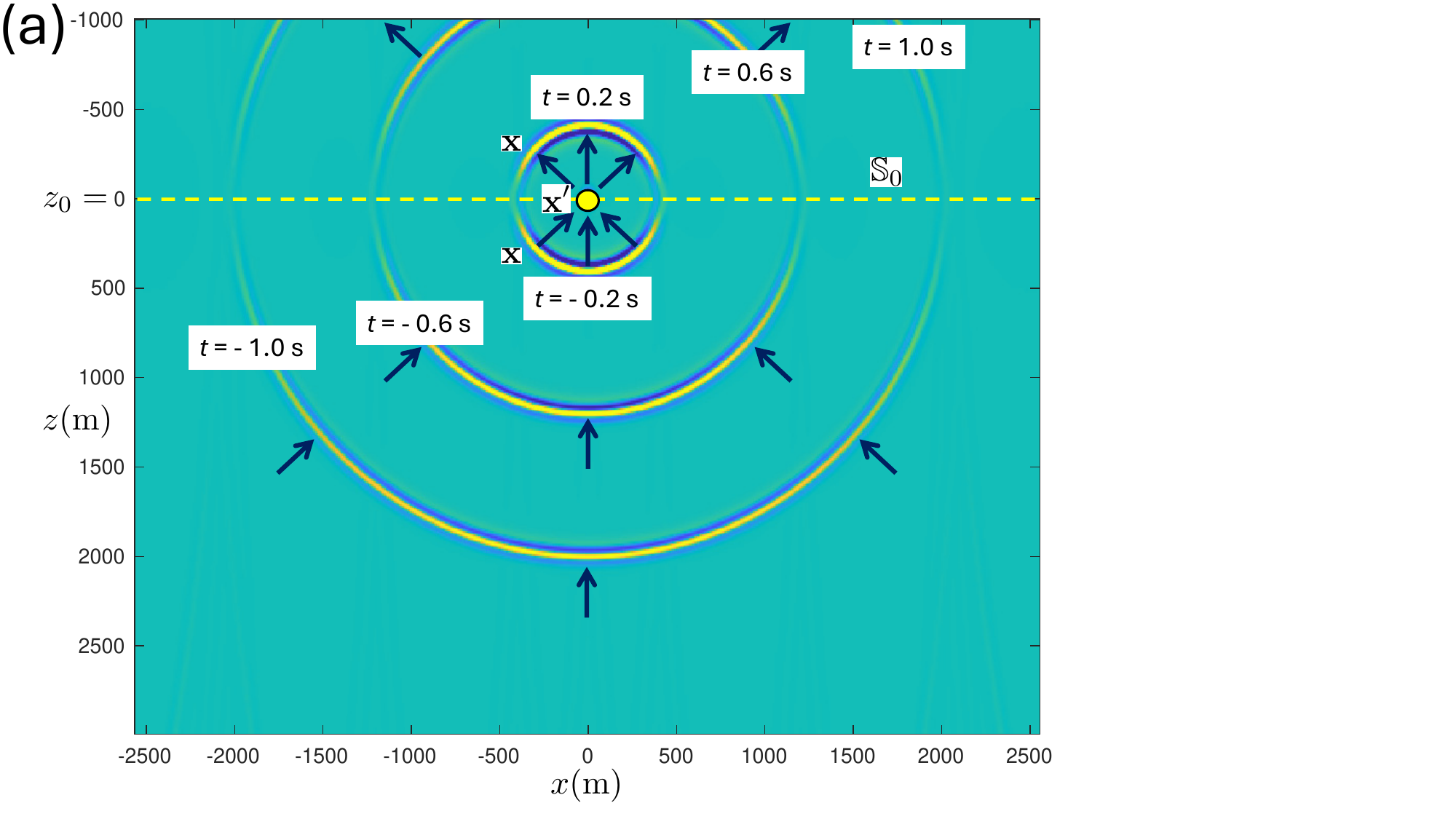}}
\centerline{\hspace{2cm}\epsfysize=4.9 cm\epsfbox{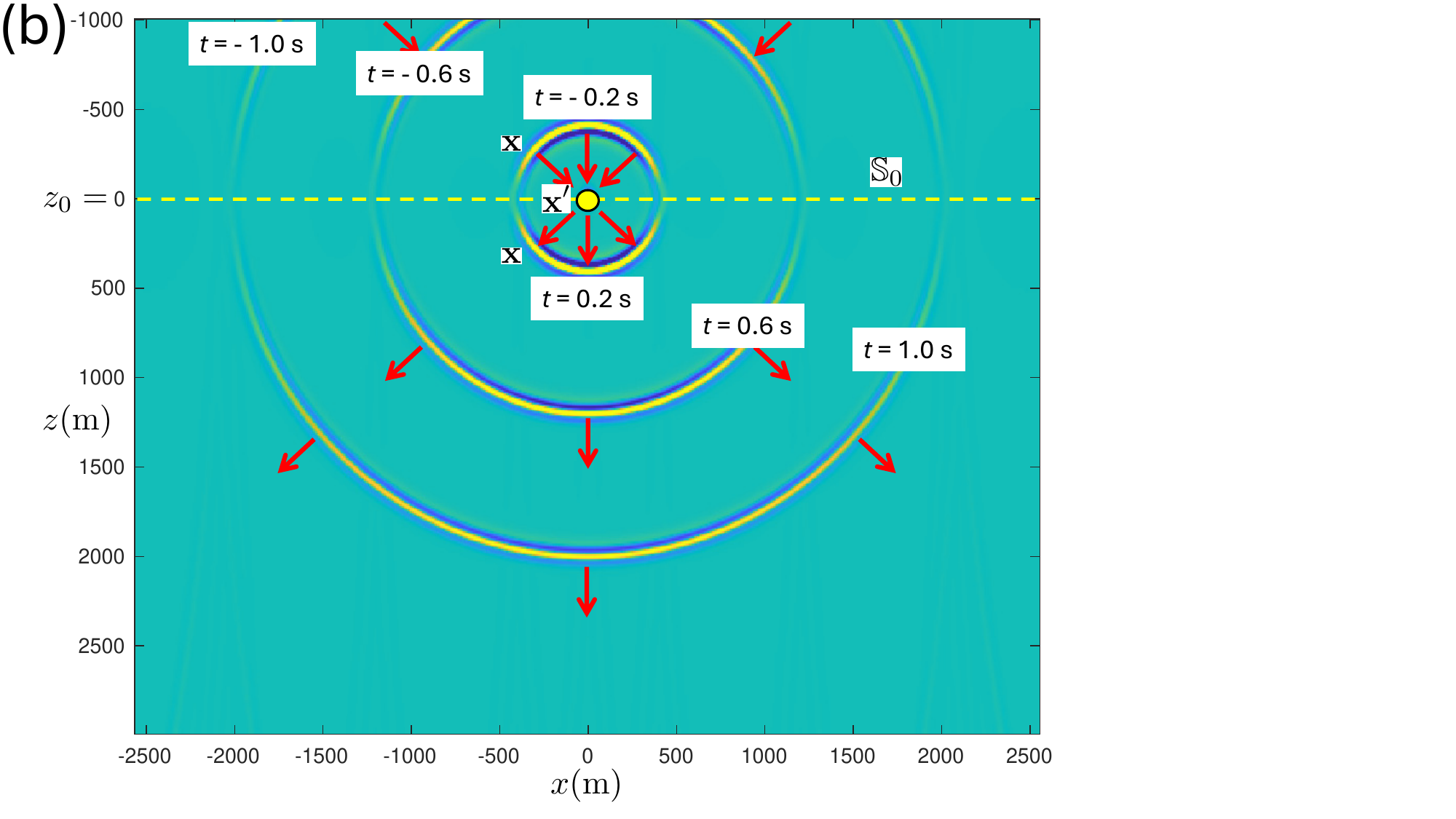}}
\caption{\it  
(a) Focusing function $F({\bf x},{\bf x}',t)$ (convolved with a Ricker wavelet to get a nicer display) and (b) its time-reversed version $F({\bf x},{\bf x}',-t)$ 
in a homogeneous medium, for a focal point at ${\bf x}'$ on $\mathbb{S}_0$ at depth $z_0$.
\rev{In the modified version of Huygens' principle for a homogeneous medium (equations (\ref{eqFoc3}) and (\ref{eqFoc4})), 
these focusing functions replace the dipole Green's function of Figure \ref{Fig2}.}
}\label{Fig14}  
\end{figure}

We define the focusing function $F({\bf x},{\bf x}',t)$ for a homogeneous lossless medium as an upward propagating wave field, 
of which the wave fronts are half-spheres (in 3D) or half-circles (in 2D)
centered at ${\bf x}'$ on $\mathbb{S}_0$ (at depth $z_0$), see Figure \ref{Fig14}a for the 2D situation. 
At negative times, the focusing function propagates  (as a function of ${\bf x}$ and $t$) upward through the lower half-space
towards $\mathbb{S}_0$, at $t=0$ it focuses at ${\bf x}={\bf x}'$ on $\mathbb{S}_0$, and at positive times it 
propagates upward through the upper half-space away from $\mathbb{S}_0$.
The time-reversed focusing function $F({\bf x},{\bf x}',-t)$, illustrated in Figure \ref{Fig14}b for the 2D situation,
propagates at negative times downward through the upper half-space towards $\mathbb{S}_0$, at $t=0$ it focuses at ${\bf x}={\bf x}'$ on $\mathbb{S}_0$,
and at positive times it propagates downward through the lower half-space away from $\mathbb{S}_0$.

In the lower half-space, we relate the upward propagating focusing function of Figure  \ref{Fig14}a to the time-reversal of the dipole Green's function of Figure \ref{Fig2} via
\begin{equation}
F({\bf x},{\bf x}',t)=2G_{\rm d}({\bf x},{\bf x}',-t),\label{eqFoc2}
\end{equation}
for ${\bf x}$ below $\mathbb{S}_0$. 
In the upper half-space, we relate it
to the dipole Green's function via
\begin{equation}
F({\bf x},{\bf x}',t)=-2G_{\rm d}({\bf x},{\bf x}',t),\label{eqFoc1}
\end{equation}
for ${\bf x}$ above $\mathbb{S}_0$. 
For ${\bf x}$ at $\mathbb{S}_0$ (hence, for $z=z'=z_0$) the focusing condition reads
\begin{equation}
F({\bf x},{\bf x}',t)|_{z=z'}=\delta({\bf x}_{\rm H}-{\bf x}_{\rm H}')\delta(t),\label{eqFoc1b}
\end{equation}
where ${\bf x}_{\rm H}$ and ${\bf x}_{\rm H}'$ denote the horizontal components of ${\bf x}$ and ${\bf x}'$, respectively,
hence ${\bf x}_{\rm H}=(x,y)$ and ${\bf x}_{\rm H}'=(x',y')$ (in 3D) or ${\bf x}_{\rm H}=x$ and ${\bf x}_{\rm H}'=x'$ (in 2D). 
In theory evanescent waves can be included in the focusing function \citep{Dukalski2022EAGE, Wapenaar2023GJI}, but to avoid instability they are usually suppressed.
This implies that the delta functions in the right-hand side of equation (\ref{eqFoc1b}) should be interpreted as band-limited delta functions.
Note that, whereas the Green's function $G_{\rm d}({\bf x},{\bf x}',t)$ is the response to a dipole source at ${\bf x}'$ on $\mathbb{S}_0$,
the focusing function $F({\bf x},{\bf x}',t)$ obeys a source-free wave equation (the right-hand side of equation (\ref{eqFoc1b}) formulates a focusing condition, not a source,
see Appendix D for details).

%
%
%

Substituting equation (\ref{eqFoc1}) into equation (\ref{eqHuyg5}) for forward extrapolation of upgoing waves to ${\bf x}$ above $\mathbb{S}_0$, 
or substituting equation (\ref{eqFoc2}) into equation (\ref{eqHuyginv3}) for inverse extrapolation of upgoing waves to ${\bf x}$ below $\mathbb{S}_0$, we obtain in both cases
\begin{equation}
p^-({\bf x},t)=\int_{\mathbb{S}_0}F({\bf x},{\bf x}',t)*p^-({\bf x}',t){\rm d}{\bf x}'.\label{eqFoc3}
\end{equation}
Hence, this expression holds for ${\bf x}$ in the upper as well as 
the lower half-space, as long as ${\bf x}$ is above the  source in the lower half-space.
Similarly, substituting equation (\ref{eqFoc2}) into equation (\ref{eqHuyg6}) for forward extrapolation of downgoing waves to ${\bf x}$ below $\mathbb{S}_0$, 
or substituting equation (\ref{eqFoc1}) into equation (\ref{eqHuyginv4}) for inverse extrapolation of downgoing waves to ${\bf x}$ above $\mathbb{S}_0$, we obtain in both cases
\begin{equation}
p^+({\bf x},t)=\int_{\mathbb{S}_0}F({\bf x},{\bf x}',-t)*p^+({\bf x}',t){\rm d}{\bf x}'.\label{eqFoc4}
\end{equation}
Hence, also this expression holds 
for ${\bf x}$ in the upper and in the lower half-space, as long as ${\bf x}$ is below the  source in the upper half-space.
In absence of sources,  equations (\ref{eqFoc3}) and (\ref{eqFoc4}) hold throughout space.

By introducing the focusing function and its time-reversal, we achieved that the four equations for forward and inverse extrapolation of upgoing and downgoing waves
through a homogeneous medium
(equations (\ref{eqHuyg5}), (\ref{eqHuyg6}), (\ref{eqHuyginv3}) and (\ref{eqHuyginv4})) are now captured by the two equations (\ref{eqFoc3}) and (\ref{eqFoc4}).
\rev{These equations formulate modified versions of Huygens' principle for a homogeneous medium.}
 Figures \ref{Fig1} and \ref{Fig3} can be seen as examples of equation (\ref{eqFoc3}) for ${\bf x}$ above and below $\mathbb{S}_0$, respectively.
For a homogeneous medium there are no further advantages of using the focusing functions instead of the dipole Green's functions.
This changes considerably for an inhomogeneous medium.

\begin{figure}
\centerline{\hspace{2cm}\epsfysize=4.9 cm\epsfbox{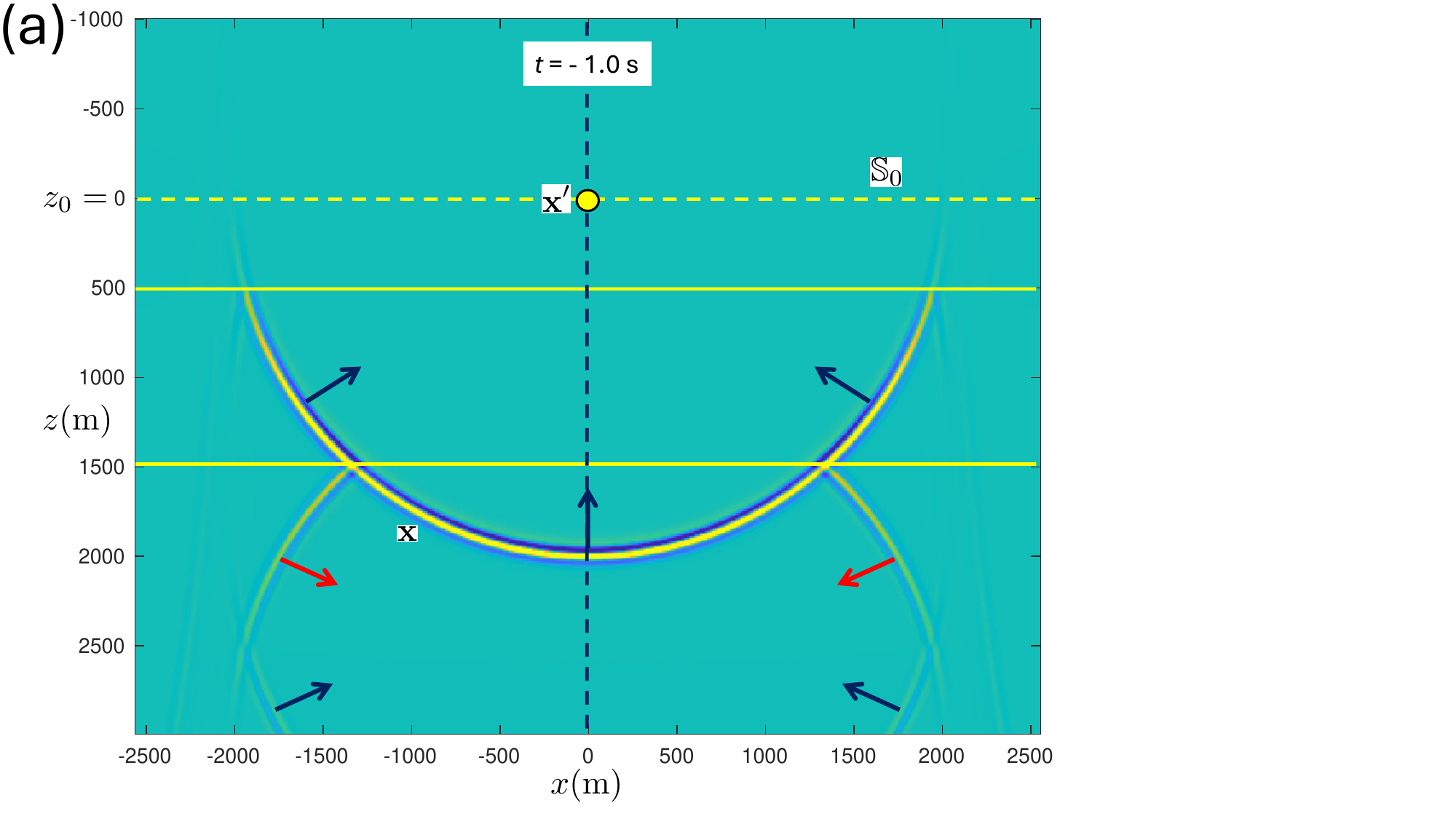}}
\centerline{\hspace{2cm}\epsfysize=4.9 cm\epsfbox{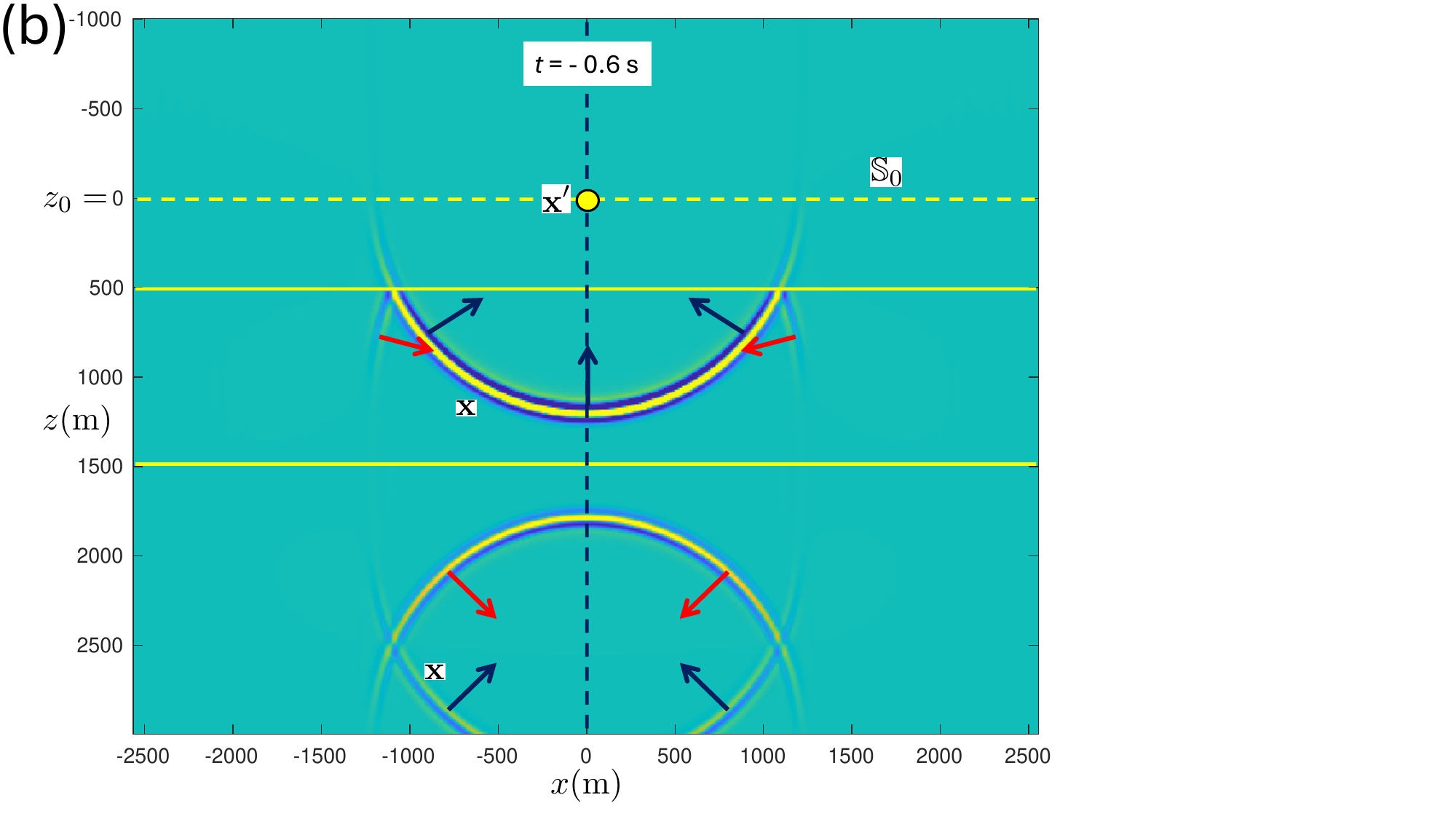}}
\centerline{\hspace{2cm}\epsfysize=4.9 cm\epsfbox{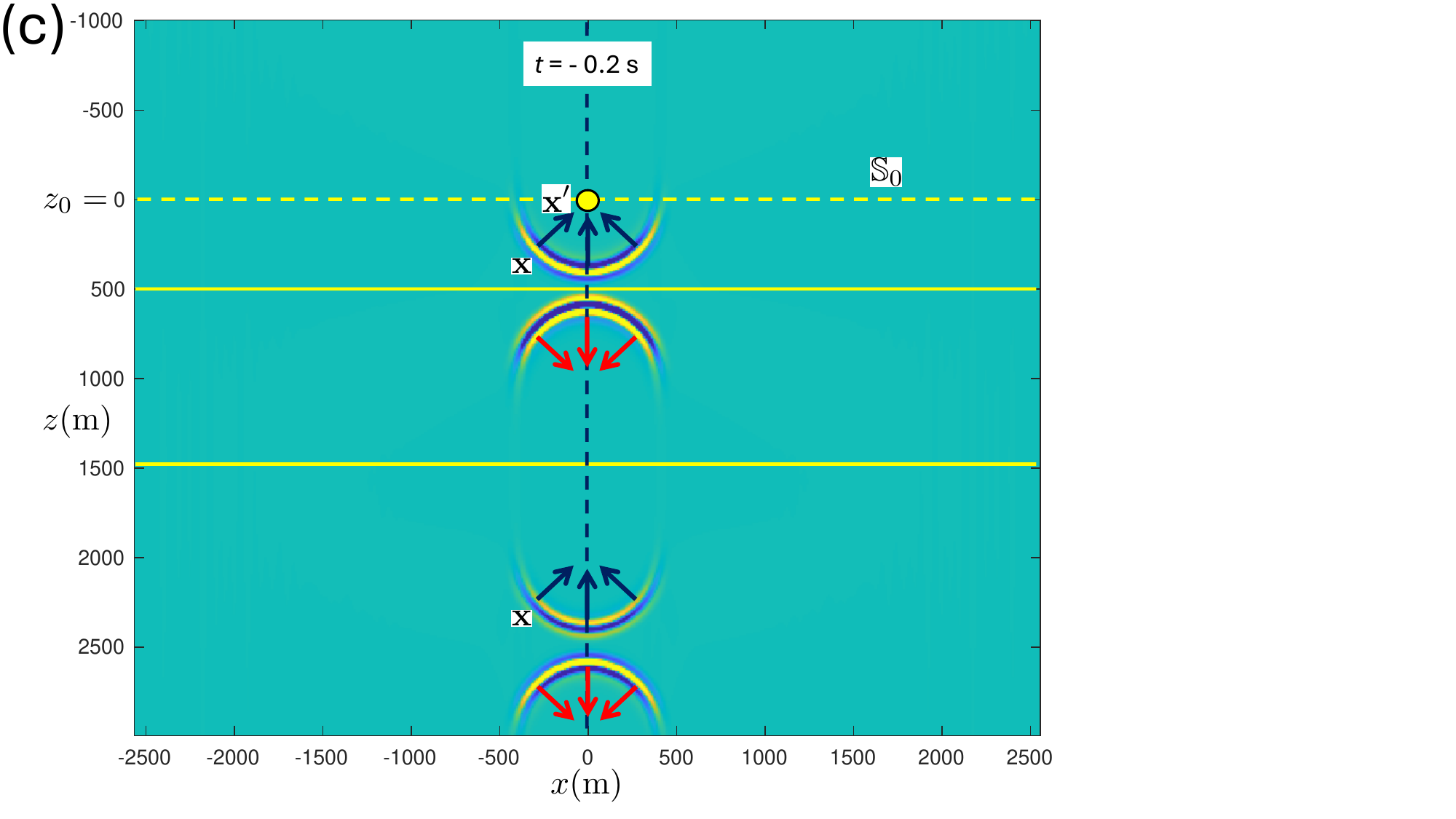}}
\centerline{\hspace{2cm}\epsfysize=4.9 cm\epsfbox{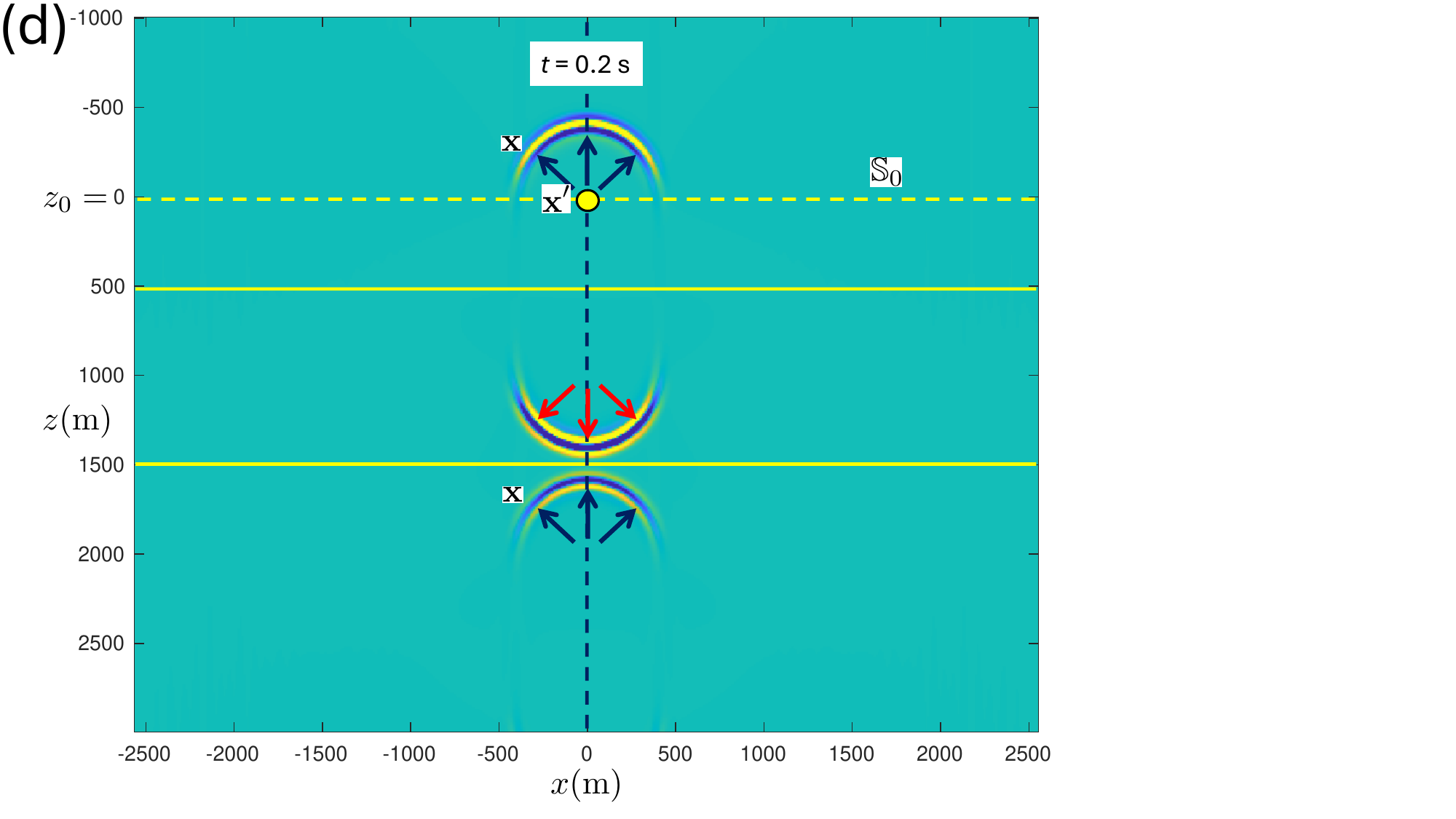}}
\caption{\it Focusing function $F({\bf x},{\bf x}',t)$ (convolved with a Ricker wavelet) in a layered medium, for a focal point at ${\bf x}'$ on $\mathbb{S}_0$ at depth $z_0$.
\rev{In the modified version of Huygens' principle for an inhomogeneous medium (equation (\ref{eq43})), this focusing function and its time-reversed version replace the dipole Green's function of Figure \ref{Fig4}.}
 }\label{Fig15}  
\end{figure}

\begin{figure}
\centerline{\hspace{2cm}\epsfysize=4.9 cm\epsfbox{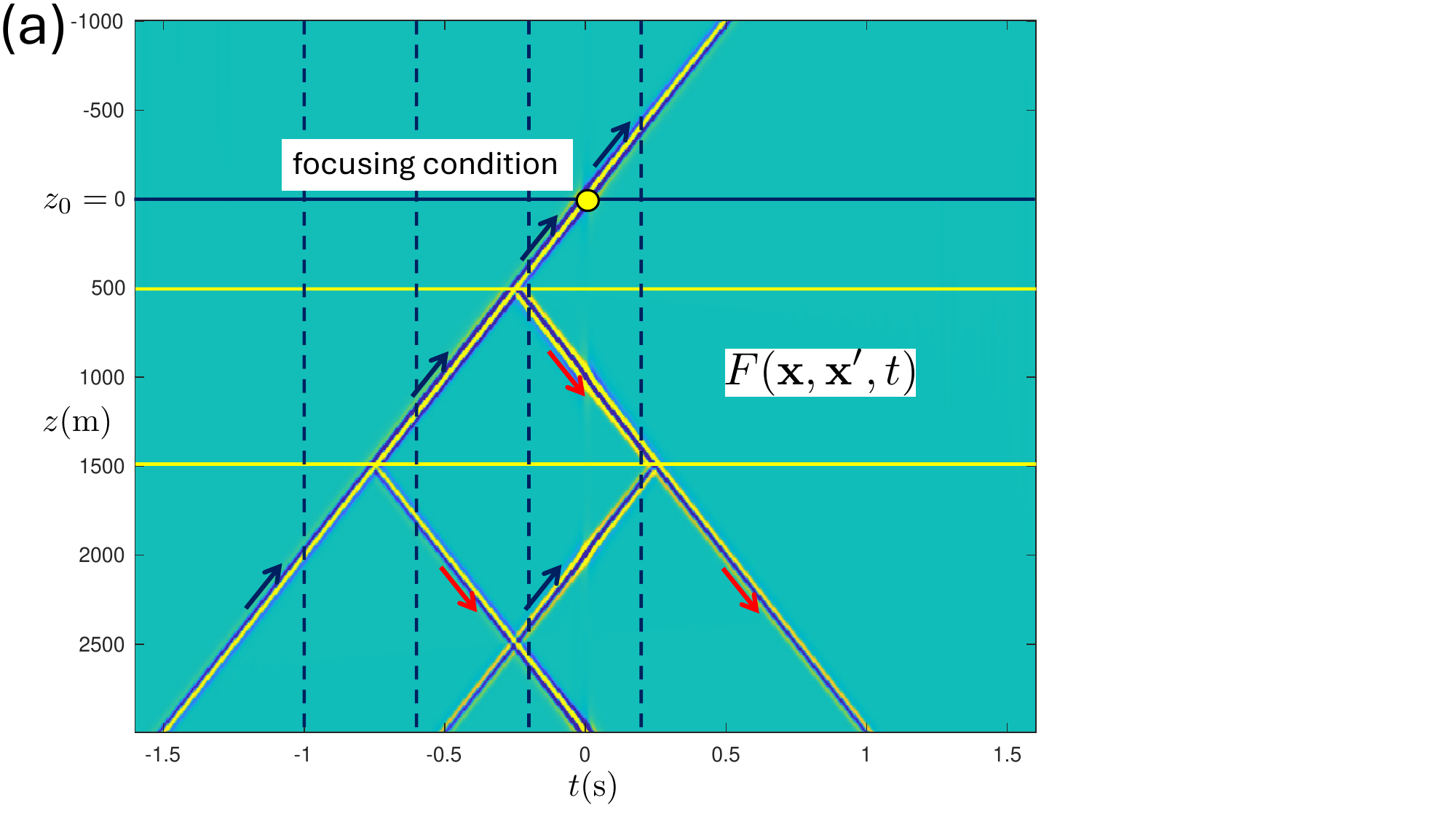}}
\centerline{\hspace{2cm}\epsfysize=4.9 cm\epsfbox{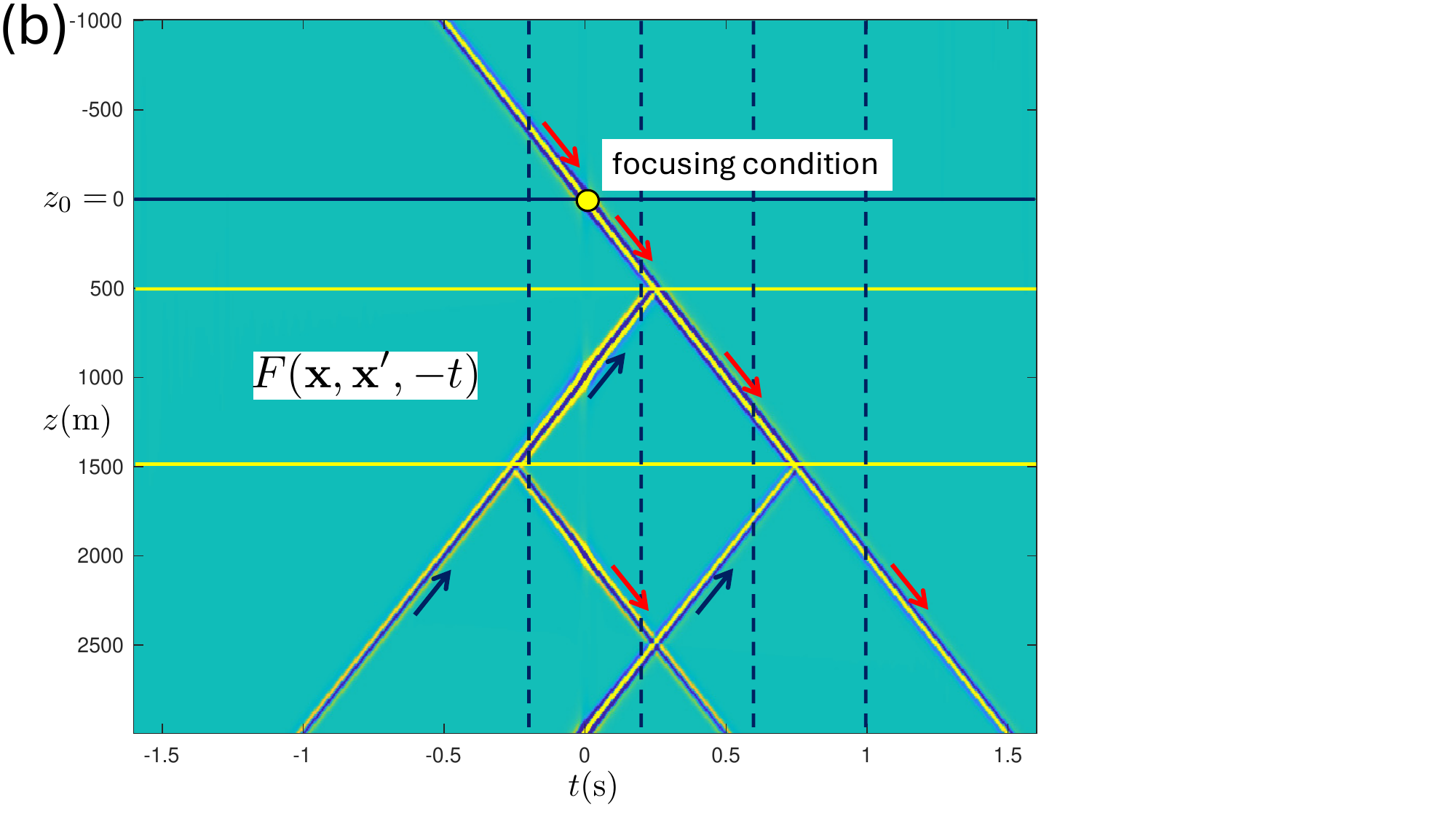}}
\centerline{\hspace{2cm}\epsfysize=4.9 cm\epsfbox{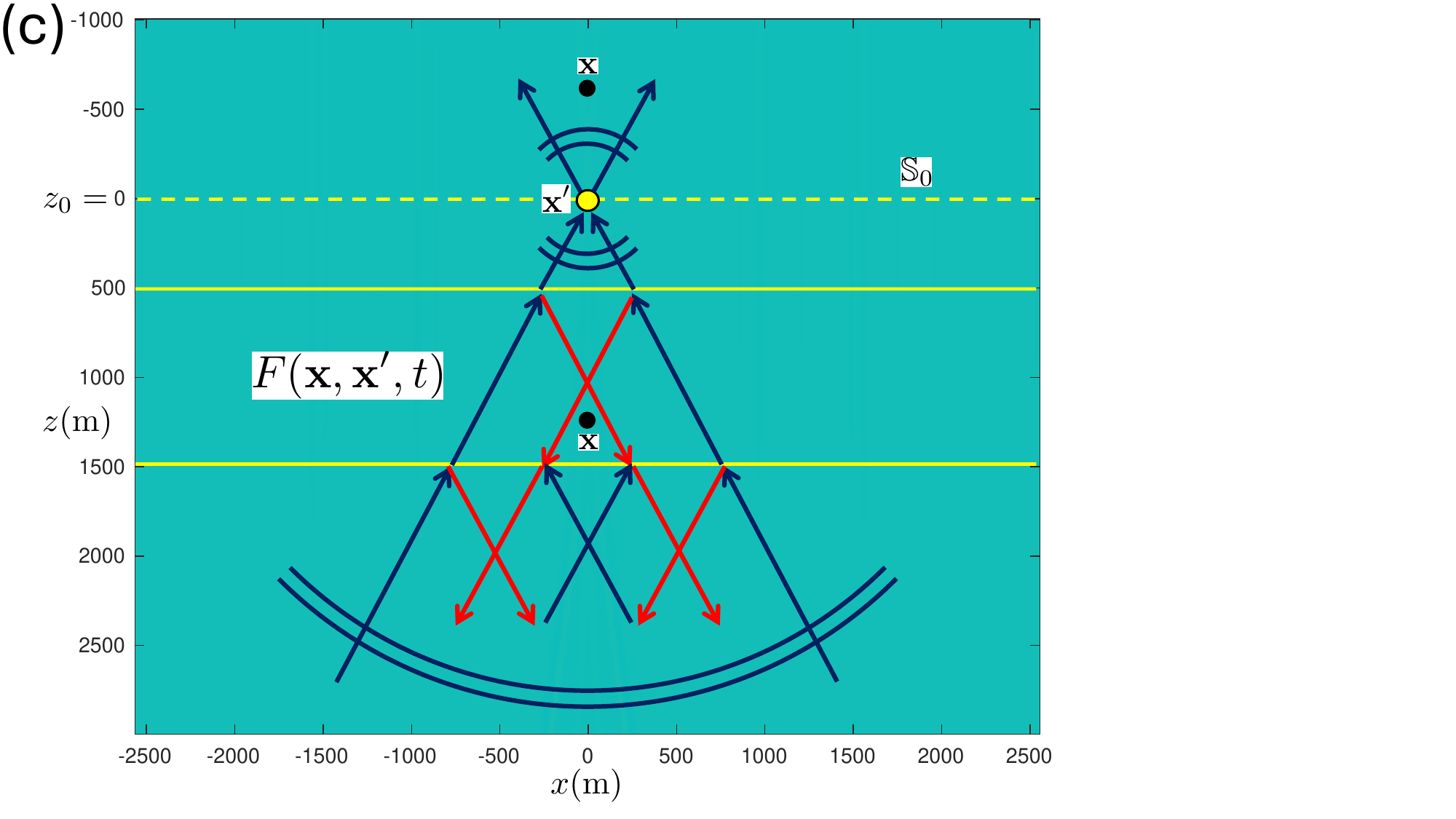}}
\centerline{\hspace{2cm}\epsfysize=4.9 cm\epsfbox{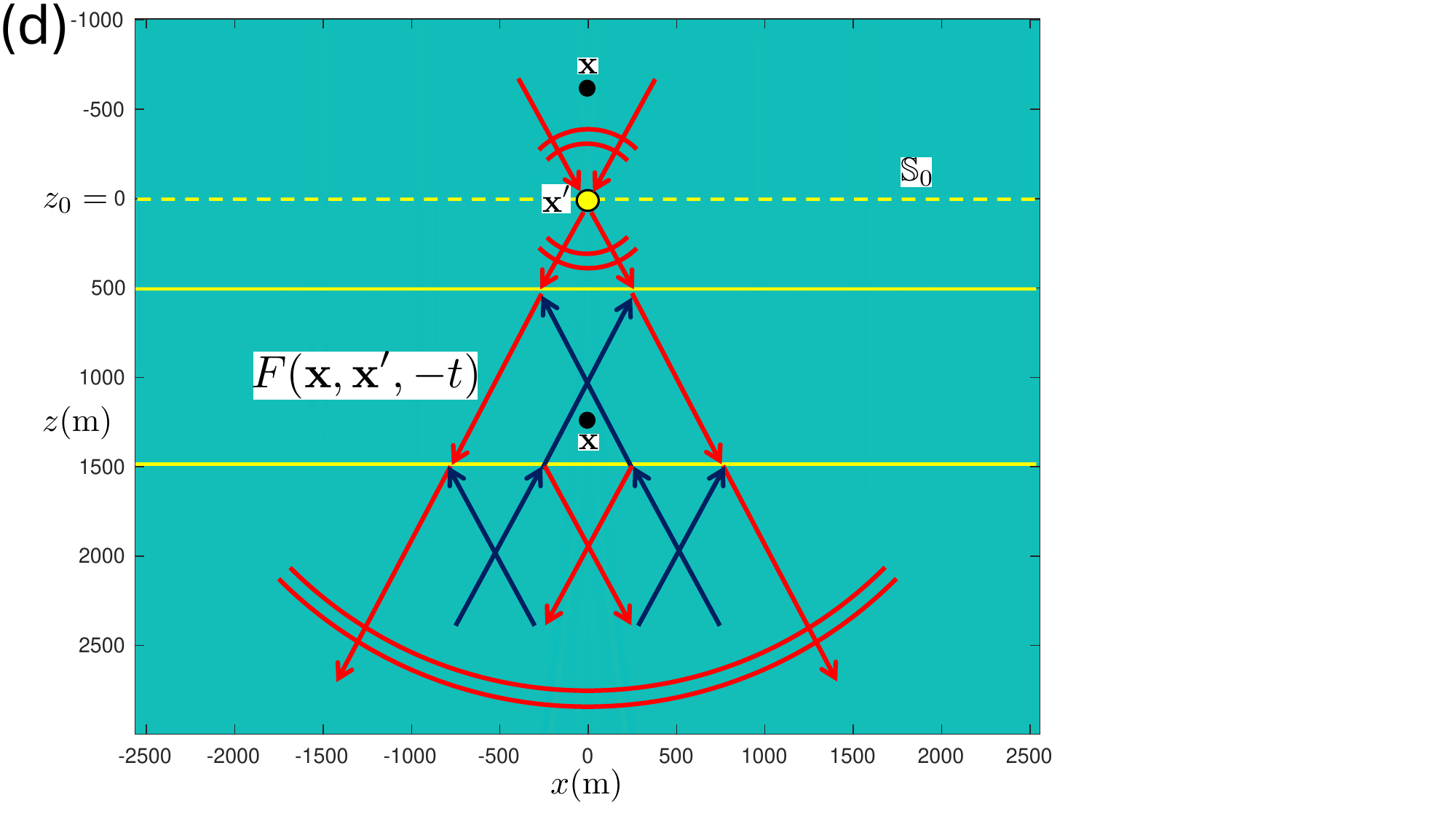}}
\caption{\it  (a),(b) Cross-sections at $x=x'=0$ and (c),(d) ray diagrams of the focusing function $F({\bf x},{\bf x}',t)$ in 
Figure \ref{Fig15} and of its time-reversed version $F({\bf x},{\bf x}',-t)$. \rev{Red and blue arrows represent downgoing and upgoing waves, respectively. 
The arcs represent some of the wave fronts at $t=\mp 1.4$ {\rm s} and $t=\mp 0.2$ {\rm s}.}}\label{Fig17}
\end{figure}

\subsection{Focusing function in an inhomogeneous medium}

We consider a medium which is inhomogeneous below the boundary $\mathbb{S}_0$ (at depth $z_0$),
\rev{with propagation velocity $c({\bf x})$ and mass density $\rho({\bf x})$.} 
At and above this boundary we assume that the medium is homogeneous,
\rev{with propagation velocity $c_0$ and mass density $\rho_0$.}
For this configuration we define the focusing function $F({\bf x},{\bf x}',t)$, with ${\bf x}'$ again denoting a focal point at $\mathbb{S}_0$ (hence, $z'=z_0$).
Throughout space, $F({\bf x},{\bf x}',t)$ obeys the source-free acoustic wave equation, with  
the condition that at $\mathbb{S}_0$ it obeys the focusing condition, formulated by equation (\ref{eqFoc1b}),
and that at and above $\mathbb{S}_0$ it propagates upward. Hence, at and above $\mathbb{S}_0$ this focusing function is the same as that for a homogeneous medium, 
discussed in the previous section; below $\mathbb{S}_0$ it is of course different. Assuming the medium is lossless throughout space, the time-reversed
focusing function $F({\bf x},{\bf x}',-t)$ obeys the same source-free wave equation as  $F({\bf x},{\bf x}',t)$. At and above $\mathbb{S}_0$ this time-reversed focusing function
propagates downward.

We illustrate $F({\bf x},{\bf x}',t)$ for the same layered medium as used for previous examples. Figures \ref{Fig15}a -- \ref{Fig15}d show snapshots at 
times $t=-1.0$ s, $t=-0.6$ s, $t=-0.2$ s and $t=0.2$ s, respectively. 
Figures \ref{Fig17}a and \ref{Fig17}b show cross-sections of $F({\bf x},{\bf x}',t)$, and of its time-reversed version $F({\bf x},{\bf x}',-t)$, 
along a vertical line through the focal point, as a function of depth $z$ and time $t$. 
The vertical dashed lines in Figure \ref{Fig17}a at $t=-1.0$ s, $t=-0.6$ s, $t=-0.2$ s and $t=0.2$ s correspond to the vertical dashed lines in the 
snapshots in Figures  \ref{Fig15}a -- \ref{Fig15}d.
Figures \ref{Fig17}c and \ref{Fig17}d show ray diagrams of $F({\bf x},{\bf x}',t)$ and $F({\bf x},{\bf x}',-t)$.
From these figures we observe that the focusing function $F({\bf x},{\bf x}',t)$ starts with upgoing waves in the half-space below the deepest interface, which are tuned in such a way
that, after interaction at the interfaces, a single upgoing wave converges to the focal point (Figures \ref{Fig15}c and \ref{Fig17}c), 
focuses at ${\bf x}={\bf x}'$ and $t=0$, and continues as a single upgoing wave, diverging from the focal point (Figures \ref{Fig15}d and \ref{Fig17}c).

Note the different character of the focusing function $F({\bf x},{\bf x}',t)$ in Figures \ref{Fig15} and \ref{Fig17} in comparison with the dipole Green's function 
$G_{\rm d}({\bf x},{\bf x}',t)$ in Figure \ref{Fig4}. Whereas the dipole Green's function obeys a causality condition related to the source at $t=0$, 
indicated by the vertical solid line in Figure \ref{Fig4}c,
the focusing function obeys a focusing condition at $z_0$, indicated by the horizontal solid lines in Figures \ref{Fig17}a and \ref{Fig17}b. 
Conversely,  the focusing function is not causal 
(it exists at negative and positive times, see Figure \ref{Fig17}a), whereas the dipole Green's function does not focus at $z_0$ 
(it contains multiple upgoing events at $z_0$, see Figures \ref{Fig4}c and \ref{Fig4}d).

\subsection{Modified Huygens' principle \rev{for an inhomogeneous medium}}

We now discuss how the focusing functions $F({\bf x},{\bf x}',t)$ and $F({\bf x},{\bf x}',-t)$ can replace the dipole Green's functions 
$G_{\rm d}({\bf x},{\bf x}',t)$ and $G_{\rm d}({\bf x},{\bf x}',-t)$ in Huygens' principle \rev{for an inhomogeneous medium}. 
Recall that we assume that the medium at and above the boundary $\mathbb{S}_0$ is
homogeneous, \rev{with propagation velocity $c_0$ and mass density $\rho_0$.}
This means that in the upper half-space we can handle downgoing and upgoing waves independent of each other. 
On the other hand, 
the medium below $\mathbb{S}_0$ is inhomogeneous, \rev{with propagation velocity $c({\bf x})$ and mass density $\rho({\bf x})$.} 
Although in inhomogeneous media, decomposition into downgoing and upgoing waves is often possible locally, 
in the following analysis we will not make use of this, so in the lower half-space we will consider the total wave field. 
For the moment we will assume that the entire medium (at, above and below $\mathbb{S}_0$) is source-free for the wave field $p({\bf x},t)$.
\rev{Hence, $p({\bf x},t)$ obeys the wave equation ${\cal L}p=0$ (with ${\cal L}$ defined in equation ($A$-2)) in the entire medium.
The focusing functions $F({\bf x},{\bf x}',t)$ and $F({\bf x},{\bf x}',-t)$ obey the same source-free wave equation.}

In the previous section we remarked that at and above  $\mathbb{S}_0$, the focusing function $F({\bf x},{\bf x}',t)$ 
 is the same as that for a homogeneous medium.
Hence, equations (\ref{eqFoc3}) and (\ref{eqFoc4}), which were derived for ${\bf x}$ in the upper and lower half-space in a homogeneous medium, still hold for ${\bf x}$ in the
homogeneous half-space above $\mathbb{S}_0$, even when the medium below $\mathbb{S}_0$ is inhomogeneous.
Defining $p({\bf x},t)=p^-({\bf x},t)+p^+({\bf x},t)$ for ${\bf x}$ at and above $\mathbb{S}_0$, we obtain from equations (\ref{eqFoc3}) and (\ref{eqFoc4})
\begin{eqnarray}
p({\bf x},t)&=&\int_{\mathbb{S}_0} F({\bf x},{\bf x}',t)*p^-({\bf x}',t){\rm d}{\bf x}'\nonumber\\
&+&\int_{\mathbb{S}_0} F({\bf x},{\bf x}',-t)*p^+({\bf x}',t){\rm d}{\bf x}',\label{eq43}
\end{eqnarray}
\rev{for ${\bf x}$ at and above $\mathbb{S}_0$. The focusing functions $F({\bf x},{\bf x}',t)$ and $F({\bf x},{\bf x}',-t)$ 
are mutually independent solutions of the same source-free wave equation as that for $p({\bf x},t)$. Hence,} 
from a mathematical viewpoint, equation (\ref{eq43}) expresses $p({\bf x},t)$ as a superposition of \rev{these independent solutions,}
with $p^-({\bf x}',t)$ and $p^+({\bf x}',t)$ being their \rev{(convolutional)} coefficients. Although we derived this equation from equations (\ref{eqFoc3}) and (\ref{eqFoc4}) 
for ${\bf x}$ at and above $\mathbb{S}_0$, the
quantities $p({\bf x},t)$, $F({\bf x},{\bf x}',t)$ and $F({\bf x},{\bf x}',-t)$ all obey the same source-free wave equation for all ${\bf x}$ (at, above and below $\mathbb{S}_0$).
Hence, if we write $p({\bf x},t)$ for ${\bf x}$ below $\mathbb{S}_0$ as a superposition of $F({\bf x},{\bf x}',t)$ and $F({\bf x},{\bf x}',-t)$, the coefficients must be the same as 
for ${\bf x}$ at and above $\mathbb{S}_0$. In other words, equation (\ref{eq43}) holds for all ${\bf x}$ throughout space.
When there are sources for the wave field $p({\bf x},t)$ in the upper  half-space, then equation (\ref{eq43}) holds for all ${\bf x}$ below the shallowest source
(following the same reasoning as for the homogeneous medium).
A more formal derivation of equation (\ref{eq43}) for all ${\bf x}$ below the shallowest source is presented in Appendix D.
From this derivation it also follows that evanescent waves are neglected at $\mathbb{S}_0$. 
This does not mean that evanescent waves are neglected altogether. Waves that are propagating at $\mathbb{S}_0$ may become evanescent in high-velocity layers
and equation (\ref{eq43}) accounts for such evanescent waves \citep{Wapenaar2021GJI}.

\begin{figure}
\centerline{\hspace{2cm}\epsfysize=4.9 cm\epsfbox{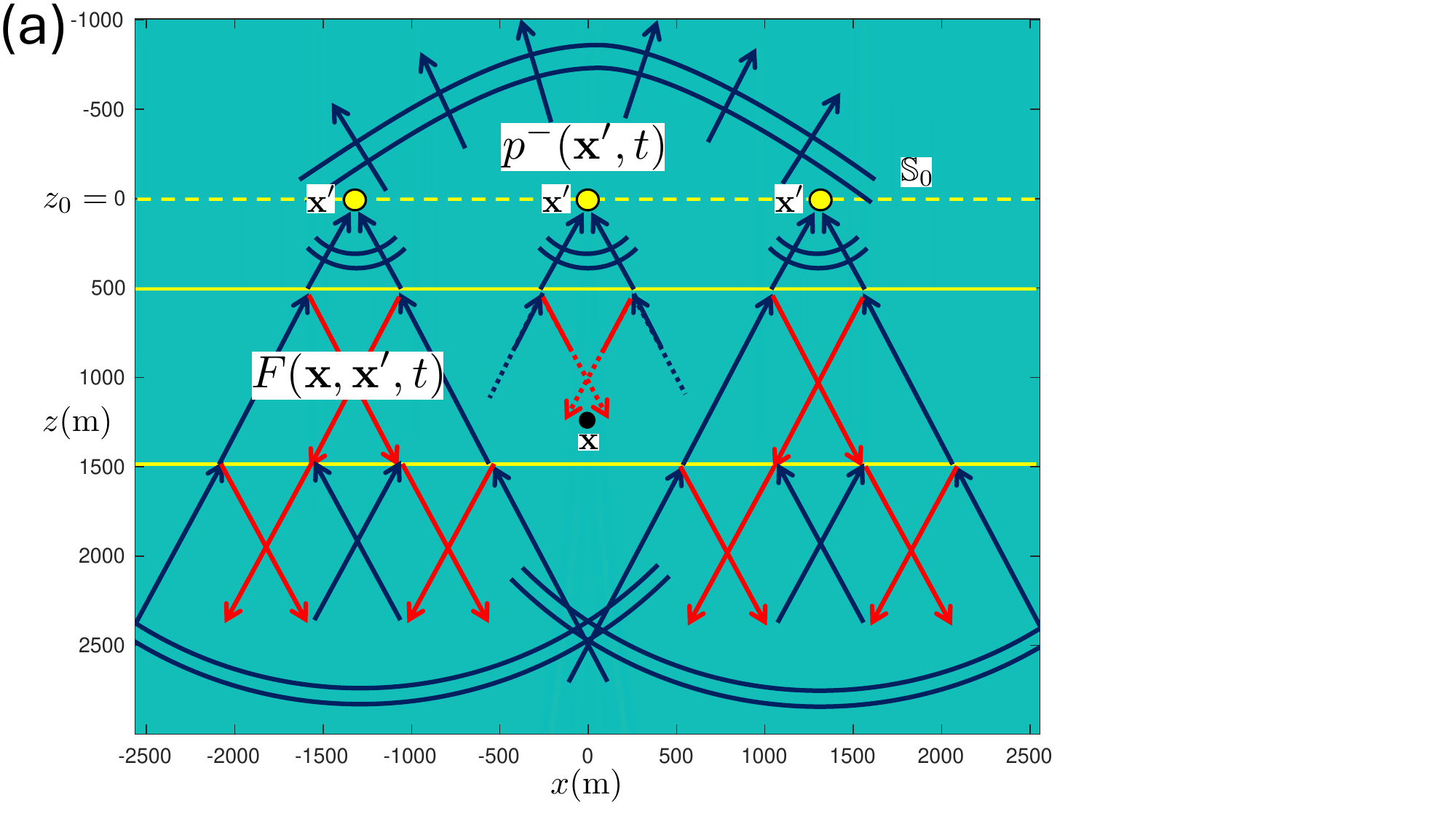}}
\centerline{\hspace{2cm}\epsfysize=4.9 cm\epsfbox{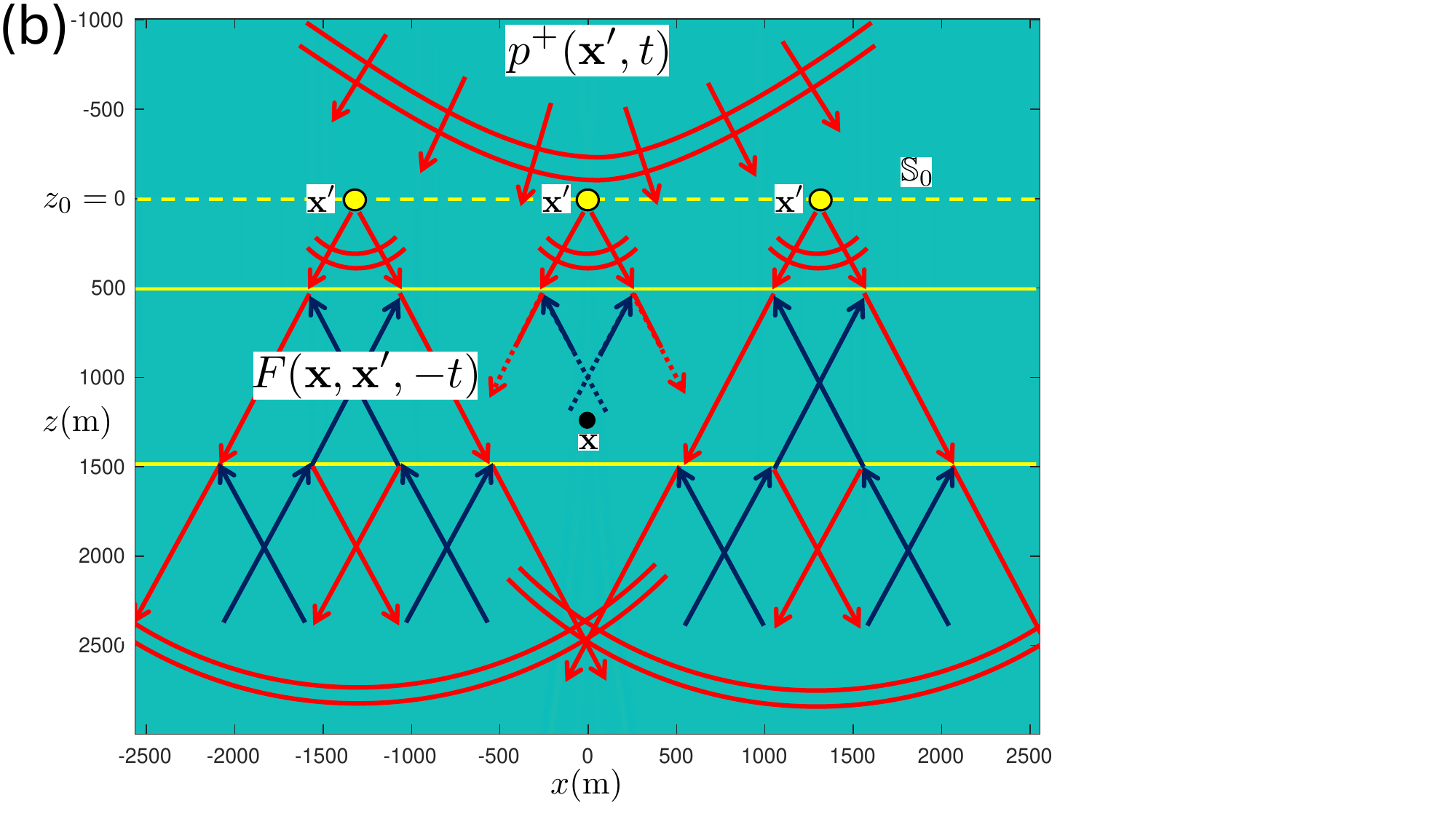}}
\caption{\it  Ray diagrams of the modified Huygens' principle, 
as formulated by equation (\ref{eq43}), \rev{with the focusing functions of Figure \ref{Fig17}}. (a) The first and (b) the second term in equation (\ref{eq43}).
}\label{Fig43}
\end{figure}

Equation (\ref{eq43}) formulates the modified Huygens' principle \rev{for an inhomogeneous medium}.
The two terms on the right-hand side  are illustrated \rev{with ray diagrams} in Figure \ref{Fig43}, for ${\bf x}$ below $\mathbb{S}_0$. 
First consider Figure \ref{Fig43}b. A downgoing wave field $p^+({\bf x}',t)$ is incident from above to the inhomogeneous lower half-space. For each ${\bf x}'$ on $\mathbb{S}_0$
it is convolved with the time-reversed focusing function $F({\bf x},{\bf x}',-t)$. 
The integral over all ${\bf x}'$ on $\mathbb{S}_0$, as formulated in the second term on the right-hand side of equation (\ref{eq43}), extrapolates the field $p^+({\bf x}',t)$ 
from $\mathbb{S}_0$ into the lower half-space. 
Since the focusing function implicitly
consists of a superposition of downgoing and upgoing waves in the lower half-space (the red and blue rays in Figure \ref{Fig43}b), the result of this integral is {\it not} the 
forward extrapolated downgoing field $p^+({\bf x},t)$ in the lower half-space (unlike in the homogeneous medium situation, as formulated by equation (\ref{eqFoc4})).
Next, consider Figure \ref{Fig43}a. Here the upgoing field $p^-({\bf x}',t)$ is convolved with the  focusing function $F({\bf x},{\bf x}',t)$ for all ${\bf x}'$ on $\mathbb{S}_0$.
The integral over all ${\bf x}'$ on $\mathbb{S}_0$ (the first term on the right-hand side of equation (\ref{eq43})), extrapolates the field $p^-({\bf x}',t)$ 
from $\mathbb{S}_0$ into the lower half-space. For similar reasons as above, this is {\it not} the inverse extrapolated upgoing 
field $p^-({\bf x},t)$ in the lower half-space. However, the superposition of the 
two integrals, as formulated by equation (\ref{eq43}), yields the total wave field $p({\bf x},t)$ in the lower half-space, including all internal \rev{multiply reflected waves}.
In comparison with equation (\ref{eqHuyginv6}), where the two integrals are taken over two different boundaries, 
in equation (\ref{eq43}) the two integrals are taken over one and the same
boundary. Hence, this makes equation (\ref{eq43}) very useful for practical situations in which a medium is often accessible from one side only, 
such as in the seismic reflection method. 
The focusing function can be retrieved from the reflection response, acquired at the same boundary, using the Marchenko method \rev{for 1D, 2D or 3D inhomogeneous media}
\citep{Broggini2012EJP, Wapenaar2014GEO, Slob2014GEO, Neut2015GJI, Meles2015GEO}.
In most papers on the Marchenko method it is assumed that  the wave field \rev{and focusing functions} inside the medium can be decomposed into downgoing and upgoing components. 
This decomposition is avoided in equation (\ref{eq43}), which opens the way to handle refracted and evanescent waves \citep{Wapenaar2021GJI, Diekmann2021PRR}.
A further discussion of the Marchenko method is beyond the scope of this paper. In the next sections we indicate applications of equation (\ref{eq43}), 
assuming the focusing function is known (either from numerical modeling or from applying the Marchenko method to the reflection response).

\subsection{Simultaneous forward and inverse wave field extrapolation through an inhomogeneous medium}

\begin{figure}
\centerline{\hspace{2cm}\epsfysize=4.9 cm\epsfbox{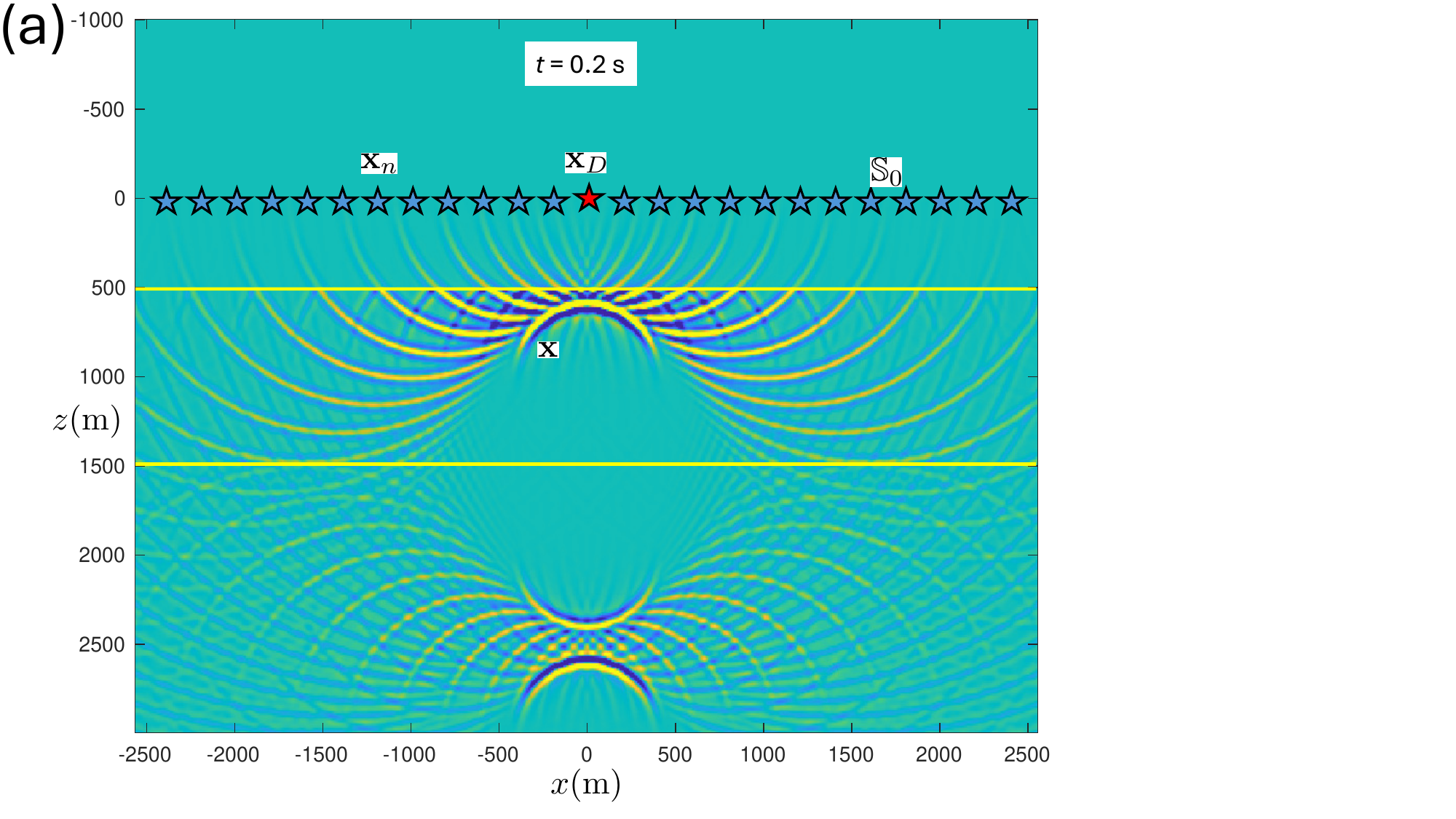}}
\centerline{\hspace{2cm}\epsfysize=4.9 cm\epsfbox{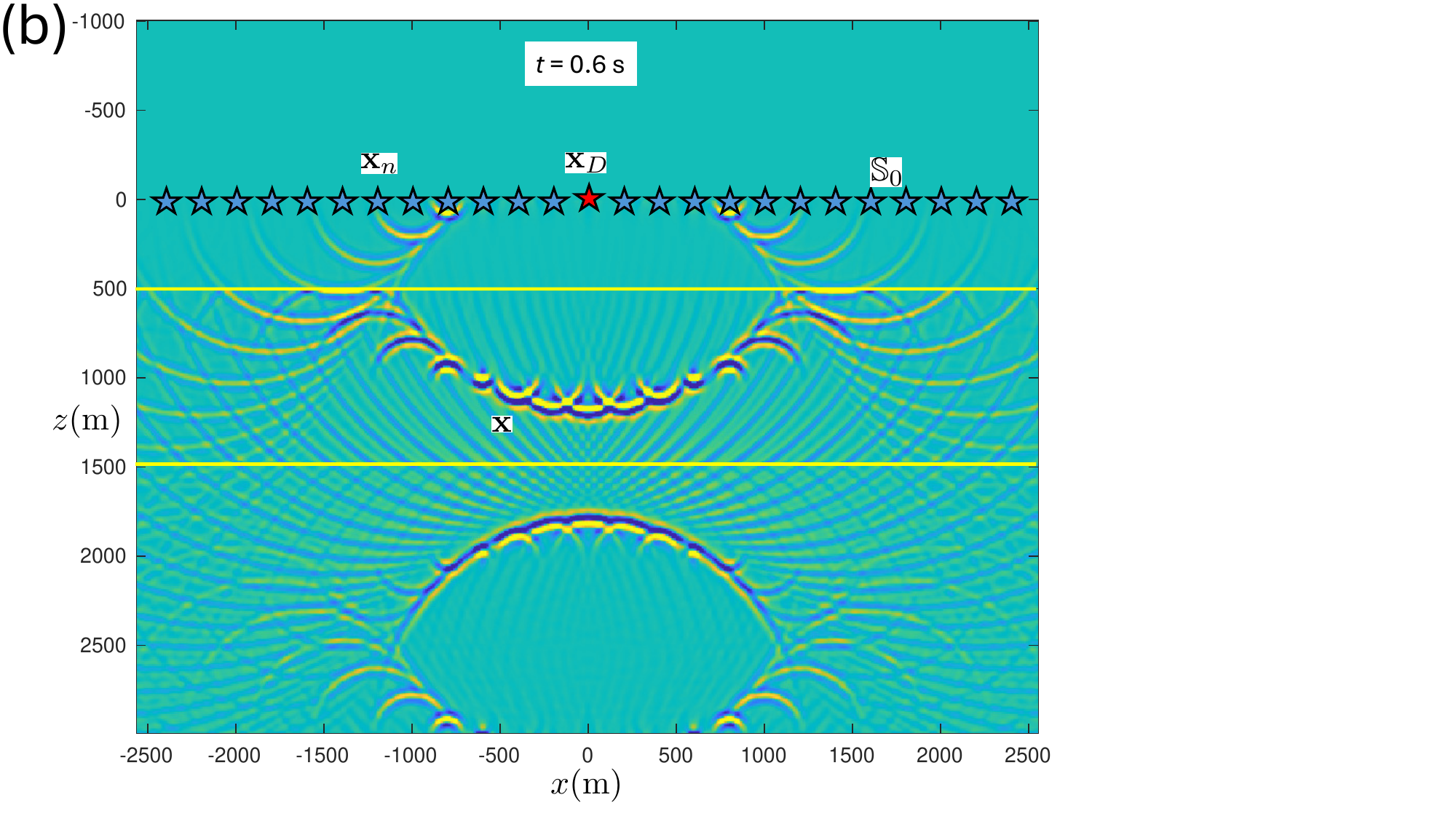}}
\centerline{\hspace{2cm}\epsfysize=4.9 cm\epsfbox{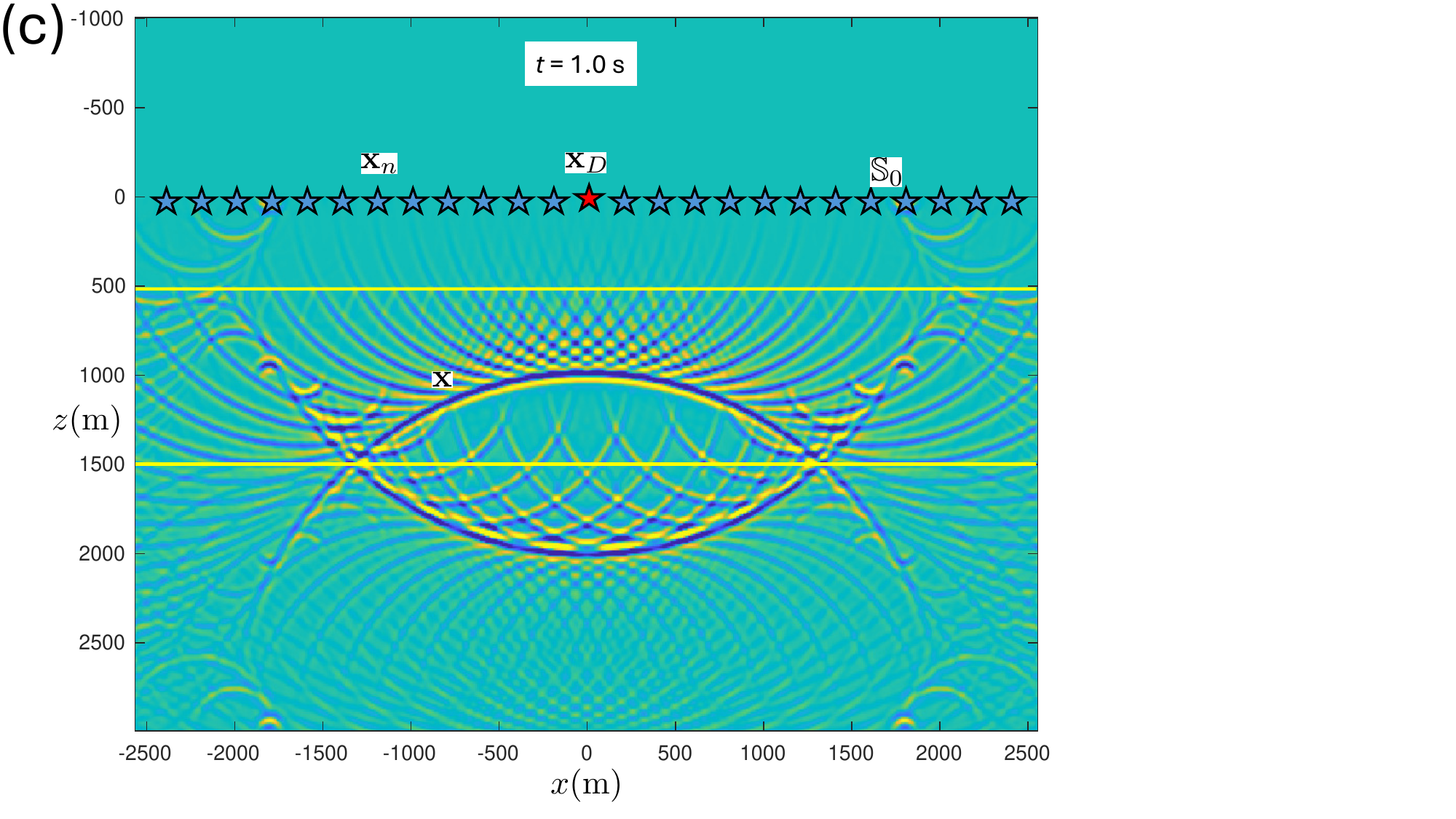}}
\centerline{\hspace{2cm}\epsfysize=4.9 cm\epsfbox{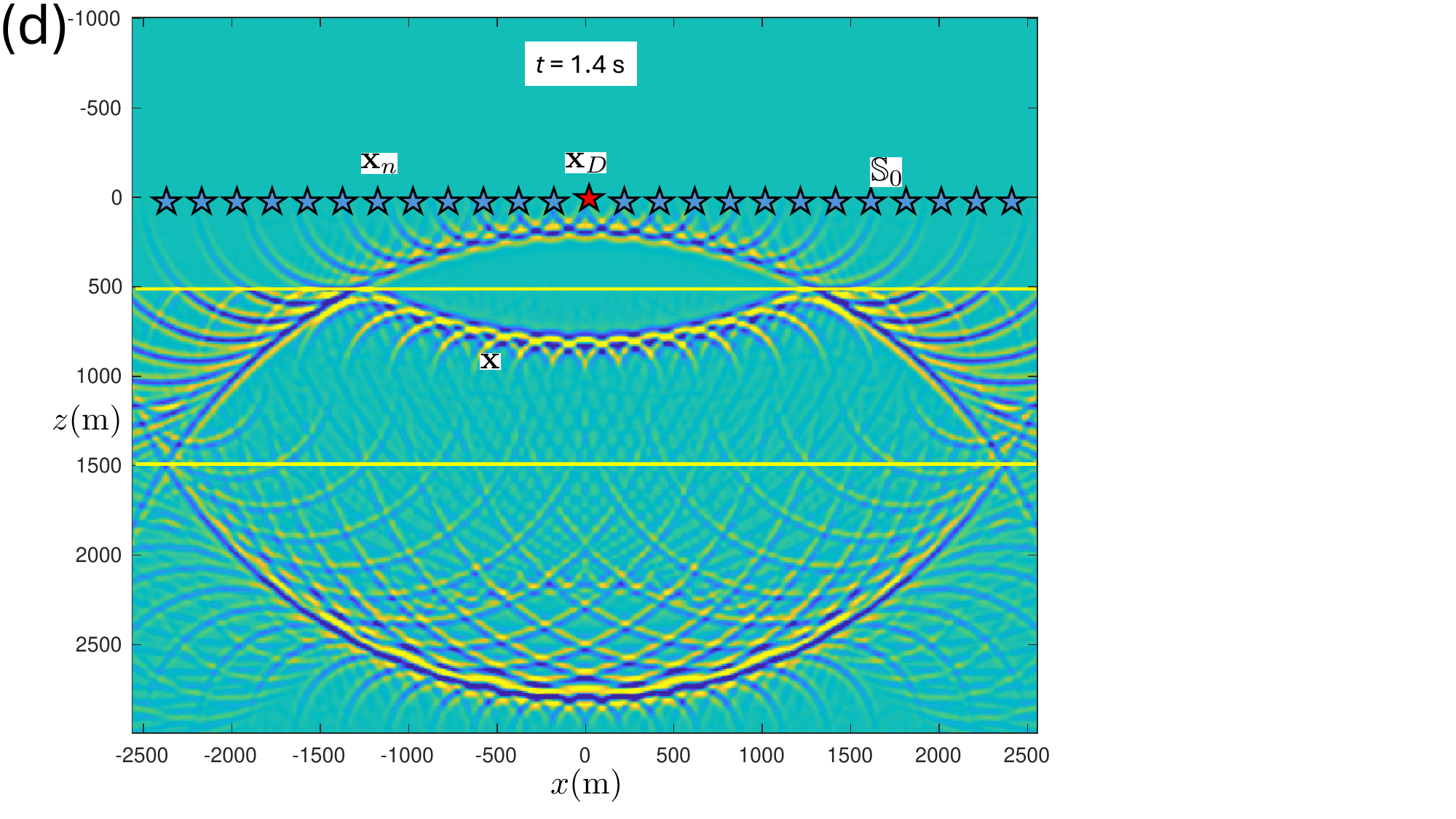}}
\caption{\it  Illustration of the first term of the modified Huygens' principle (equation (\ref{eq45}), \rev{with the focusing function of Figure \ref{Fig15}}), applied to the discretized reflection response $R({\bf x}_n,{\bf x}_D,t)*s(t)$.}\label{Fig21}
\end{figure}

\begin{figure}
\centerline{\hspace{2cm}\epsfysize=4.9 cm\epsfbox{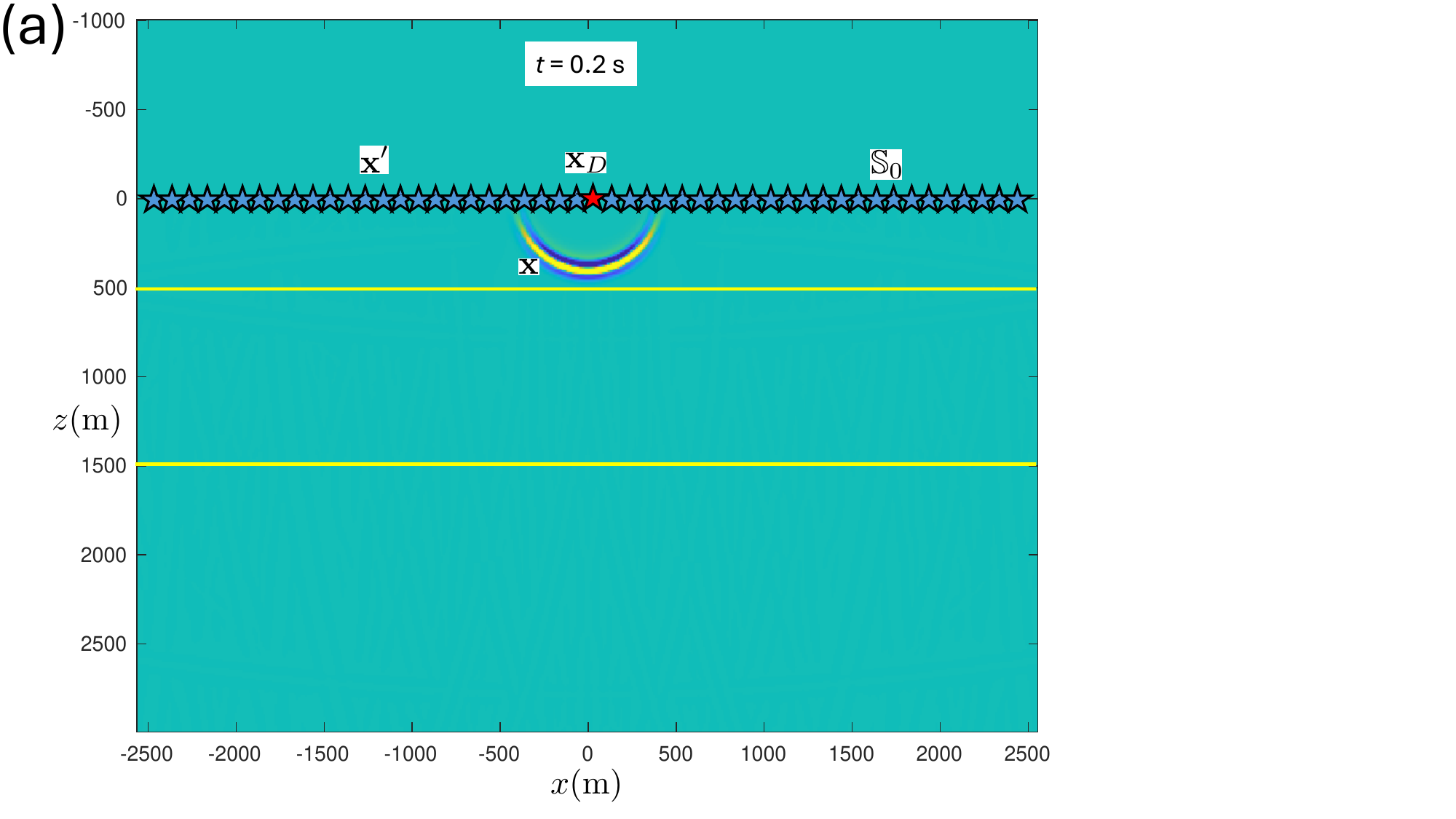}}
\centerline{\hspace{2cm}\epsfysize=4.9 cm\epsfbox{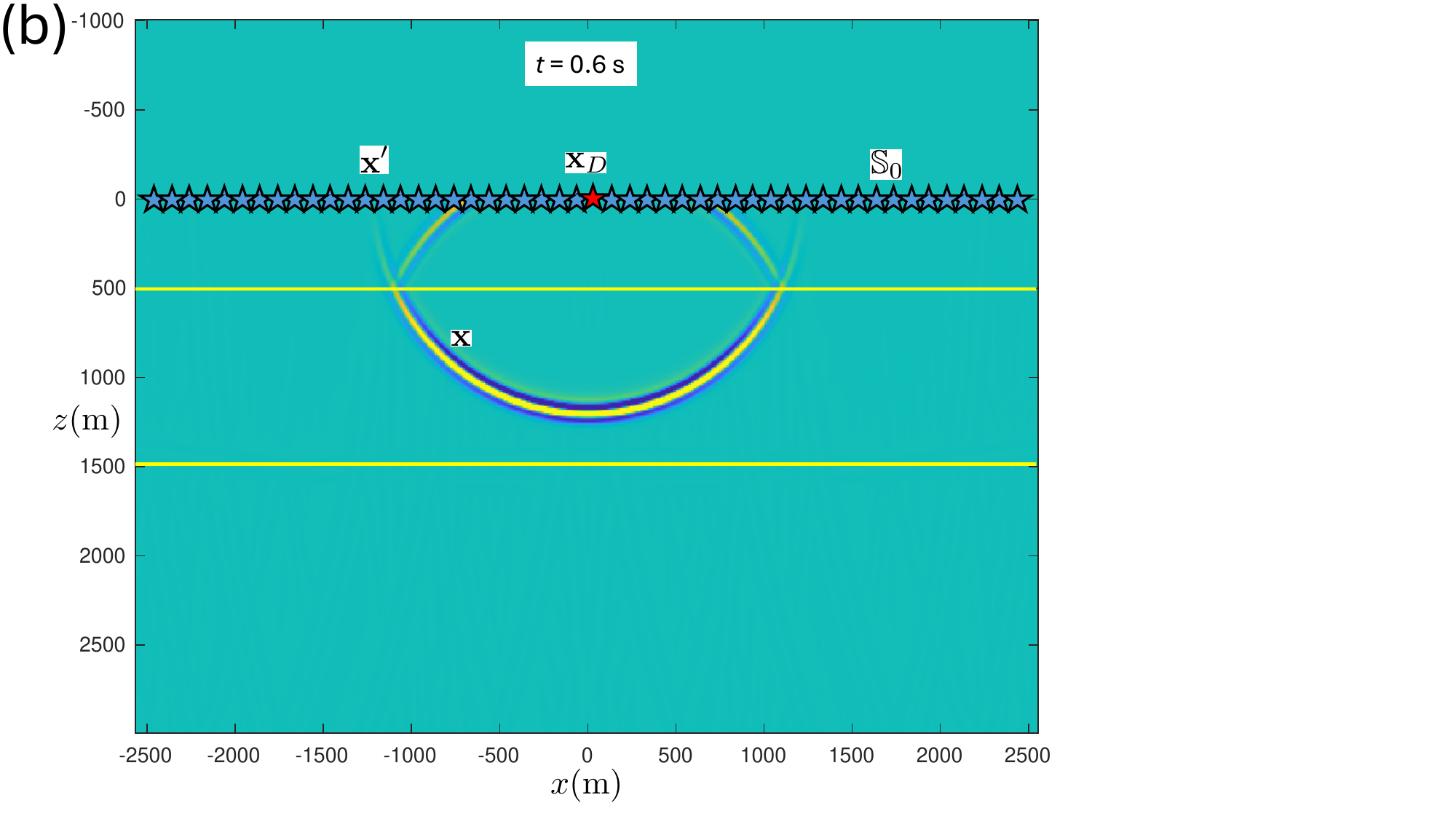}}
\centerline{\hspace{2cm}\epsfysize=4.9 cm\epsfbox{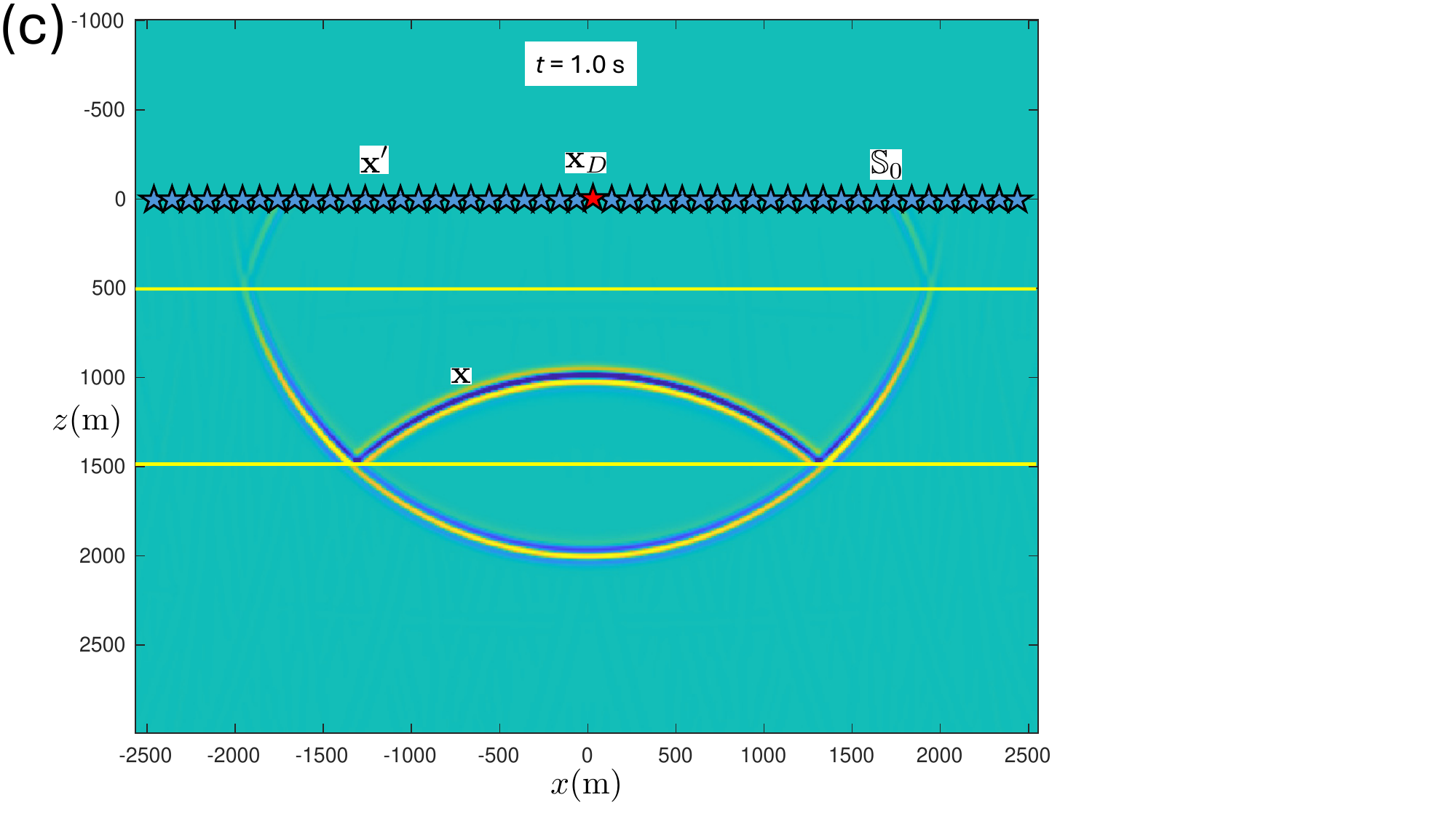}}
\centerline{\hspace{2cm}\epsfysize=4.9 cm\epsfbox{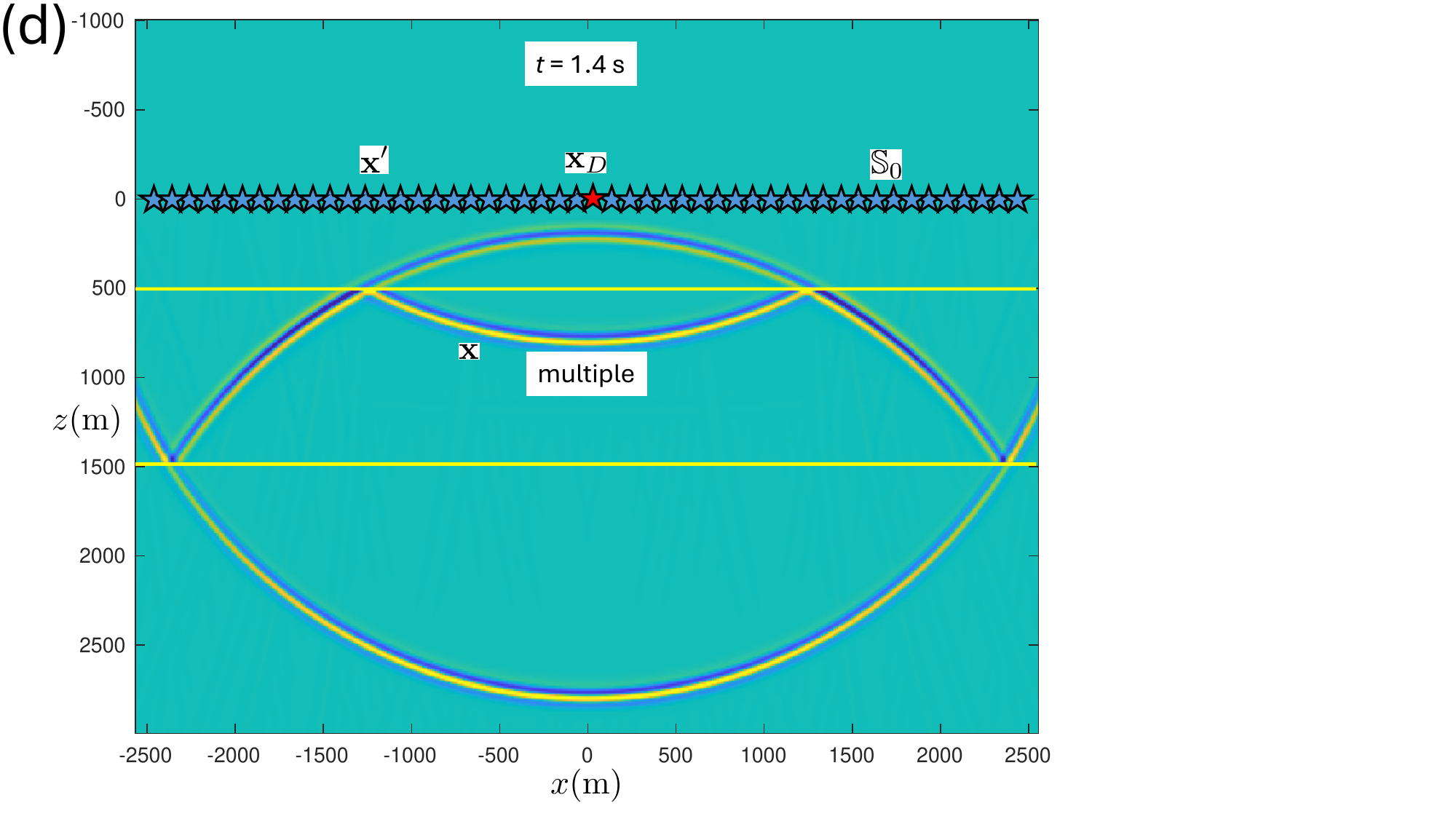}}
\caption{\it  Illustration of the modified Huygens' principle (both terms of equation (\ref{eq44})), applied to the continuous reflection response $R({\bf x}',{\bf x}_D,t)*s(t)$.}\label{Fig22}
\end{figure}

Since equation (\ref{eq43}) yields the total wave field in the inhomogeneous half-space below $\mathbb{S}_0$ from downgoing and upgoing waves at $\mathbb{S}_0$,
we can interpret it as an expression for simultaneous forward and inverse wave field extrapolation. 
We illustrate this for the situation of reflection measurements at the boundary $\mathbb{S}_0$.
We define the reflection response $R({\bf x}',{\bf x}'',t)$ of the inhomogeneous lower half-space via the relation
\begin{equation}
p^-({\bf x}',t)=\int_{\mathbb{S}_0}R({\bf x}',{\bf x}'',t)*p^+({\bf x}'',t){\rm d}{\bf x}'',\label{eqR}
\end{equation}
for ${\bf x}'$ and ${\bf x}''$ at $\mathbb{S}_0$.
We consider a dipole source in the homogeneous upper half-space at ${\bf x}_D=({\bf x}_{{\rm H},D},z_0-\epsilon)$ 
(with ${\bf x}_{{\rm H},D}$  denoting the horizontal component(s) of ${\bf x}_D$), 
 at a vanishing distance $\epsilon$ above $\mathbb{S}_0$. 
 This is the source for the wave field $p({\bf x},t)$.
 We scale it such that for the downgoing field at ${\bf x}''$ on $\mathbb{S}_0$ we have 
\begin{equation}
p^+({\bf x}'',t)=\delta({\bf x}_H''-{\bf x}_{{\rm H},D})s(t),\label{eq25}
\end{equation}
where $s(t)$ is the source wavelet.
Substitution of equation (\ref{eq25})  into equation (\ref{eqR}) gives 
\begin{equation}
p^-({\bf x}',t)=R({\bf x}',{\bf x}_D,t)*s(t), \label{eq26}
\end{equation}
for ${\bf x}'$ at $\mathbb{S}_0$. Substitution of equations (\ref{eq25}) and (\ref{eq26}) into equation (\ref{eq43}) yields
\begin{eqnarray}
p({\bf x},t)&=&\int_{\mathbb{S}_0} F({\bf x},{\bf x}',t)*R({\bf x}',{\bf x}_D,t)*s(t){\rm d}{\bf x}'\nonumber\\
&&+ \,\,F({\bf x},{\bf x}_D,-t)*s(t).\label{eq44}
\end{eqnarray}
Since the source at ${\bf x}_D$ lies just above $\mathbb{S}_0$, this expression holds only for ${\bf x}$ at and below $\mathbb{S}_0$.
First we evaluate a discretized version of only the first term on the right-hand side of equation (\ref{eq44}), i.e.,
\begin{equation}
\sum_{n=-N}^N F({\bf x},{\bf x}_n,t)*R({\bf x}_n,{\bf x}_D,t)*s(t),\label{eq45}
\end{equation}
with ${\bf x}_n=(n\Delta x,z_0)$, $\Delta x=200$ m and $N=50$. For this 2D example we choose ${\bf x}_{{\rm H},D}=x_D=0$.
The result is shown in Figures \ref{Fig21}a--\ref{Fig21}d, for $t=0.2$ s, $t=0.6$ s, $t=1.0$ s, and $t=1.4$ s,
respectively. The envelopes of the superposed waves converge to wave fronts, but it is not yet obvious how they are connected to the desired response $p({\bf x},t)$
(which is the response to a dipole source at ${\bf x}_D$).
Next, we replace the summation by an integration and we add the term $F({\bf x},{\bf x}_D,-t)*s(t)$, i.e., we add Figures \ref{Fig15}c, \ref{Fig15}b and \ref{Fig15}a
to the converged versions of Figures \ref{Fig21}a, \ref{Fig21}b and \ref{Fig21}c. The results are shown in Figure \ref{Fig22}.
This figure clearly shows the desired response to the dipole source at ${\bf x}_D$, observed at all ${\bf x}$ in the lower half-space, including internal \rev{multiply reflected waves} 
(compare Figures \ref{Fig22}a and \ref{Fig22}d
with the directly modeled dipole Green's function  in the lower half-space in Figures \ref{Fig4}a and \ref{Fig4}b at $t=0.2$ s and $t=1.4$ s, respectively).

Extrapolation of reflection data with focusing functions finds applications in acoustic and seismic imaging schemes, accounting for internal \rev{multiply reflected waves}
\citep{Ravasi2016GJI, Jia2018GEO, Staring2020GP, Brackenhoff2022GP}. In those applications the focusing functions are obtained with the Marchenko method
\rev{from numerically modeled or field reflection responses of 2D and 3D inhomogeneous media.}

\subsection{Retrieval of the homogeneous Green's function in an inhomogeneous medium}

In equation (\ref{eqHuyginv9}) we introduced the homogeneous Green's function $G_{\rm h}({\bf x},{\bf x}_S,t)=G({\bf x},{\bf x}_S,t)+G({\bf x},{\bf x}_S,-t)$
for an inhomogeneous lossless medium.
Both terms on the right-hand side obey a wave equation with a source at ${\bf x}_S$, but these sources cancel each other, implying that
the wave equation for $G_{\rm h}({\bf x},{\bf x}_S,t)$ is source-free, see Appendix A-2.
Hence, when in equation (\ref{eq43}) we replace $p({\bf x},t)$  by $G_{\rm h}({\bf x},{\bf x}_S,t)$, it will hold throughout space,
independent of the position of ${\bf x}_S$. With this replacement, 
the upgoing and downgoing wave fields $p^-({\bf x}',t)$ and $p^+({\bf x}',t)$ on the right-hand side of equation (\ref{eq43}) need to be replaced 
by $G_{\rm h}^-({\bf x}',{\bf x}_S,t)$ and $G_{\rm h}^+({\bf x}',{\bf x}_S,t)$, respectively. Assuming again that ${\bf x}_S$ lies below $\mathbb{S}_0$,
taking into account that ${\bf x}'$ is situated at $\mathbb{S}_0$ and that the half-space above $\mathbb{S}_0$ is homogeneous, it follows that the Green's function $G({\bf x}',{\bf x}_S,t)$
propagates upward for ${\bf x}'$ at $\mathbb{S}_0$ (see Figure \ref{Fig4a}). As a result, the time-reversed Green's function $G({\bf x}',{\bf x}_S,-t)$
propagates downward for ${\bf x}'$ at $\mathbb{S}_0$. 
Since these two terms constitute the total homogeneous Green's function, it follows that
 $G_{\rm h}^-({\bf x}',{\bf x}_S,t)=G({\bf x}',{\bf x}_S,t)$ and $G_{\rm h}^+({\bf x}',{\bf x}_S,t)=G({\bf x}',{\bf x}_S,-t)$ for ${\bf x}'$ at $\mathbb{S}_0$.
With these substitutions, equation (\ref{eq43}) becomes
\begin{eqnarray}
G_{\rm h}({\bf x},{\bf x}_S,t)&=&\int_{\mathbb{S}_0} F({\bf x},{\bf x}',t) * G({\bf x}',{\bf x}_S,t){\rm d}{\bf x}'\nonumber\\
&+&\int_{\mathbb{S}_0} F({\bf x},{\bf x}',-t) * G({\bf x}',{\bf x}_S,-t){\rm d}{\bf x}',\label{eq49}
\end{eqnarray}
for all ${\bf x}$ throughout space. Compare this with equation (\ref{eqHuyginv8}), where the two integrals are taken over two different boundaries. 
In equation (\ref{eq49}) both integrals are taken over the same boundary. Moreover, the
 second integral  is merely the time-reversal of the first integral.

\begin{figure}
\centerline{\hspace{2cm}\epsfysize=4.9 cm\epsfbox{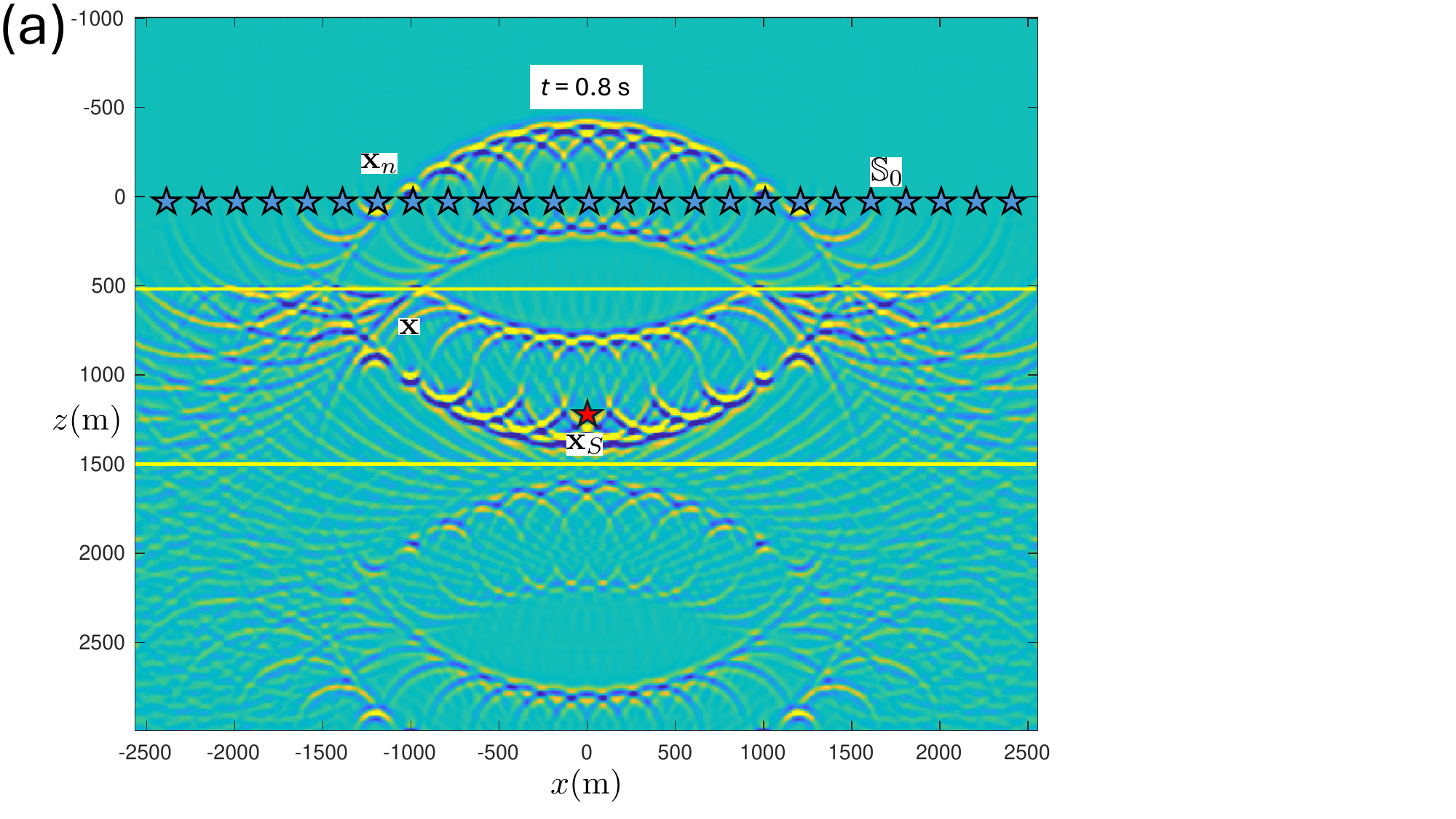}}
\centerline{\hspace{2cm}\epsfysize=4.9 cm\epsfbox{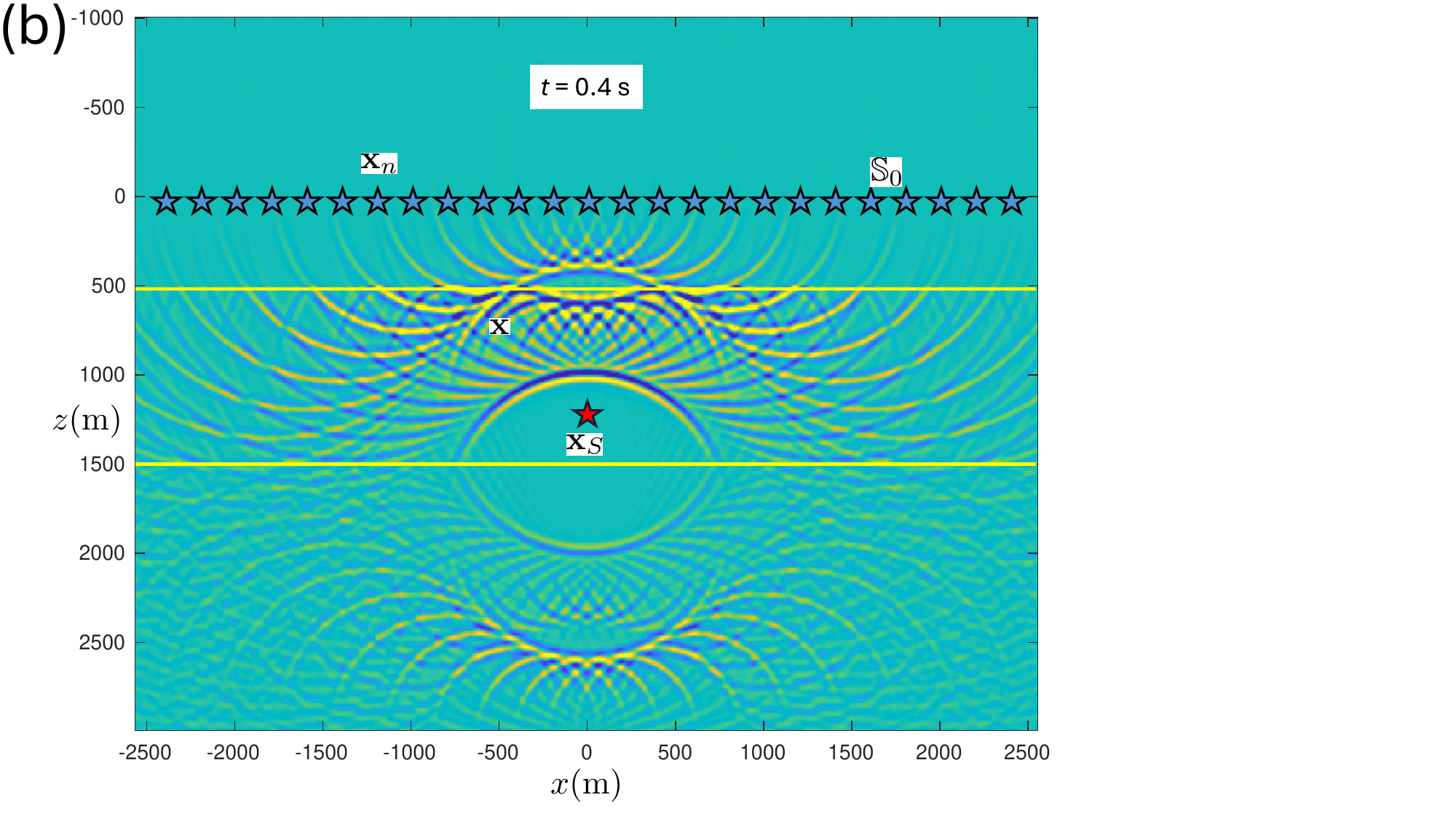}}
\centerline{\hspace{2cm}\epsfysize=4.9 cm\epsfbox{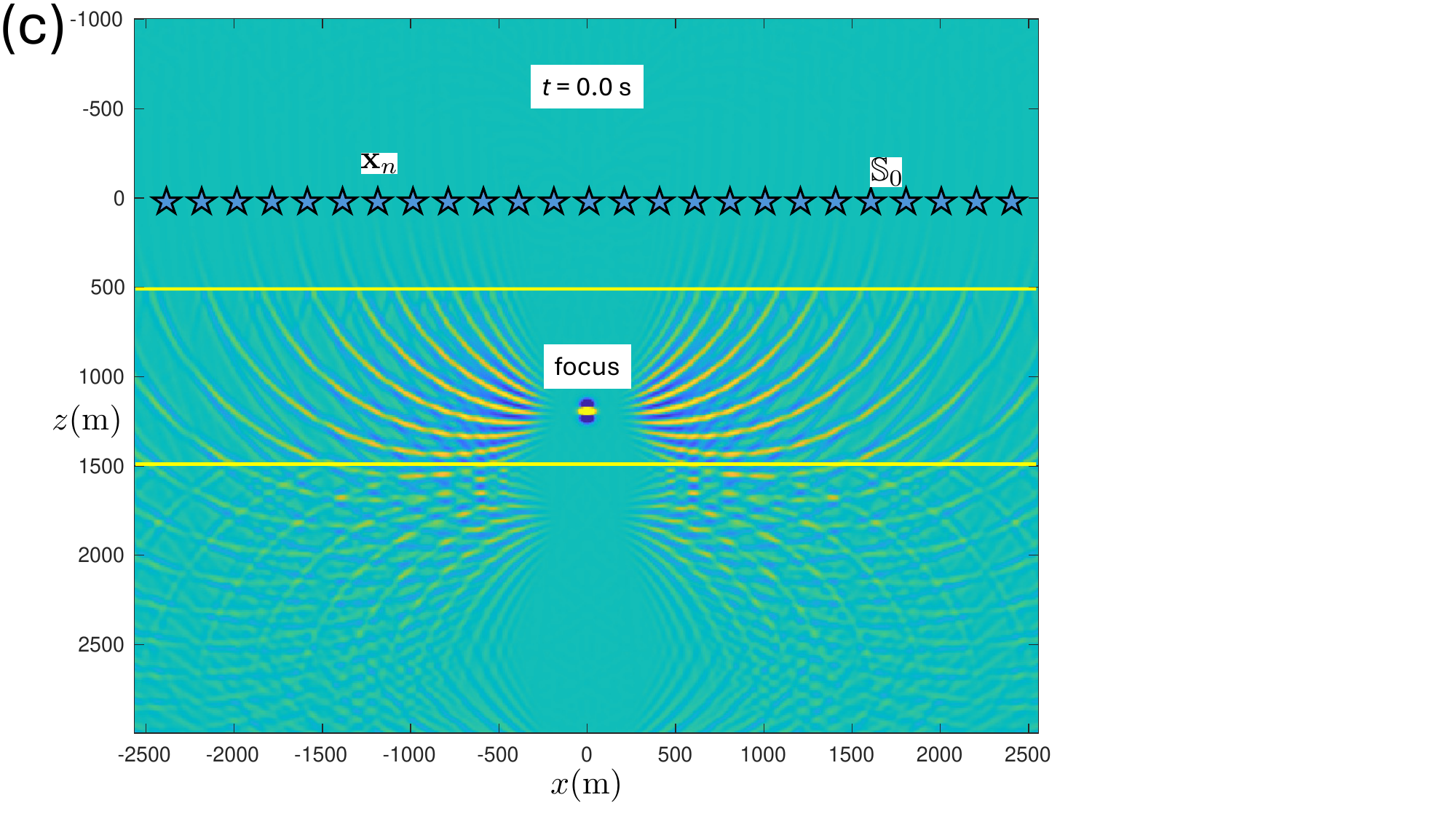}}
\centerline{\hspace{2cm}\epsfysize=4.9 cm\epsfbox{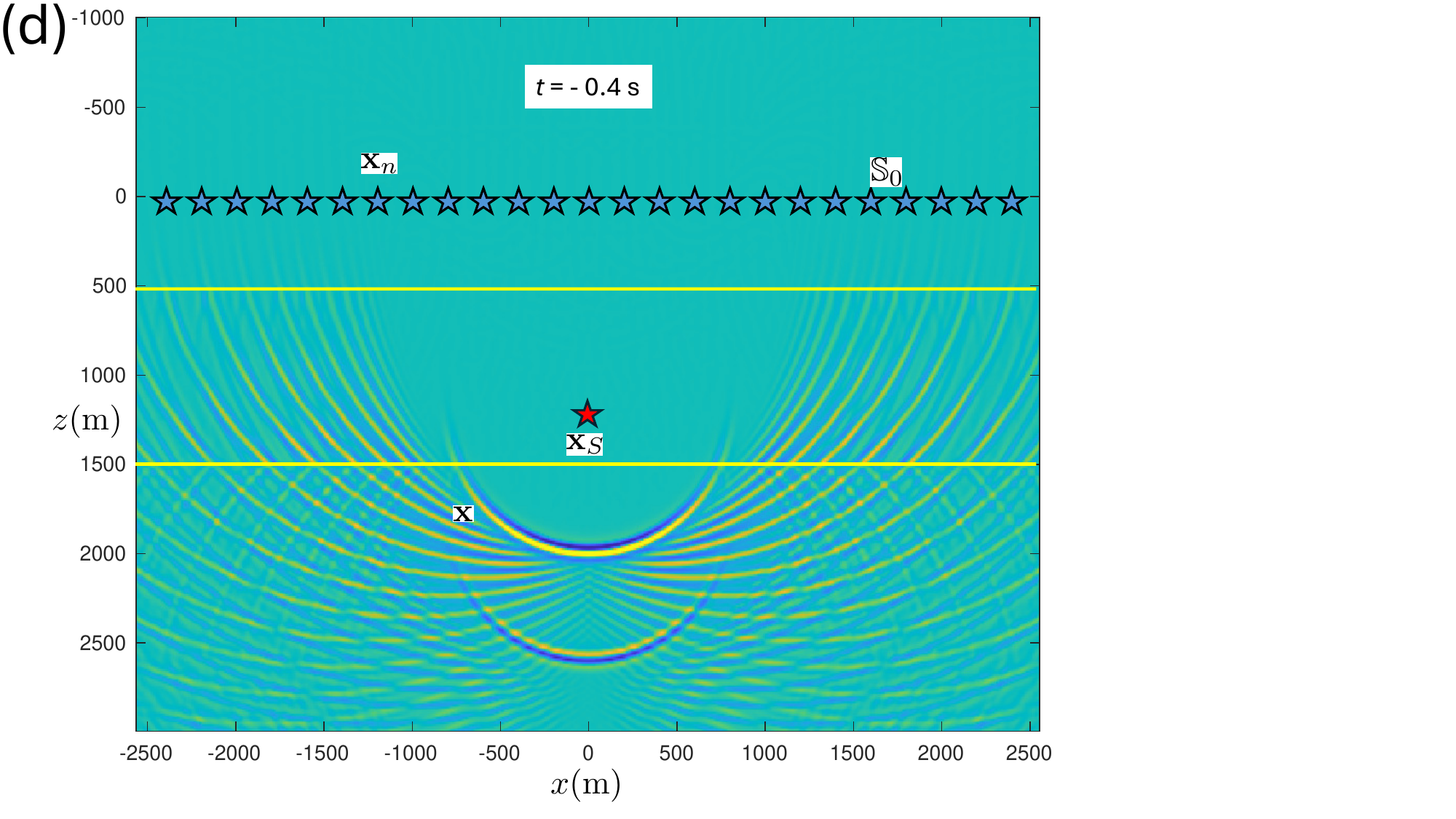}}
\caption{\it  Illustration of  the first term of the modified Huygens' principle  (equation (\ref{eq50}), \rev{with the focusing function of Figure \ref{Fig15}}), applied to the discretized Green's function $G({\bf x}_n,{\bf x}_S,t)$.}\label{Fig18}
\end{figure}

\begin{figure}
\centerline{\hspace{2cm}\epsfysize=4.9 cm\epsfbox{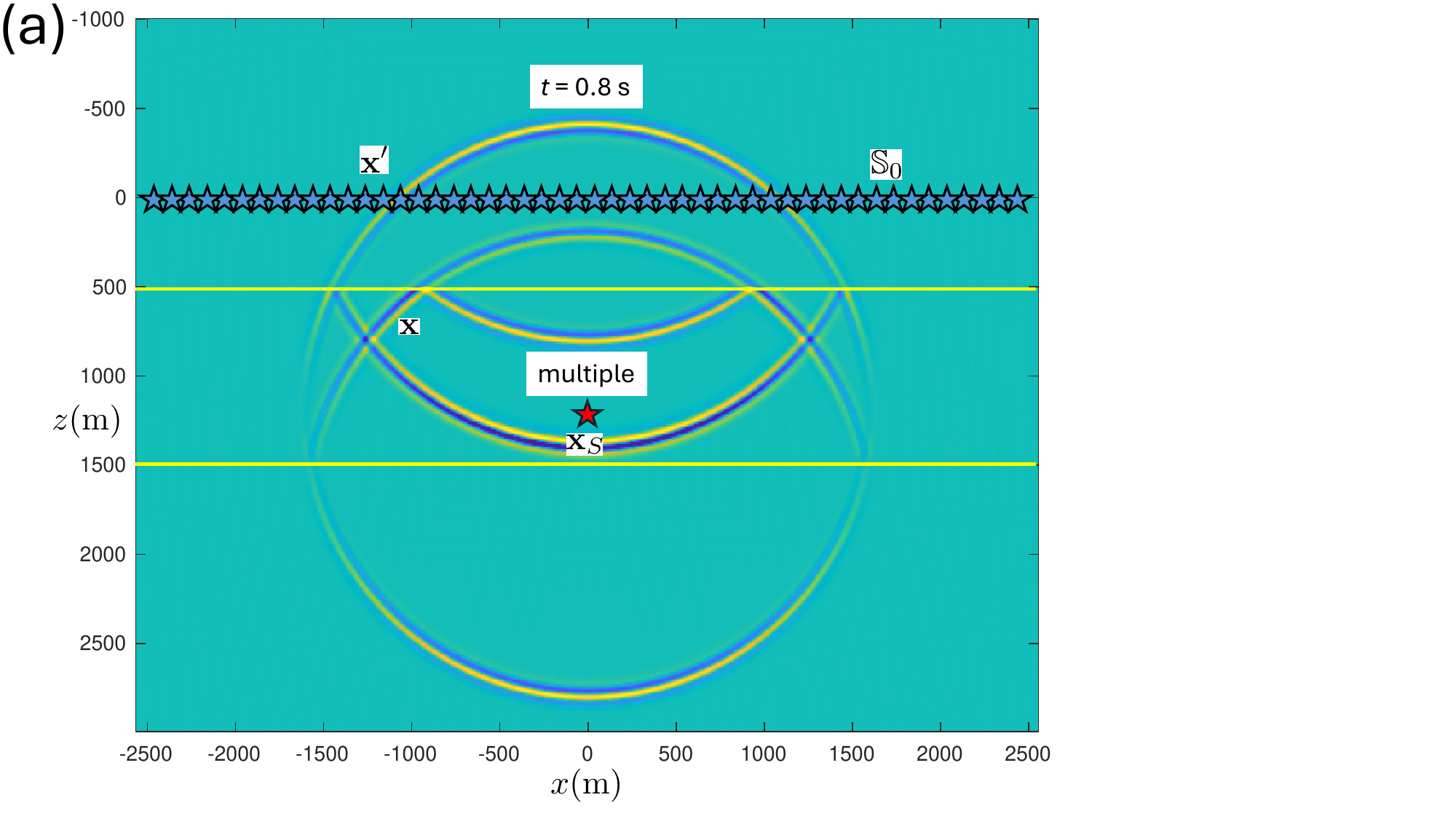}}
\centerline{\hspace{2cm}\epsfysize=4.9 cm\epsfbox{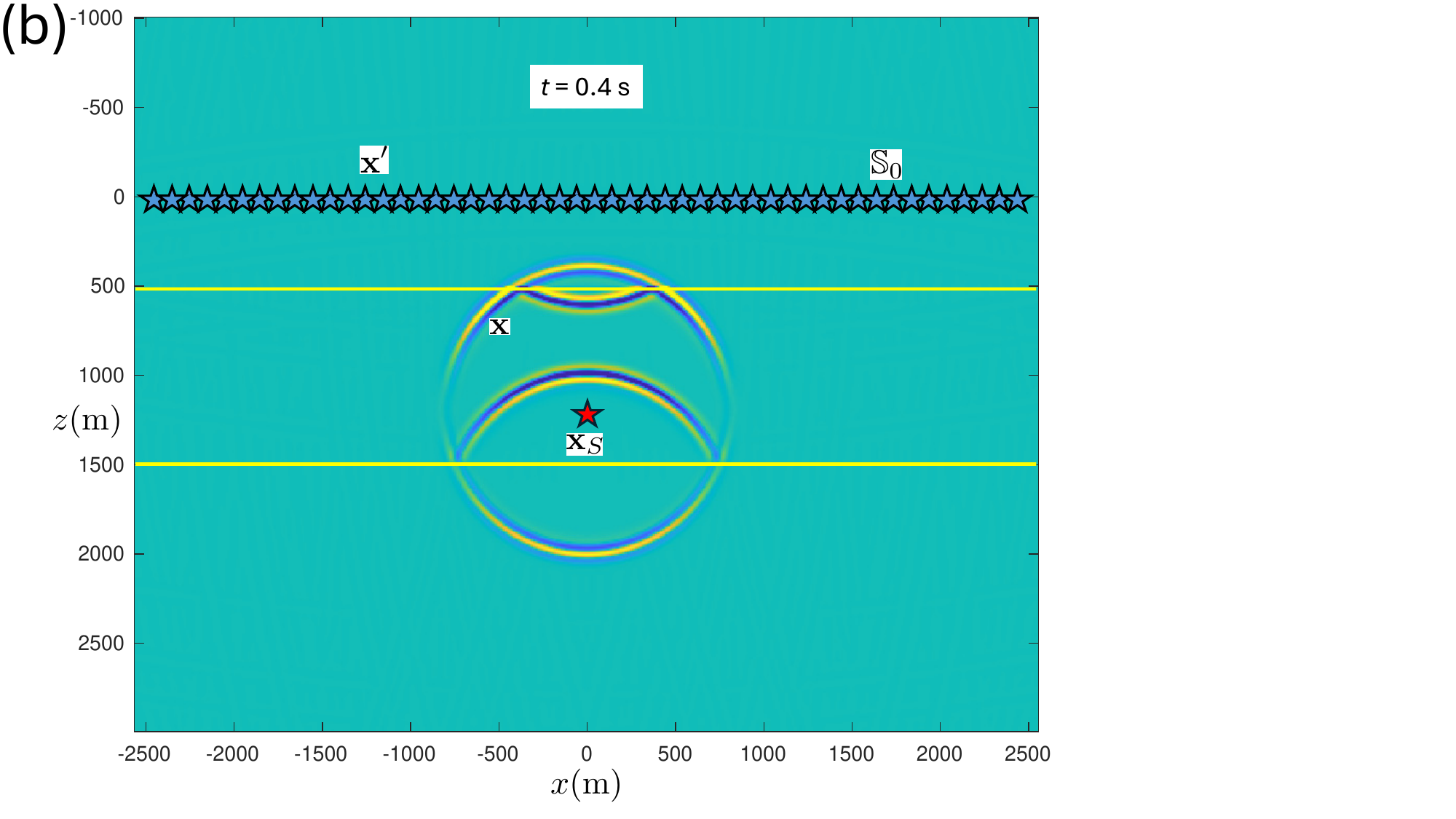}}
\centerline{\hspace{2cm}\epsfysize=4.9 cm\epsfbox{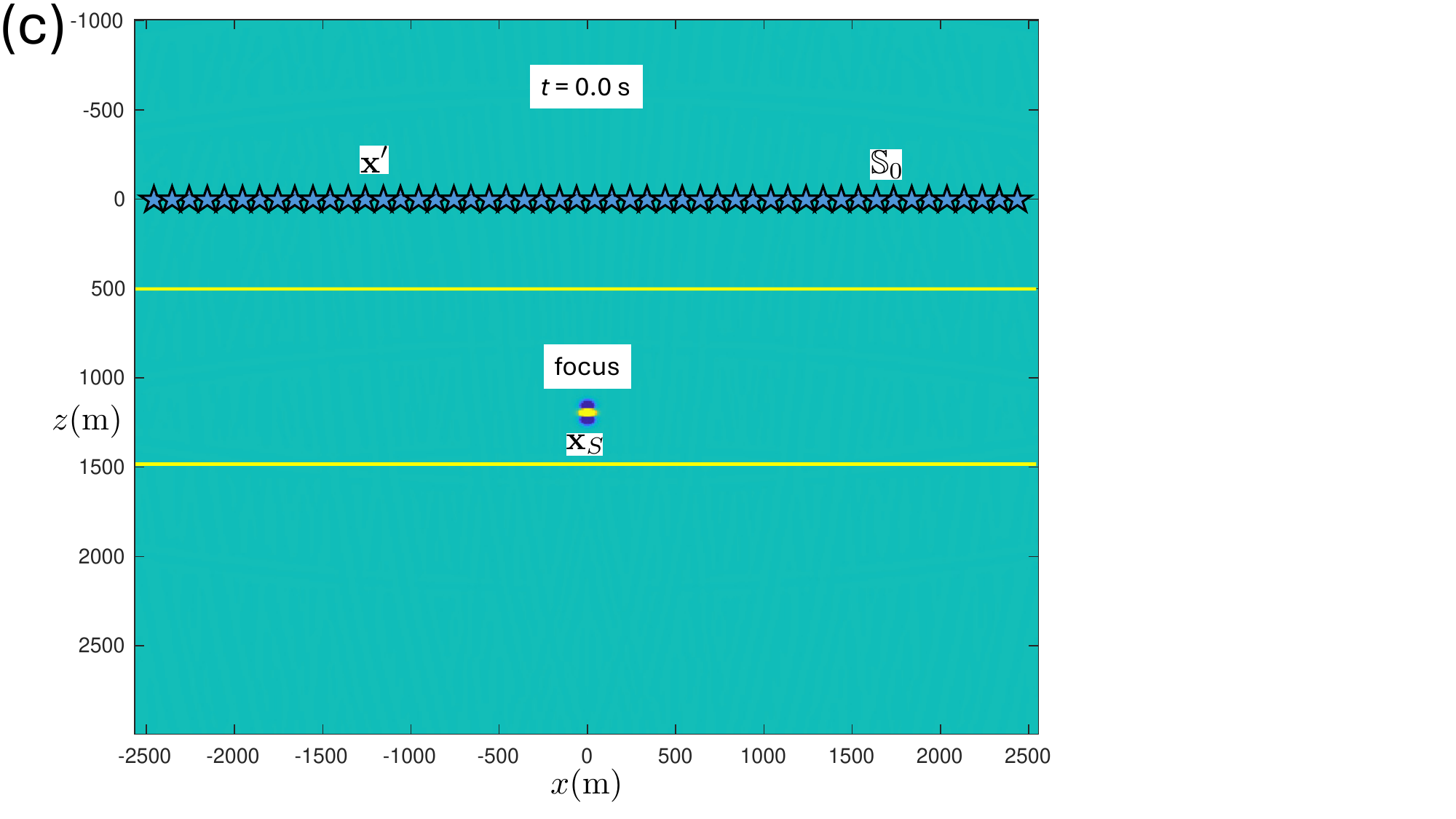}}
\centerline{\hspace{2cm}\epsfysize=4.9 cm\epsfbox{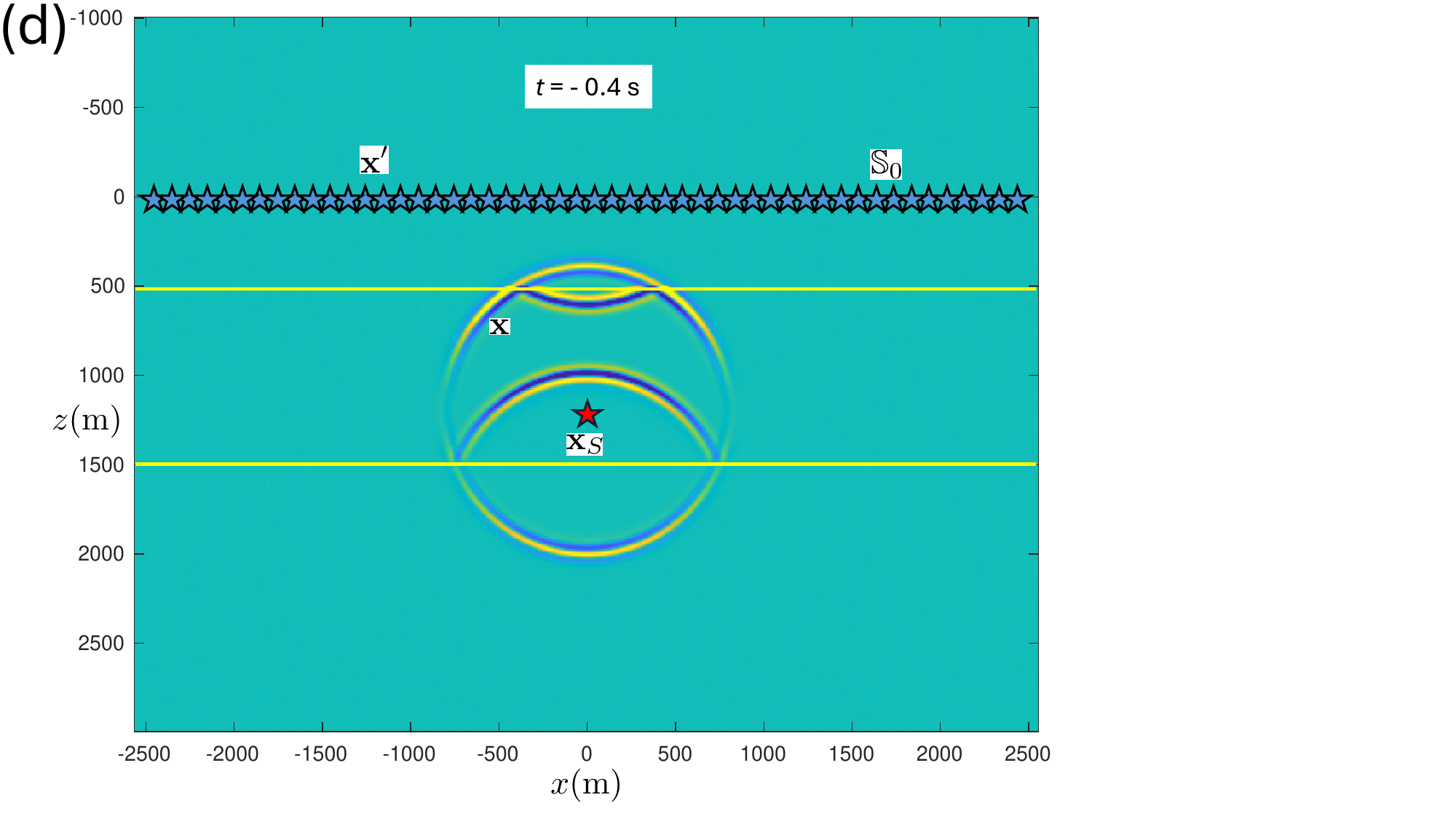}}
\caption{\it  Illustration of the modified Huygens' principle (both terms of equation (\ref{eq49})), applied to the continuous Green's function $G({\bf x}',{\bf x}_S,t)$ and
its time-reversal $G({\bf x}',{\bf x}_S,-t)$.}\label{Fig19}
\end{figure}

We illustrate equation (\ref{eq49})  for the same layered medium as in the previous examples and with ${\bf x}_S=(0,1200)$ m.
First we evaluate a discretized version of only the first term on the right-hand side of equation (\ref{eq49}), i.e.,
\begin{equation}
\sum_{n=-N}^N F({\bf x},{\bf x}_n,t)*G({\bf x}_n,{\bf x}_S,t),\label{eq50}
\end{equation}
with ${\bf x}_n=(n\Delta x,z_0)$, $\Delta x=200$ m, $N=50$, and with $G({\bf x}_n,{\bf x}_S,t)$ tapered at large propagation angles. 
The result is shown in Figures \ref{Fig18}a--\ref{Fig18}d, for $t=0.8$ s, $t=0.4$ s, $t=0.0$ s, and $t=-0.4$ s,
respectively. 
Next, we replace the summation by an integration and \rev{(following equation (\ref{eq49}))} 
we add its time-reversal, i.e., we superpose the converged versions of Figures \ref{Fig18}b \rev{(at $t=0.4$ s)} and \ref{Fig18}d \rev{(at $t=-0.4$ s)}, etc.
The result is shown in Figure \ref{Fig19}.
This figure shows the retrieved homogeneous Green's function $G_{\rm h}({\bf x},{\bf x}_S,t)$.
Note that the result in Figure  \ref{Fig19} is indistinguishable from that in Figure \ref{Fig6}.
However, whereas Figure \ref{Fig6} was obtained from an integral over two boundaries, 
using the traditional Huygens' principle with time-reversed dipole Green's functions,
Figure \ref{Fig19} is the result of an integral over a single boundary and its time-reversal, using  the modified Huygens' principle with focusing functions.

Retrieval of the homogeneous Green's function from \rev{wave field} observations at a single boundary finds applications in holographic imaging and inverse scattering 
\citep{Wapenaar2016RS, Diekmann2021PRR} and in monitoring of induced acoustic \citep{Neut2017JASA} and seismic sources \citep{Brackenhoff2019SE}.
In those applications the focusing functions are obtained with the Marchenko method from \rev{numerically modeled or field reflection responses of 2D inhomogeneous media}.

\section{Conclusions}

Huygens' principle stands as a milestone in the history of wave theory. 
Originally formulated to explain the propagation of light, 
it has found many applications in optics, acoustics, geophysics, etc. Central in the mathematical formulation of Huygens' principle, due to 19th century physicists 
\rev{Fresnel}, Kirchhoff, Helmholtz, Rayleigh and others, is the Green's function, which formalizes the responses to the secondary sources in Huygens' principle.
Many of present-day applications of Huygens' principle use time-reversed Green's functions. 
These time-reversed Green's functions are acausal and as such do not describe physical responses to secondary sources. However, they play a fundamental role in
algorithms for backpropagation, imaging, inversion, seismic interferometry, etc.

We have demonstrated with numerical examples that the traditional Huygens' principle with time-reversed Green's functions has limitations when
the medium is inhomogeneous. In particular, when measurements are available only at a single boundary, \rev{internal multiply reflected waves} are incorrectly handled.
To remedy this, Huygens' principle has been modified by replacing the Green's functions by focusing functions. 
For a homogeneous medium this replacement does not make much difference,
but for an inhomogeneous medium the \rev{improvement} is considerable. Using the modified Huygens' principle with focusing functions, 
the limitations of the traditional Huygens' principle with time-reversed Green's functions are evaded.
This has been \rev{demonstrated} with numerical examples for a simple horizontally layered medium, but note that the modified principle holds for an arbitrary inhomogeneous medium.
The only assumption is that evanescent waves can be ignored at the boundary.
No assumptions are made about up-down decomposition inside the inhomogeneous lower half-space.

Similar as the time-reversed Green's functions \rev{in Huygens' principle}, the focusing functions \rev{in the modified Huygens' principle}
do not describe physical responses to secondary sources. However, they play an important role in novel algorithms for acoustic and seismic imaging and inverse scattering, 
for monitoring of induced acoustic and seismic sources, etc. In all these cases, the focusing functions can be obtained from the reflection response with the Marchenko method,
taking internal \rev{multiply reflected waves} properly into account.

\rev{\section{Supporting Information}
Animations associated with the figures in this paper are available and can be accessed via the following URL: https://gitlab.com/geophysicsdelft/OpenSource
in the directory .../huygens
}


\section{Appendix A: Monopole and dipole Green's functions}

\subsection{A--1: Monopole Green's function}

The  wave equation for the acoustic pressure $p({\bf x},t)$ in an inhomogeneous lossless
medium with propagation velocity $c({\bf x})$ and mass density $\rho({\bf x})$ reads 
\begin{equation}
{\cal L}p=-\partial_tq,\nonumber\eqno{(A\mbox{--}1)}
\end{equation}
with $\partial_t$ standing for $\partial/\partial t$, wave operator ${\cal L}({\bf x},t)$ defined as 
\begin{equation}
{\cal L}={\bf \nabla}\cdot\frac{1}{\rho}{\bf \nabla} - \frac{1}{\rho c^2}\partial_t^2\nonumber\eqno{(A\mbox{--}2)}
\end{equation}
and source function $q({\bf x},t)$ being the volume-injection rate density. We define the Green's function ${\cal G}({\bf x},{\bf x}',t)$ as the solution of
\begin{equation}
{\cal L}{\cal G}=-\delta({\bf x}-{\bf x}')\delta(t),\nonumber\eqno{(A\mbox{--}3)}
\end{equation}
with $\delta({\bf x})=\delta(x)\delta(y)\delta(z)$ and causality condition
\begin{equation}
{\cal G}({\bf x},{\bf x}',t)=0,\quad\mbox{for}\quad t<0.\nonumber\eqno{(A\mbox{--}4)}
\end{equation}
Hence, ${\cal G}({\bf x},{\bf x}',t)$ is the response to an impulse at ${\bf x}'$ and $t=0$, observed at ${\bf x}$ as a function of $t$.
Similarly, we define $G({\bf x},{\bf x}',t)$ as the response to an impulsive volume-injection rate source $q({\bf x},t)=\delta({\bf x}-{\bf x}')\delta(t)$,
hence, it obeys 
\begin{equation}
{\cal L}G=-\delta({\bf x}-{\bf x}')\partial_t\delta(t)\nonumber\eqno{(A\mbox{--}5)}
\end{equation}
and a causality condition similar to equation ($A$--4). 
We apply the operator $\partial_t$ to both sides of equation ($A$--3) and use the property $\partial_t{\cal L}={\cal L}\partial_t$ (since $\rho({\bf x})$ and $c({\bf x})$
are independent of $t$).
Comparing the result with equation ($A$--5), it follows that
$G({\bf x},{\bf x}',t)$ and ${\cal G}({\bf x},{\bf x}',t)$ are mutually related via
\begin{equation}
G({\bf x},{\bf x}',t)=\partial_t{\cal G}({\bf x},{\bf x}',t).\nonumber\eqno{(A\mbox{--}6)}
\end{equation}
Note that the source terms in equations ($A$--3) and ($A$--5) represent monopole sources at ${\bf x}'$, hence, we call  $G$ and ${\cal G}$ monopole Green's functions.

For the special case of a homogeneous medium we have for the 3D situation
\begin{equation}
{\cal G}({\bf x},{\bf x}',t)=\rho\frac{\delta(t-|{\bf x}-{\bf x}'|/c)}{4\pi|{\bf x}-{\bf x}'|}\nonumber\eqno{(A\mbox{--}7)}
\end{equation}
and for the 2D situation
\begin{equation}
{\cal G}({\bf x},{\bf x}',t)=\rho\frac{H(t-|{\bf x}-{\bf x}'|/c)}{2\pi\sqrt{t^2-|{\bf x}-{\bf x}'|^2/c^2}},\nonumber\eqno{(A\mbox{--}8)}
\end{equation}
where $H(t)$ is the Heaviside function. 
According to equation ($A$-6), explicit expressions for $G({\bf x},{\bf x}',t)$ in a homogeneous medium
are obtained by taking the time-derivative of the expressions in the right-hand sides of equations ($A$-7) and ($A$-8). 

\subsection{A--2: Homogeneous Green's function}

We consider again an  inhomogeneous lossless medium,  with propagation velocity $c({\bf x})$ and mass density $\rho({\bf x})$.
Since operator ${\cal L}$, defined in equation ($A$-2), contains only even order time-derivatives and the source term in wave equation ($A$-3) is an even function of time,
the time-reversed Green's function ${\cal G}({\bf x},{\bf x}',-t)$ also obeys this wave equation.
We define the homogeneous Green's function ${\cal G}_{\rm h}({\bf x},{\bf x}',t)$ as
\begin{equation}
{\cal G}_{\rm h}({\bf x},{\bf x}',t)={\cal G}({\bf x},{\bf x}',t)-{\cal G}({\bf x},{\bf x}',-t)\nonumber\eqno{(A\mbox{--}9)}
\end{equation}
\citep{Porter70JOSA, Oristaglio89IP}. 
Since equation ($A$-3) holds for both terms in the right-hand side of equation($A$-9), their difference obeys the homogeneous differential equation ${\cal L}{\cal G}_{\rm h}=0$ 
(hence the name ``homogeneous Green's function'' for ${\cal G}_{\rm h}$).

Since the source term in wave equation ($A$-5) is an odd function of time, the opposite time-reversed Green's function $-G({\bf x},{\bf x}',-t)$ also obeys this wave equation. 
Hence, the homogeneous Green's function $G_{\rm h}({\bf x},{\bf x}',t)$, defined as
\begin{equation}
G_{\rm h}({\bf x},{\bf x}',t)=G({\bf x},{\bf x}',t)+G({\bf x},{\bf x}',-t),\nonumber\eqno{(A\mbox{--}10)}
\end{equation}
obeys the homogeneous differential equation ${\cal L}G_{\rm h}=0$.
From equation ($A$-6) and the definitions of ${\cal G}_{\rm h}$ and $G_{\rm h}$, it follows that these homogeneous Green's functions are mutually related via
\begin{equation}
G_{\rm h}({\bf x},{\bf x}',t)=\partial_t{\cal G}_{\rm h}({\bf x},{\bf x}',t).\nonumber\eqno{(A\mbox{--}11)}
\end{equation}

\subsection{A--3: Dipole Green's function}

Next, we define a Green's function $G_{\rm d}({\bf x},{\bf x}',t)$ as the solution of
\begin{equation}
{\cal L}G_{\rm d}=\frac{1}{\rho({\bf x}')}\partial_z\delta({\bf x}-{\bf x}')\delta(t),\nonumber\eqno{(A\mbox{--}12)}
\end{equation}
with  a causality condition similar to equation ($A$--4), and $\partial_z$ standing for $\partial/\partial z$. Note that
\begin{equation}
\partial_z\delta(z-z')=\lim_{\Delta z\to 0}\frac{\delta(z+\frac{\Delta z}{2}-z')-\delta(z-\frac{\Delta z}{2}-z')}{\Delta z},\nonumber\eqno{(A\mbox{--}13)}
\end{equation}
hence, the right-hand side of equation ($A$--12) represents a vertically oriented dipole source at ${\bf x}'$. Therefore we call $G_{\rm d}$ a dipole Green's function.
We define $\partial_z'$ as an operator for differentiation with respect to $z'$.
We apply this operator to both sides of equation ($A$--3) and use the properties $\partial_z'{\cal L}={\cal L}\partial_z'$ 
(since $\rho({\bf x})$ and $c({\bf x})$ are independent of $z'$) and $\partial_z'\delta(z-z')=-\partial_z\delta(z-z')$.
Comparing the result with equation ($A$--12), 
it follows that $G_{\rm d}({\bf x},{\bf x}',t)$ and ${\cal G}({\bf x},{\bf x}',t)$ are mutually related via
\begin{equation}
G_{\rm d}({\bf x},{\bf x}',t)=\frac{1}{\rho({\bf x}')}\partial_z'{\cal G}({\bf x},{\bf x}',t).\nonumber\eqno{(A\mbox{--}14)}
\end{equation}

\section{Appendix B: Forward wave field extrapolation}

We define the temporal Fourier transform of a space- and time-dependent quantity $p({\bf x},t)$ as 
\begin{equation}
\hat p({\bf x},\omega)=\int_{-\infty}^\infty p({\bf x},t)\exp(i\omega t){\rm d}t,\nonumber\eqno{(B\mbox{--}1)}
\end{equation}
where $i$ is the imaginary unit and $\omega$  the angular frequency (we consider positive $\omega$ only). 
With this definition, differentiation with respect to $t$ in the time domain is replaced by multiplication with $-i\omega$
in the frequency domain, hence, wave equations ($A$--1) and ($A$--3) transform to
\begin{equation}
\hat{\cal L}\hat p=i\omega \hat q\nonumber\eqno{(B\mbox{--}2)}
\end{equation}
and
\begin{equation}
\hat{\cal L}\hat{\cal G}=-\delta({\bf x}-{\bf x}'),\nonumber\eqno{(B\mbox{--}3)}
\end{equation}
respectively, with operator $\hat{\cal L}({\bf x},\omega)$ defined as 
\begin{equation}
\hat{\cal L}={\bf \nabla}\cdot\frac{1}{\rho}{\bf \nabla} + \frac{\omega^2}{\rho c^2}.\nonumber\eqno{(B\mbox{--}4)}
\end{equation}
Consider the quantity ${\bf \nabla}\cdot\{\hat{\cal G}(\frac{1}{\rho}{\bf \nabla}\hat p) - \hat p(\frac{1}{\rho}{\bf \nabla}\hat{\cal G})\}$, apply the product rule for differentiation
and simplify the result using wave equations ($B$--2) and ($B$--3). This yields
\begin{equation}
{\bf \nabla}\cdot\{\hat{\cal G}\frac{1}{\rho}{\bf \nabla}\hat p - \hat p\frac{1}{\rho}{\bf \nabla}\hat{\cal G}\}=
i\omega\hat{\cal G}\hat q+\hat p\delta({\bf x}-{\bf x}').\nonumber\eqno{(B\mbox{--}5)}
\end{equation}
Integrate both sides of this equation
over a domain ${\mathbb{V}}$ with boundary $\mathbb{S}$ and outward pointing normal vector ${\bf n}$ and apply the theorem of Gauss to the left-hand side.
Use the source-receiver reciprocity relation $\hat{\cal G}({\bf x},{\bf x}',\omega)=\hat{\cal G}({\bf x}',{\bf x},\omega)$ and subsequently modify the notation by 
replacing all ${\bf x}$ by ${\bf x}'$ and vice-versa (meaning that effectively $\hat{\cal G}({\bf x},{\bf x}',\omega)$ remains unchanged).  This yields
the Kirchhoff-Helmholtz integral representation
\citep{Morse53Book, Bleistein84Book}
%
%
%
\begin{eqnarray}
\chi_{\mathbb{V}}({\bf x})\hat p({\bf x},\omega)&=&\oint_{\mathbb{S}}\frac{1}{\rho({\bf x}')}\Bigl(\hat{\cal G}({\bf x},{\bf x}',\omega){\bf \nabla}'\hat p({\bf x}',\omega)\nonumber\\
&&-\hat p({\bf x}',\omega){\bf \nabla}'\hat{\cal G}({\bf x},{\bf x}',\omega)\Bigr)\cdot{\bf n}{\rm d}{\bf x}'\nonumber\\
&&-\int_{\mathbb{V}}i\omega \hat{\cal G}({\bf x},{\bf x}',\omega)\hat q({\bf x}',\omega){\rm d}{\bf x}',\nonumber\hspace{.4cm}\mbox{($B$--6)}
\end{eqnarray}
with operator ${\bf \nabla}'$ acting on ${\bf x}'$ and  $\chi_{\mathbb{V}}({\bf x})$ being the characteristic function for ${\mathbb{V}}$, defined as
\begin{equation}
\chi_{\mathbb{V}}({\bf x})=\begin{cases}
&1\quad \mbox{for}\quad {\bf x} \,\mbox{in}\, {\mathbb{V}},\\
&\frac12 \quad \mbox{for}\quad {\bf x} \,\mbox{on}\, {\mathbb{S}},\\
&0 \quad \mbox{for}\quad {\bf x} \,\mbox{outside}\, {\mathbb{V}} \cup {\mathbb{S}}.
\end{cases}\nonumber\eqno{(B\mbox{--}7)}
\end{equation}

\begin{figure}
\centerline{\hspace{2cm}\epsfysize=5. cm\epsfbox{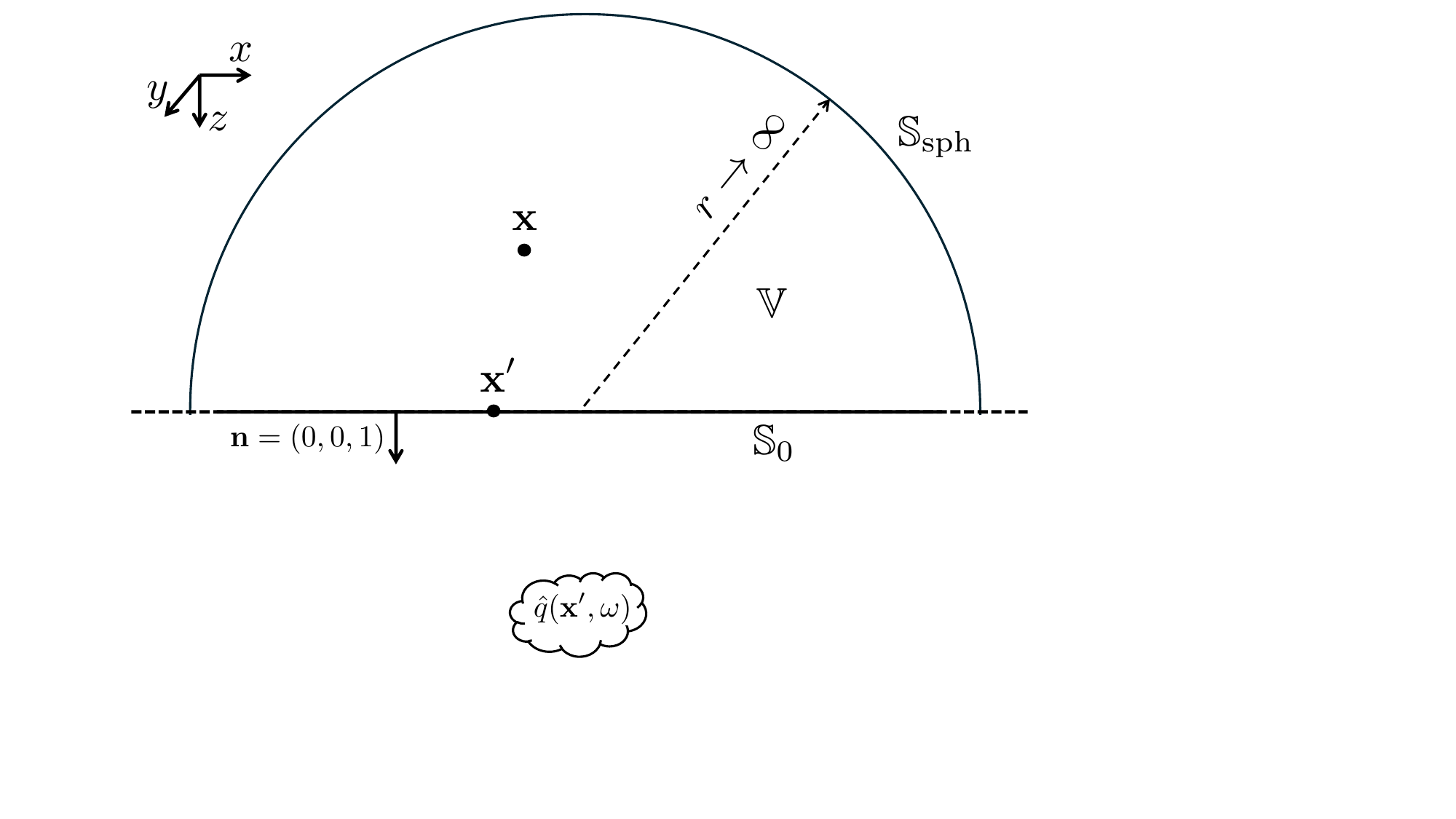}}
\vspace{-0.8cm}
{\bf Figure B--1.} {\it Configuration for forward wave field extrapolation (side view).}
\end{figure}

We use equation ($B$--6) to derive an expression for forward wave field extrapolation 
in the configuration of Figure $B$--1 \citep{Berkhout85Book, Frazer85GJRAS}.
The closed boundary $\mathbb{S}$  consists of an infinite  horizontal boundary ${\mathbb{S}_0}$ (at $z=z_0$)
and a half-sphere $\mathbb{S}_{\rm sph}$ with infinite radius ($r\to\infty$) in the upper half-space (above ${\mathbb{S}_0}$).
The upper half-space is homogeneous; the lower half-space may be inhomogeneous.
We choose the source distribution $\hat q({\bf x}',\omega)$ in the lower half-space (below $\mathbb{S}_0$) hence outside $\mathbb{V}$.
This implies that the volume integral on the right-hand side of equation ($B$--6) vanishes. 
The boundary integral over the half-sphere with infinite radius also vanishes (Sommerfeld radiation condition). 
Hence, we are left with a boundary integral over $\mathbb{S}_0$. At this boundary the outward pointing normal vector
${\bf n}$ equals $(0,0,1)$, hence, ${\bf \nabla}'\cdot{\bf n}=\partial_z'$ at $\mathbb{S}_0$.
Using the equation of motion $\frac{1}{\rho({\bf x}')}\partial_z'\hat p({\bf x}',\omega)=i\omega\hat v_z({\bf x}',\omega)$, where $\hat v_z$ is the vertical component of the particle velocity,
and the Fourier transforms of equations ($A$--6) and ($A$--14), we thus obtain
\begin{eqnarray}
\chi_{\mathbb{V}}({\bf x})\hat p({\bf x},\omega)&=&-\int_{\mathbb{S}_0}\Bigl(\hat G({\bf x},{\bf x}',\omega)\hat v_z({\bf x}',\omega)\nonumber\\
&&+\,\hat G_{\rm d}({\bf x},{\bf x}',\omega)\hat p({\bf x}',\omega)\Bigr){\rm d}{\bf x}'.\nonumber\hspace{.7cm}\mbox{($B$--8)}
\end{eqnarray}
For ${\bf x}$ in the lower half-space we have $\chi_{\mathbb{V}}({\bf x})=0$, hence, the integral on the right-hand side vanishes for this situation.
For ${\bf x}$ in the upper half-space we have $\chi_{\mathbb{V}}({\bf x})=1$, hence, for this situation equation ($B$-8)  describes forward wave field extrapolation
from the horizontal boundary $\mathbb{S}_0$ to any point ${\bf x}$ above this boundary.
Since the upper half-space is homogeneous, the actual wave field ($\hat p$ and $\hat v_z$) is upward propagating at $\mathbb{S}_0$.
Taking the entire medium homogeneous for the Green's functions
($\hat G$ and $\hat G_{\rm d}$), then the two terms under the integral give equal contributions, hence, equation ($B$-8) can be replaced by
\begin{equation}
\hat p({\bf x},\omega)=-2\int_{\mathbb{S}_0}\hat G_{\rm d}({\bf x},{\bf x}',\omega)\hat p({\bf x}',\omega){\rm d}{\bf x}',\nonumber\eqno{(B\mbox{--}9)}
\end{equation}
for ${\bf x}$ in the upper half-space \citep{Berkhout89GEO}. 
Equation ($B$-9) is a Rayleigh integral (\citet{Rayleigh78Book} derived expressions like this to describe the radiation of sources, distributed over a plane).
Transforming this back to the time domain and using the causality condition of the Green's function gives
\begin{equation}
p({\bf x},t)=-2\int_{\mathbb{S}_0}\int_0^\infty G_{\rm d}({\bf x},{\bf x}',t')p({\bf x}',t-t'){\rm d}t'{\rm d}{\bf x}',\nonumber\eqno{(B\mbox{--}10)}
\end{equation}
for ${\bf x}$ in the upper half-space. 
The time integral in this expression is a convolution, which can be written in a shorter notation using the convolution symbol $*$. Equation ($B$-10) thus becomes
\begin{equation}
p({\bf x},t)=-2\int_{\mathbb{S}_0}G_{\rm d}({\bf x},{\bf x}',t)*p({\bf x}',t){\rm d}{\bf x}',\nonumber\eqno{(B\mbox{--}11)}
\end{equation}
for ${\bf x}$ in the upper half-space.

\section{Appendix C: Inverse wave field extrapolation}

The Fourier transform of the time-reversed Green's function ${\cal G}({\bf x},{\bf x}',-t)$ is given by $\hat {\cal G}^*({\bf x},{\bf x}',\omega)$, 
where the superscript $*$ denotes complex conjugation. Since for a lossless medium $\hat {\cal G}^*$ obeys the same wave equation as $\hat {\cal G}$ (equation ($B$-3)),
equation ($B$-6) remains valid when we replace $\hat {\cal G}$ by $\hat {\cal G}^*$ \citep{Bojarski83JASA}, hence
%
%
\begin{eqnarray}
\chi_{\mathbb{V}}({\bf x})\hat p({\bf x},\omega)&=&\oint_{\mathbb{S}}\frac{1}{\rho({\bf x}')}\Bigl(\hat{\cal G}^*({\bf x},{\bf x}',\omega){\bf \nabla}'\hat p({\bf x}',\omega)\nonumber\\
&&-\hat p({\bf x}',\omega){\bf \nabla}'\hat{\cal G}^*({\bf x},{\bf x}',\omega)\Bigr)\cdot{\bf n}{\rm d}{\bf x}'\nonumber\\
&&-\int_{\mathbb{V}}i\omega \hat{\cal G}^*({\bf x},{\bf x}',\omega)\hat q({\bf x}',\omega){\rm d}{\bf x}'.\nonumber\hspace{.2cm}\mbox{($C$--1)}
\end{eqnarray}
\begin{figure}
\centerline{\hspace{2cm}\epsfysize=5. cm\epsfbox{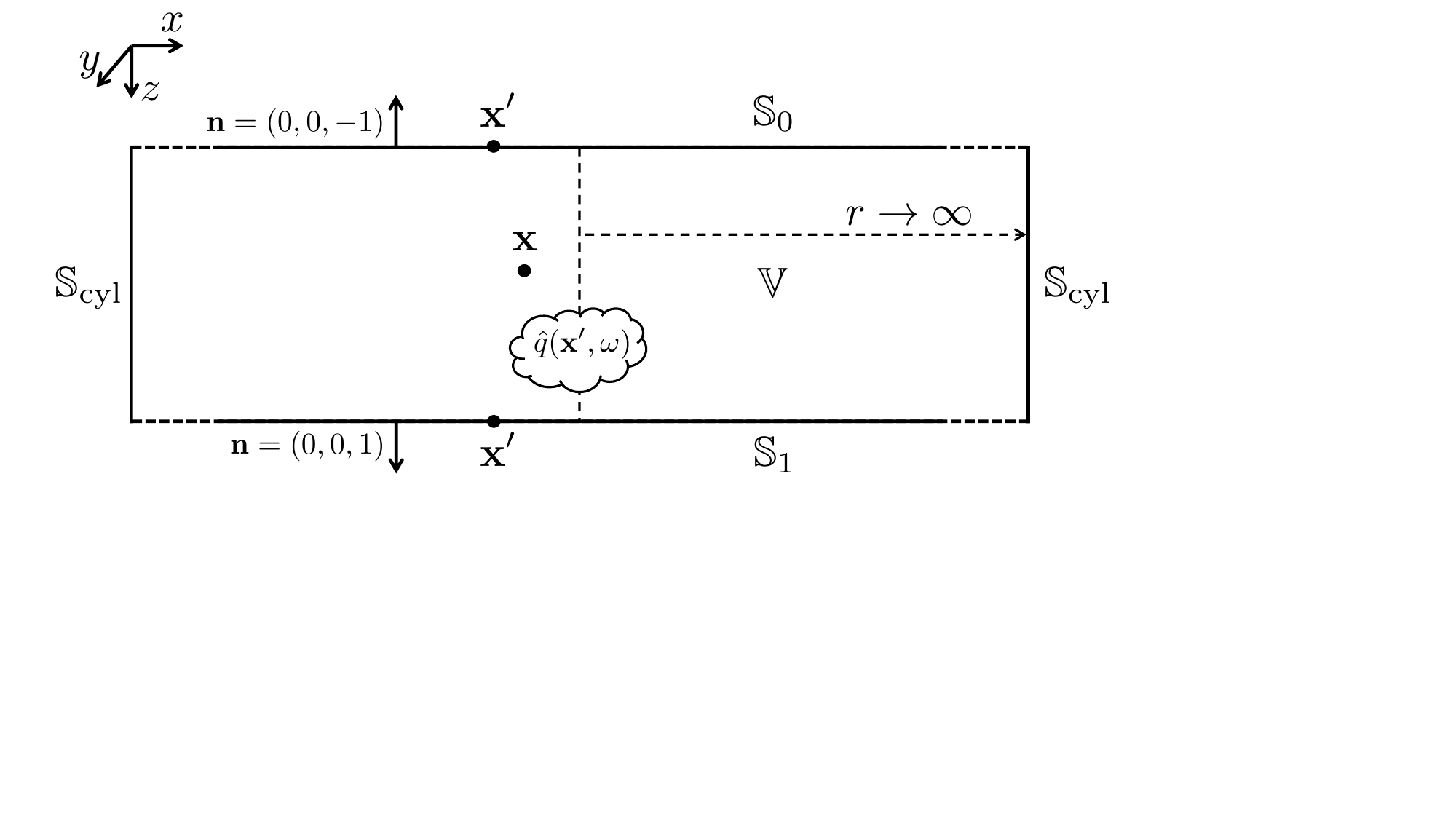}}
\vspace{-1.8cm}
{\bf Figure C--1.} {\it Configuration for inverse wave field extrapolation (side view).}
\end{figure}

We use equation ($C$--1) to derive an expression for inverse wave field extrapolation in the configuration of Figure $C$--1. 
 The closed boundary $\mathbb{S}$ consists of two infinite horizontal boundaries ${\mathbb{S}_0}$ (at $z=z_0$) and $\mathbb{S}_1$ (at $z=z_1$), connected by 
 a cylindrical surface ${\mathbb{S}_{\rm cyl}}$ with infinite radius ($r\to\infty$).
 The half-spaces above $\mathbb{S}_0$ and below $\mathbb{S}_1$ are homogeneous; the medium between these boundaries may be inhomogeneous.
We choose the source distribution $\hat q({\bf x}',\omega)$ between the two boundaries $\mathbb{S}_0$ and $\mathbb{S}_1$, hence inside $\mathbb{V}$.
The boundary integral over the cylindrical surface with infinite radius vanishes because its surface area increases with $r$, but its integrand decays with $1/r^2$. 
The outward pointing normal vector ${\bf n}$ equals $(0,0,-1)$ at $\mathbb{S}_0$ and $(0,0,1)$ at $\mathbb{S}_1$, hence,
 ${\bf \nabla}'\cdot{\bf n}=-\partial_z'$ at $\mathbb{S}_0$ and ${\bf \nabla}'\cdot{\bf n}=\partial_z'$ at $\mathbb{S}_1$.
Using again the equation of motion 
and the Fourier transforms of equations ($A$--6) and ($A$--14), we thus obtain
%
%
\begin{eqnarray}
\hat p({\bf x},\omega)&=&-\int_{\mathbb{S}_0}\Bigl(\hat G^*({\bf x},{\bf x}',\omega)\hat v_z({\bf x}',\omega)\nonumber\\
&&-\,\hat G_{\rm d}^*({\bf x},{\bf x}',\omega)\hat p({\bf x}',\omega)\Bigr){\rm d}{\bf x}'\nonumber\\
&&+\int_{\mathbb{S}_1}\Bigl(\hat G^*({\bf x},{\bf x}',\omega)\hat v_z({\bf x}',\omega)\nonumber\\
&&-\,\hat G_{\rm d}^*({\bf x},{\bf x}',\omega)\hat p({\bf x}',\omega)\Bigr){\rm d}{\bf x}'\nonumber\\
&&-\int_{\mathbb{V}}\hat G^*({\bf x},{\bf x}',\omega)\hat q({\bf x}',\omega){\rm d}{\bf x}',\nonumber\hspace{1.3cm}\mbox{($C$--2)}
\end{eqnarray}
for ${\bf x}$ in $\mathbb{V}$. Since the medium above $\mathbb{S}_0$ and below $\mathbb{S}_1$ is homogeneous (for the actual wave field and for the Green's function),
the two terms under the boundary integrals give equal contributions, hence, equation ($C$-2) can be replaced by 
%
%
\begin{eqnarray}
\hat p({\bf x},\omega)&=&2\int_{\mathbb{S}_0}\hat G_{\rm d}^*({\bf x},{\bf x}',\omega)\hat p({\bf x}',\omega){\rm d}{\bf x}'\nonumber\\
&&-2\int_{\mathbb{S}_1}\hat G_{\rm d}^*({\bf x},{\bf x}',\omega)\hat p({\bf x}',\omega){\rm d}{\bf x}'\nonumber\\
&&-\int_{\mathbb{V}}\hat G^*({\bf x},{\bf x}',\omega)\hat q({\bf x}',\omega){\rm d}{\bf x}',\nonumber\hspace{1.3cm}\mbox{($C$--3)}
\end{eqnarray}
for ${\bf x}$ in $\mathbb{V}$. Going from equation ($C$-2) to equation ($C$-3), it is assumed that evanescent waves at $\mathbb{S}_0$ and $\mathbb{S}_1$
can be ignored \citep{Wapenaar89GEO}. Transforming equation ($C$-3) back to the time domain 
and choosing a point source $q({\bf x}',t)=\delta({\bf x}'-{\bf x}_S)s(t)$ (with ${\bf x}_S$ in $\mathbb{V}$)
gives
%
%
\begin{eqnarray}
p({\bf x},t)&=&2\int_{\mathbb{S}_0} G_{\rm d}({\bf x},{\bf x}',-t) * p({\bf x}',t){\rm d}{\bf x}'\nonumber\\
&&-2\int_{\mathbb{S}_1} G_{\rm d}({\bf x},{\bf x}',-t) * p({\bf x}',t){\rm d}{\bf x}'\nonumber\\
&&-G({\bf x},{\bf x}_S,-t) * s(t),\nonumber\hspace{2.4cm}\mbox{($C$--4)}
\end{eqnarray}
for ${\bf x}$ in $\mathbb{V}$. Inverse extrapolation is often approximated by the first term only \citep{Schneider78GEO, Berkhout85Book}, hence
\begin{equation}
\langle p({\bf x},t)\rangle=2\int_{\mathbb{S}_0} G_{\rm d}({\bf x},{\bf x}',-t) * p({\bf x}',t){\rm d}{\bf x}',\nonumber\eqno{(C\mbox{--}5)}
\end{equation}
or, writing the integrand as a correlation integral,
\begin{equation}
\langle p({\bf x},t)\rangle=2\int_{\mathbb{S}_0}\int_0^\infty G_{\rm d}({\bf x},{\bf x}',t')p({\bf x}',t+t'){\rm d}t'{\rm d}{\bf x}',\nonumber\eqno{(C\mbox{--}6)}
\end{equation}
for ${\bf x}$ below $\mathbb{S}_0$ and above the source, see the main text for a further discussion.

\section{Appendix D: Extrapolation with focusing functions}

Consider a configuration, consisting of an inhomogeneous lossless medium below $\mathbb{S}_0$ (at $z=z_0$), 
with propagation velocity $c({\bf x})$ and mass density $\rho({\bf x})$,
and a homogeneous lossless medium \rev{at and} above $\mathbb{S}_0$, with propagation velocity $c_0$ and mass density $\rho_0$.
In the space-frequency domain, the acoustic pressure $\hat p({\bf x},\omega)$ in this configuration obeys wave equation ($B$-2), 
with operator $\hat{\cal L}({\bf x},\omega)$ defined in equation ($B$-4).
In the following we assume that the source distribution $\hat q({\bf x},\omega)$ is restricted to the upper half-space. 
Hence, for all ${\bf x}$ below the source distribution, $\hat p({\bf x},\omega)$ obeys the source-free wave equation
\begin{equation}
\hat{\cal L}\hat p=0.\nonumber\eqno{(D\mbox{--}1)}
\end{equation}
For the same configuration we define the Fourier-transformed focusing function $\hat F({\bf x},{\bf x}',\omega)$, with ${\bf x}'$ denoting a focal point
at $\mathbb{S}_0$ (hence, $z'=z_0$). This focusing function obeys the same source-free wave equation throughout space, hence
\begin{equation}
\hat{\cal L}\hat F=0.\nonumber\eqno{(D\mbox{--}2)}
\end{equation}
Moreover, $\hat F({\bf x},{\bf x}',\omega)$ is defined such that 
for ${\bf x}$ at $\mathbb{S}_0$ it obeys the Fourier transform of the focusing condition of equation (\ref{eqFoc1b}), hence
\begin{equation}
\hat F({\bf x}_{\rm H},z_0,{\bf x}_{\rm H}',z_0,\omega)=\delta({\bf x}_{\rm H}-{\bf x}_{\rm H}'),\nonumber\eqno{(D\mbox{--}3)}
\end{equation}
with ${\bf x}_{\rm H}$ and ${\bf x}_{\rm H}'$ being the horizontal components of ${\bf x}$ and ${\bf x}'$, respectively.
Finally, for ${\bf x}$ at and above $\mathbb{S}_0$ this focusing function propagates upward. 
Note that $\hat F^*({\bf x},{\bf x}',\omega)$, which is the Fourier transform of the time-reversed focusing function, obeys the same wave equation and the same focusing condition as 
$\hat F({\bf x},{\bf x}',\omega)$; for ${\bf x}$ at and above $\mathbb{S}_0$ it propagates downward.

For all ${\bf x}$ below the source distribution, we write $\hat p({\bf x},\omega)$ as a superposition of the mutually independent focusing functions, according to
\begin{eqnarray}
\hat p({\bf x},\omega)&=&\int_{\mathbb{S}_0} \hat F({\bf x},{\bf x}',\omega)\hat a({\bf x}',\omega){\rm d}{\bf x}'\nonumber\\
&+&\int_{\mathbb{S}_0} \hat F^*({\bf x},{\bf x}',\omega)\hat b({\bf x}',\omega){\rm d}{\bf x}',\nonumber\hspace{1.5cm}\mbox{($D$--4)}
\end{eqnarray}
where $\hat a({\bf x}',\omega)$ and $\hat b({\bf x}',\omega)$ are coefficients which still need to be determined. 
\rev{In the main text we use physical arguments to show that these coefficients are the upgoing and downgoing parts of the wave field at $\mathbb{S}_0$. 
Here we show this via a more formal derivation.}
Using the equation of motion, we obtain a similar expression for the vertical component of the particle velocity from equation ($D$-4), according to
\begin{eqnarray}
\hat v_z({\bf x},\omega)&=&\frac{1}{i\omega\rho({\bf x})}\partial_z\hat p({\bf x},\omega)\nonumber\\
&=&\frac{1}{i\omega\rho({\bf x})}\int_{\mathbb{S}_0} \partial_z\hat F({\bf x},{\bf x}',\omega)\hat a({\bf x}',\omega){\rm d}{\bf x}'\nonumber\\
&+&\frac{1}{i\omega\rho({\bf x})}\int_{\mathbb{S}_0} \partial_z\hat F^*({\bf x},{\bf x}',\omega)\hat b({\bf x}',\omega){\rm d}{\bf x}'.\nonumber\hspace{.1cm}\mbox{($D$--5)}
\end{eqnarray}
We solve the coefficients $\hat a({\bf x}',\omega)$ and $\hat b({\bf x}',\omega)$ from the boundary conditions for $\hat p({\bf x},\omega)$ and $\hat v_z({\bf x},\omega)$ at $\mathbb{S}_0$.
Choosing ${\bf x}$ at $\mathbb{S}_0$ and using equation ($D$-3), we obtain from equation ($D$-4)
\begin{equation}
\hat p({\bf x}_{\rm H},z_0,\omega)=\hat a({\bf x}_{\rm H},z_0,\omega)+\hat b({\bf x}_{\rm H},z_0,\omega).\nonumber\eqno{(D\mbox{--}6)}
\end{equation}
From equation ($D$-5) we obtain for ${\bf x}$ at $\mathbb{S}_0$
\begin{eqnarray}
\hat v_z({\bf x}_{\rm H},z_0,\omega)
&=&\frac{1}{i\omega\rho_0}\int_{\mathbb{S}_0} \partial_z\hat F({\bf x},{\bf x}',\omega)|_{z=z_0}\hat a({\bf x}',\omega){\rm d}{\bf x}'\nonumber\\
&+&\frac{1}{i\omega\rho_0}\int_{\mathbb{S}_0} \partial_z\hat F^*({\bf x},{\bf x}',\omega)|_{z=z_0}\hat b({\bf x}',\omega){\rm d}{\bf x}'.\nonumber\\
&&\nonumber\hspace{4.5cm}\mbox{($D$--7)}
\end{eqnarray}
We define the spatial Fourier transform of $\hat v_z({\bf x}_{\rm H},z_0,\omega)$ as
\begin{equation}
\tilde v_z({\bf k}_{\rm H},z_0,\omega)=\int_{\mathbb{S}_0}\hat v_z({\bf x},\omega)\exp(-i{\bf k}_{\rm H}\cdot{\bf x}_{\rm H}){\rm d}{\bf x},\nonumber\eqno{(D\mbox{--}8)}
\end{equation}
where ${\bf k}_{\rm H}=(k_x,k_y)$ (in 3D) or ${\bf k}_{\rm H}=k_x$ (in 2D). 
\rev{We apply this transformation} to both sides of equation ($D$-7). \rev{Since the focusing function is upward propagating at and above $\mathbb{S}_0$ 
and the medium is homogeneous at and above $\mathbb{S}_0$, we can use the following one-way wave equation for the Fourier transform of the focusing function at $z=z_0$
%
\begin{equation}
\partial_z\tilde F({\bf k}_{\rm H},z,{\bf x}_{\rm H}', z_0,\omega)|_{z=z_0}=-ik_z\tilde F({\bf k}_{\rm H},z_0,{\bf x}_{\rm H}',z_0,\omega),\nonumber\eqno{(D\mbox{--}9)}
\end{equation}
}
where the vertical wavenumber $k_z$ is defined as
\begin{equation}
k_z=\begin{cases}
\sqrt{\frac{\omega^2}{c_0^2}-{\bf k}_{\rm H}\cdot{\bf k}_{\rm H}},\quad&\mbox{for}\quad{\bf k}_{\rm H}\cdot{\bf k}_{\rm H}\le\frac{\omega^2}{c_0^2},\\
i\sqrt{{\bf k}_{\rm H}\cdot{\bf k}_{\rm H}-\frac{\omega^2}{c_0^2}},\quad&\mbox{for}\quad{\bf k}_{\rm H}\cdot{\bf k}_{\rm H}>\frac{\omega^2}{c_0^2}.
\end{cases}\nonumber\eqno{(D\mbox{--}10)}
\end{equation}
The two cases in the latter equation correspond to propagating and evanescent waves, respectively.
\rev{With this, we obtain for the Fourier transform of equation ($D$-7)}
\begin{eqnarray}
\tilde v_z({\bf k}_{\rm H},z_0,\omega)
&=&\frac{-k_z}{\omega\rho_0}\int_{\mathbb{S}_0} \tilde F({\bf k}_{\rm H},z_0,{\bf x}',\omega)\hat a({\bf x}',\omega){\rm d}{\bf x}'\nonumber\\
&+&\frac{k_z^*}{\omega\rho_0}\int_{\mathbb{S}_0} \tilde F^*(-{\bf k}_{\rm H},z_0,{\bf x}',\omega)\hat b({\bf x}',\omega){\rm d}{\bf x}'.\nonumber\\
&&\nonumber\hspace{4.3cm}\mbox{($D$--11)}
\end{eqnarray}
Applying the spatial Fourier transformation to equation ($D$-3) we obtain
\begin{equation}
\tilde F({\bf k}_{\rm H},z_0,{\bf x}_{\rm H}',z_0,\omega)=\exp(-i{\bf k}_{\rm H}\cdot{\bf x}_{\rm H}').\nonumber\eqno{(D\mbox{--}12)}
\end{equation}
\rev{Substituting} this into equation ($D$-11) \rev{and using equation ($D$-8) to define the spatial Fourier transforms of $\hat a$ and $\hat b$} gives
\begin{equation}
\tilde v_z({\bf k}_{\rm H},z_0,\omega)
=\frac{-k_z}{\omega\rho_0}\tilde a({\bf k}_{\rm H},z_0,\omega)+\frac{k_z^*}{\omega\rho_0}\tilde b({\bf k}_{\rm H},z_0,\omega).\nonumber\eqno{(D\mbox{--}13)}
\end{equation}
Equation ($D$-13) can be combined with the spatial Fourier transform of equation ($D$-6) into the following matrix-vector equation
\begin{equation}
\begin{pmatrix}\tilde p({\bf k}_{\rm H},z_0,\omega)\\ \tilde v_z({\bf k}_{\rm H},z_0,\omega) \end{pmatrix}=
\begin{pmatrix} 1 & 1 \\ \frac{k_z^*}{\omega\rho_0} & -\frac{k_z}{\omega\rho_0} \end{pmatrix}
\begin{pmatrix}\tilde b({\bf k}_{\rm H},z_0,\omega)\\ \tilde a({\bf k}_{\rm H},z_0,\omega) \end{pmatrix}.\nonumber\eqno{(D\mbox{--}14)}
\end{equation}
For propagating waves, i.e., for ${\bf k}_{\rm H}\cdot{\bf k}_{\rm H}\le\frac{\omega^2}{c_0^2}$, the vertical wavenumber  $k_z$ is real-valued, see equation ($D$-10).
Hence, for propagating waves at depth $z_0$ 
we may replace $k_z^*$ by $k_z$ in equation ($D$-14). We then recognize this equation as the well-known expression for wave field composition at depth $z_0$
\citep{Corones75JMAA, Ursin83GEO, Fishman84JMP}, with 
$\tilde b({\bf k}_{\rm H},z_0,\omega)=\tilde p^+({\bf k}_{\rm H},z_0,\omega)$ and $\tilde a({\bf k}_{\rm H},z_0,\omega)=\tilde p^-({\bf k}_{\rm H},z_0,\omega)$, 
where the superscripts $+$ and $-$ refer to downward and upward propagation.
For evanescent waves at depth $z_0$, i.e., for ${\bf k}_{\rm H}\cdot{\bf k}_{\rm H}>\frac{\omega^2}{c_0^2}$, 
\rev{we have $k_z^*=-k_z$, hence,} 
this interpretation of equation ($D$-14) breaks down.
\rev{However, assuming evanescent waves are negligible at depth $z_0$, we may extend the relations 
$\tilde b({\bf k}_{\rm H},z_0,\omega)=\tilde p^+({\bf k}_{\rm H},z_0,\omega)$ and $\tilde a({\bf k}_{\rm H},z_0,\omega)=\tilde p^-({\bf k}_{\rm H},z_0,\omega)$ to all ${\bf k}_{\rm H}$. We thus}
obtain in the space-frequency domain
$\hat b({\bf x}_{\rm H}',z_0,\omega)=\hat p^+({\bf x}_{\rm H}',z_0,\omega)$ and $\hat a({\bf x}_{\rm H}',z_0,\omega)=\hat p^-({\bf x}_{\rm H}',z_0,\omega)$.
Substituting this into equation ($D$-4) yields
\begin{eqnarray}
\hat p({\bf x},\omega)&=&\int_{\mathbb{S}_0} \hat F({\bf x},{\bf x}',\omega)\hat p^-({\bf x}',\omega){\rm d}{\bf x}'\nonumber\\
&+&\int_{\mathbb{S}_0} \hat F^*({\bf x},{\bf x}',\omega)\hat p^+({\bf x}',\omega){\rm d}{\bf x}',\nonumber\hspace{1.1cm}\mbox{($D$--15)}
\end{eqnarray}
for all ${\bf x}$ below the source distribution.
Equation ($D$-15) describes extrapolation of the wave field from $\mathbb{S}_0$ to any point ${\bf x}$ below the source distribution.
Since the source distribution is restricted to the upper half-space, equation ($D$-15) 
holds for the entire lower half-space and for a part of the upper half-space below the shallowest source.
\rev{The lower half-space may in general be inhomogeneous, with propagation velocity $c({\bf x})$ and mass density $\rho({\bf x})$
(we used the one-way wave equation ($D$-9) in the wavenumber-frequency domain only at $\mathbb{S}_0$, 
where the medium is laterally invariant,
to derive that $\hat a$ and $\hat b$ in equations ($D$-4) and ($D$-5) can be replaced by $\hat p^-$ and $\hat p^+$).
Finally, transforming equation ($D$-15)}
back to the space-time domain yields equation (\ref{eq43}) for all ${\bf x}$ below the shallowest source.

\end{spacing}
\end{document}